\documentclass[longauth]{aa} 

\newcommand\hmmax{0}
\newcommand\bmmax{0}

\usepackage{graphicx}
\usepackage{txfonts}
\usepackage[colorlinks=true,
            linkcolor=blue,
            urlcolor=blue,
            citecolor=blue]{hyperref}
\usepackage{booktabs}
\usepackage[flushleft]{threeparttable}
\usepackage{amsmath}
\usepackage{microtype}
\usepackage{upgreek}
\usepackage{textcomp}
\usepackage{url}
\usepackage{bm}

\makeatletter
\renewcommand*\aa@pageof{, page \thepage{} of \pageref*{LastPage}}
\makeatother

\def\ptrad{\texttt{petitRADTRANS}\xspace}
\def\prt{\texttt{pRT}\xspace}

\def\star{_*}

\def\f1{f_{\rm I}}

\def\Tbot{\ensuremath{T_{\rm bot}}\xspace}
\def\kzz{\ensuremath{K_{\rm zz}}\xspace}

\def\water{H$_{2}$O\xspace}
\def\methane{CH$_{4}$\xspace}
\def\ammonia{NH$_{3}$\xspace}
\def\cotwo{CO$_{2}$\xspace}

\def\htwos{H$_{2}$S\xspace}

\def\beq{\begin{equation}}
\def\eeq{\end{equation}}

\def\t2{\tau_{\rm II}}

\def\sigmas0{\Sigma_{\rm s,0}}

\def\({\left(}
\def\){\right)}
\def\<{\left<}
\def\>{\right>}

\defcitealias{ackerman_precipitating_2001}{AM01}
\defcitealias{molliere_retrieving_2020}{M20}
\defcitealias{zhang_elemental_2023}{Z23}
\defcitealias{guillot_radiative_2010}{G10}
\defcitealias{mccarthy_simp_2025}{M25}

\begin{document} 

   \title{The JWST weather report: Retrieving temperature variations, auroral heating, and static cloud coverage on SIMP-0136}
    \titlerunning{The JWST weather report for SIMP-0136}
    \authorrunning{Nasedkin et al.}
   \subtitle{}

   \author{E.~Nasedkin\inst{\ref{tcd},\ref{mpia}}\fnmsep\thanks{Corresponding Author, \href{mailto:nasedkin@tcd.ie}{nasedkin@tcd.ie}}
   \and M. Schrader\inst{\ref{tcd}}
   \and J. M. Vos\inst{\ref{tcd}}
   \and B. Biller\inst{\ref{edinburgh}, \ref{CES}}
   \and B. Burningham\inst{\ref{hertfordshire}}
   \and N. B. Cowan\inst{\ref{mcgill}}
   \and J. K. Faherty\inst{\ref{AMNH}}
   \and E. Gonzales\inst{\ref{SFSU}}
   \and M. B. Lam\inst{\ref{tcd}}
   \and A. M. McCarthy\inst{\ref{boston}}
   \and P. S. Muirhead\inst{\ref{boston}}
   \and C. O'Toole\inst{\ref{tcd}}
   \and M. K. Plummer\inst{\ref{AFA}}
   \and G. Su\'{a}rez\inst{\ref{AMNH}}
   \and X. Tan\inst{\ref{shanghai}}
   \and C. Visscher\inst{\ref{dordt},\ref{boulder}}
   \and N. Whiteford\inst{\ref{AMNH}}
   \and Y. Zhou\inst{\ref{virginia}}
 }
\institute{ 
   School of Physics, Trinity College Dublin, The University of Dublin, Dublin 2, Ireland
\label{tcd}     \and
   Max-Planck-Institut f\"ur Astronomie, K\"onigstuhl 17, 69117 Heidelberg, Germany
\label{mpia}     \and
    Institute for Astronomy, University of Edinburgh, Royal Observatory, Edinburgh EH9 3HJ,UK
\label{edinburgh}\and
    Centre for Exoplanet Science, University of Edinburgh, Edinburgh, UK
\label{CES}\and
    Centre for Astrophysics Research, Department of Physics, Astronomy and Mathematics, University of Hertfordshire, Hatfield AL10 9AB, UK
\label{hertfordshire}\and
    Department of Physics and Department of Earth \& Planetary Sciences, McGill University, Montréal, QC H3A 2A7, Canada
\label{mcgill} \and
    Department of Astrophysics, American Museum of Natural History, New York, NY 10024, USA
\label{AMNH} \and
    Department of Physics and Astronomy, San Francisco State University, 1600 Holloway Ave., San Francisco, CA 94132, USA
\label{SFSU} \and
    Department of Astronomy \& The Institute for Astrophysical Research, Boston University, 725 Commonwealth Avenue, Boston, MA 02215, USA
\label{boston} \and
    Department of Physics and Meteorology, United States Air Force Academy, 2354 Fairchild Drive, CO 80840, USA
\label{AFA} \and
    Tsung-Dao Lee Institute \& School of Physics and Astronomy, Shanghai Jiao Tong University, Shanghai 201210, People's Republic of China
\label{shanghai}\and
    Chemistry \& Planetary Sciences, Dordt University, Sioux Center IA 51250
\label{dordt} \and
    Center for Extrasolar Planetary Systems, Space Science Institute, Boulder, CO 80301
\label{boulder} \and
    University of Virginia, 530 McCormick Road, Charlottesville, VA 22904, USA
\label{virginia}
  }

   \date{Received 02-05-2025; accepted 30-07-2025}

  \abstract{
    SIMP-0136 is a T2.5 brown dwarf whose young age ($200\pm50$~Myr) and low mass ($15\pm3$~M$_{\rm Jup}$) make it an ideal analogue for the directly imaged exoplanet population.
    With a 2.4 hour period, it is known to be variable in both the infrared (IR) and the radio, which has been attributed to changes in the cloud coverage and the presence of an aurora, respectively.
    To quantify the changes in the atmospheric state that drive this variability, we obtained time-series spectra of SIMP-0136 covering one full rotation with both NIRSpec/PRISM and the MIRI/LRS on board JWST.
    We performed a series of time-resolved atmospheric retrievals using \texttt{petitRADTRANS}  to measure changes in the temperature structure, chemistry, and cloudiness.
    We inferred the presence of a ${\sim}250$ K thermal inversion above 10 mbar of SIMP-0136 at all phases and we propose that this inversion is due to the deposition of energy into the upper atmosphere by an aurora.
    Statistical tests were performed to determine which parameters were driving the observed spectroscopic variability.
    The primary contribution was due to changes in the temperature profile at pressures deeper than 10 mbar, which resulted in variation of the effective temperature from 1243 K to 1248 K. 
    This changing effective temperature was also correlated to observed changes in the abundances of \cotwo and \htwos, while all other chemical species were consistent with being homogeneous throughout the atmosphere.
    Patchy silicate clouds were required to fit the observed spectra, but the cloud properties were not found to systematically vary with longitude.
    This work paints a portrait of an L-T transition object, where the primary variability mechanisms are magnetic and thermodynamic in nature, rather than due to inhomogeneous cloud coverage. 
      }
   

   \maketitle

\section{Introduction}
\begin{table*}[t]
    \centering
    \caption{Observation log.}\label{tab:observations}
    \begin{tabular}{l|llllllll}
        \toprule
        \textbf{Instrument} & \textbf{Subarray} & \textbf{Readout}& \textbf{Mode} & \textbf{Start Time [UTC]}  & \textbf{DIT [s]} & \textbf{Groups} & \textbf{Int.} & \textbf{Exp.}\\
        \midrule
        NIRSpec/PRISM   & \texttt{SUB512} & \texttt{NRSRAPID} & BOTS &  23 July 2023 18:40:56  &  1.8 & 5726 & 1 & 1 \\
        MIRI/LRS  & \texttt{SLITLESSPRISM} & \texttt{FASTR1} & ...  &  23 July 2023 22:05:11  & 19.2 & 575 & 1 & 1 \\
        \bottomrule
    \end{tabular}
    
    \vspace{0.5em}
    \raggedright
    Data were acquired as part of GO 3548 (PI: Vos), and originally published in \citetalias{mccarthy_simp_2025}.
\end{table*}

The broad wavelength coverage and superb stability of \textit{JWST} has enabled transformative studies of the dynamic processes in brown dwarf and exoplanet atmospheres \citep{biller_weather_2024,mccarthy_simp_2025}. 
In the absence of stellar contamination present when studying their young, directly imaged exoplanet cousins, isolated brown dwarfs can be observed with exquisite spectrophotometric precision.
For nearby brown dwarfs, this allows us to measure changes in their emission spectrum over time.
This variability is due to changes in brightness over the 2D emitting surface of the object, which are thought to be caused by changing cloud coverage, thermal structure, chemistry, and the presence of upper-atmosphere aurorae \citep{radigan_brightness_2014,robinson_temperature_2014,tremblin_rotational_2020,tan_atmospheric_2021,faherty_emission_2024}.
\textit{Spitzer} and \textit{Hubble} observed light curves for dozens of brown dwarfs, finding that the variability amplitude depended on the spectral type, age, and viewing geometry \citep[e.g.][]{metchev_wow_2015,yang_extrasolarstorms_2016,vos_viewing_2017,lew_cloudatlas_2020,vos_greatworlds_2022}.
This variability monitoring enabled the first steps towards three-dimensional (3D) studies of substellar atmospheres, allowing for the identification of bands and spots \citep{apai_zonesbands_2017}, as well as pressure-dependent weather \citep{apai_hst_2013,yang_extrasolarstorms_2016}. 
However, it is only in the era of \textit{JWST} that broad wavelength, spectroscopic light curves have been observed, allowing us to relate the dynamic atmospheric processes to changes across the entire emission spectrum \citep{biller_weather_2024,mccarthy_simp_2025, chen_variability_2025}.

\object{SIMP J013656.5+093347.3} (hereafter SIMP-0136) is a nearby \citep[6.12$\pm$0.02 pc,][]{gaia_edr3_2020}, well-studied exoplanet analogue, ideally suited for variability studies due to its brightness, large variability amplitude of about 2.5\%, and 2.4 hour rotation period \citep{artigau_discovery_2006,artigau_photometric_2009}.
With a spectral type of T2.5$\pm$0.5 \citep{artigau_discovery_2006} and an effective temperature of around 1100K \citep{zhang_coconuts_2020a}, SIMP-0136 shares many similar atmospheric properties to directly imaged exoplanets, such as HR 8799 b or AF Lep b \citep[][]{marois_direct_2008,balmer_aflep_2025}.
As a member of the Carina-Near Association, SIMP-0136 is estimated to be 200$\pm$50 Myr old \citep{gagne_simp_2017}.
Likewise, its relatively low surface gravity also implies a relatively youthful object, with $\log g = 4.46\pm0.09$ \citep{zhang_coconuts_2020a}.
Its mass has been estimated to range from $12.7\pm1$ M$_{\rm Jup}$ \citep{gagne_banyan_2018} to $17.8 \pm 11.9$ M$_{\rm Jup}$ \citep{vos_patchy_2023}.
Its inclination was measured by \cite{vos_viewing_2017} to be nearly equator-on, with $i = 80_{-12}^{+10}\,^{\circ}$.
In the most in-depth atmospheric analysis of SIMP-0136 to-date, \cite{vos_patchy_2023} performed atmospheric retrievals on a broad wavelength spectrum consisting of Spex PRISM \citep{burgasser_subtle_2008}, Akari \citep{sorahana_akari_2012}, and Spitzer/IRS \citep{filipazzo_irs_2015} data. 
They identified layers of patchy silicate and iron clouds and robustly inferred the atmospheric temperature structure and composition.
The presence of an aurora has been identified as a mechanism to explain the observations of pulsed radio emission \citep{kao_aurora_2016,kao_aurora_2018}.
The interaction of the aurora with the upper atmosphere has also been suggested to contribute to the observed near-infrared (NIR) variability \citep{morley_spectral_2014,mccarthy_simp_2025}.

In the NIR, the variability amplitude is mildly wavelength-dependent, with a weaker amplitude in the 1.4 $\upmu$m water absorption feature \citep{apai_hst_2013,lew_cloudatlas_2020}. 
\cite{mccarthy_multiple_2024} found a phase shift of $39.9^{+3.6\,\circ}_{-1.1}$ between the $J$ and $Ks$ band light curves.
\cite{yang_extrasolarstorms_2016} demonstrated that the different wavelength ranges correspond to different pressures probed by the emission spectrum.
Both groups attributed the chromatic variability effects to differing cloud coverage at different levels of the atmosphere, while \cite{apai_zonesbands_2017} and \cite{plummer_waves_2024} determined that planetary scale (e.g. Kelvin or Rossby) waves are likely the primary driver of the variability.
Recently, \cite{mccarthy_simp_2025} (hereafter \citetalias{mccarthy_simp_2025}) published the first \textit{JWST} observations of SIMP-0136, from 0.8-11 $\upmu$m using NIRSpec/PRISM and MIRI/LRS.
The authors proposed that the distinct variability patterns could be tied to variations in the cloud patchiness, high-altitude hotspots, and changes in the carbon chemistry.

Given this observed wavelength dependent variability, it is crucial to quantify how changes in the atmospheric state drive the observed changes in the emission spectrum.
In the era of \textit{JWST}, atmospheric retrievals \citep[e.g.][]{madhusudhan_seager_2009,benneke_retrieval_2012,molliere_petitradtrans_2019,hood_fire_2023,grant_quartz_2023,kuhnle_depletion_2024, hoch_silicates_2025} have become one of the primary tools for characterising the atmospheres of exoplanets and brown dwarfs.
Atmospheric retrievals typically use a 1D parametric forward model to fit the observed spectrum.
They are highly dependent on the wavelength coverage of the data, as different sources of opacity only impact the spectrum at particular wavelengths.
\cite{burningham_cloud_2021} and \cite{vos_greatworlds_2022} demonstrated the importance of broad wavelength coverage, combining NIR and MIR observations to robustly characterise silicate clouds.

To date, most retrieval studies have treated extrasolar objects as one-dimensional (1D).
However, from 3D general circulation models (GCMs), it is clear that exoplanets and brown dwarfs vary across their surface and as a function of depth in the atmosphere.  \cite{blecic_implication_2017} highlighted the implications of this for atmospheric retrievals of transiting planets, showing that at best retrievals will find the average thermal structure of the object.  
Some attempts have been made to extend retrieval frameworks to 3D for transiting planets, which exhibit strong day-to-night contrasts due to the incident stellar radiation \citep[e.g.][]{macdonald_trident_2022,dobbsdixon_gcmomotivated_2022}.
Without any incident stellar light, isolated brown dwarfs do not exhibit such diurnal contrasts and the variation in the emission spectrum across the surface is typically small, of the order of a few percent at most \citep{showman_atmospheric_2013,tan_atmospheric_2021, lee_dynamicallII_2024}.
In this work, we extended the \ptrad retrieval framework \citep{molliere_petitradtrans_2019,nasedkin_atmospheric_2024} to perform time-resolved retrievals of emission spectra, similar to \cite{gandhi_supjupv_2025},  to tie the atmospheric properties to the observed variability.

This study is structured as follows.
In Section \ref{sec:observations}, we  outline the JWST observations, as well as our novel approach to processing the time-series data.
Section \ref{sec:methods} presents the details of our retrieval framework, presenting the parameterisations used for the thermal structure, chemistry, and clouds.
We present the results of this study in Section \ref{sec:results}, highlighting the measured atmospheric properties of SIMP-0136 and how they vary over time.
These results are placed in the context of existing theoretical models and \textit{JWST} observations in Section \ref{sec:discussion} and the conclusions are summarised in Section \ref{sec:conclusions}.
Additional information is included in the appendices, including a comparison of SIMP-0136 to self-consistent forward models and evolutionary models (Appendix \ref{app:gridfits}), an in-depth demonstration of the sensitivity of the retrievals to various mechanisms driving the variability (Appendix \ref{app:forward}) and additional results from the retrieval analysis (Appendix \ref{app:retrievals}).

\section{Observations}\label{sec:observations}

\begin{figure*}[t]
    \centering
    \includegraphics[width=0.47\linewidth]{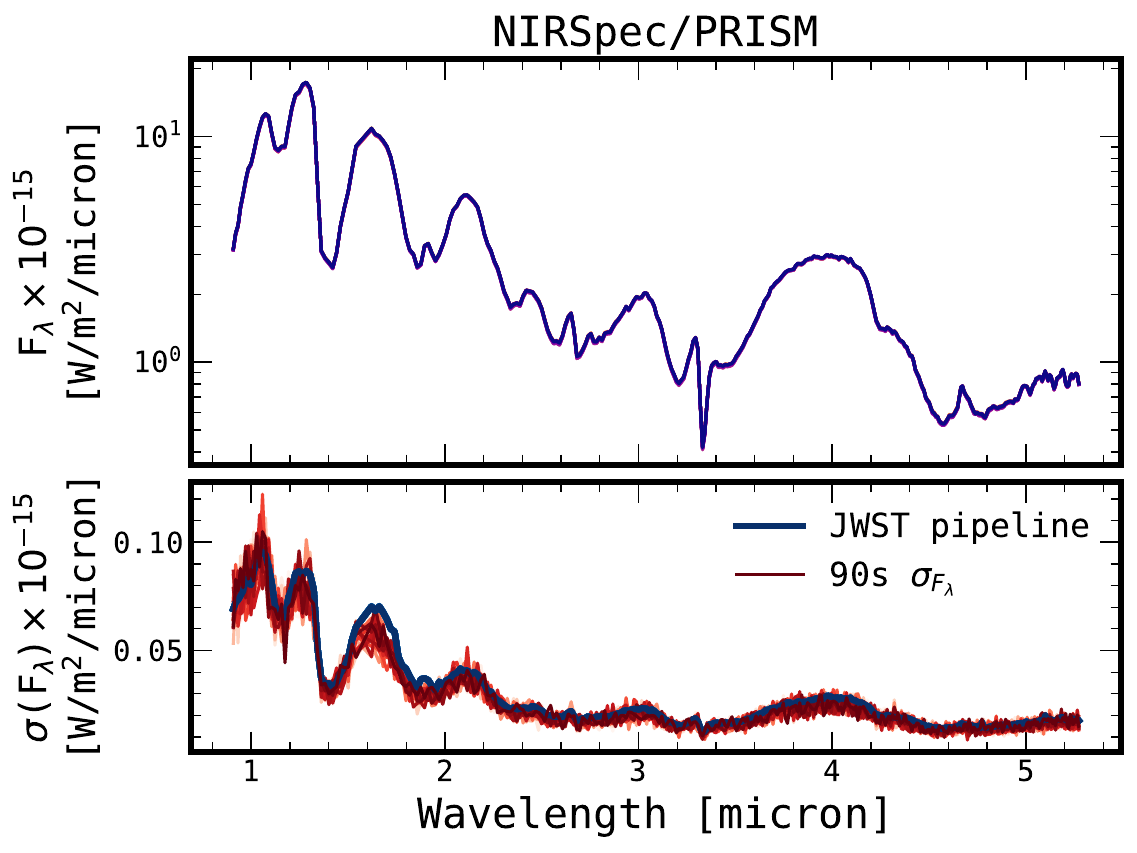}
    \includegraphics[width=0.438\linewidth]{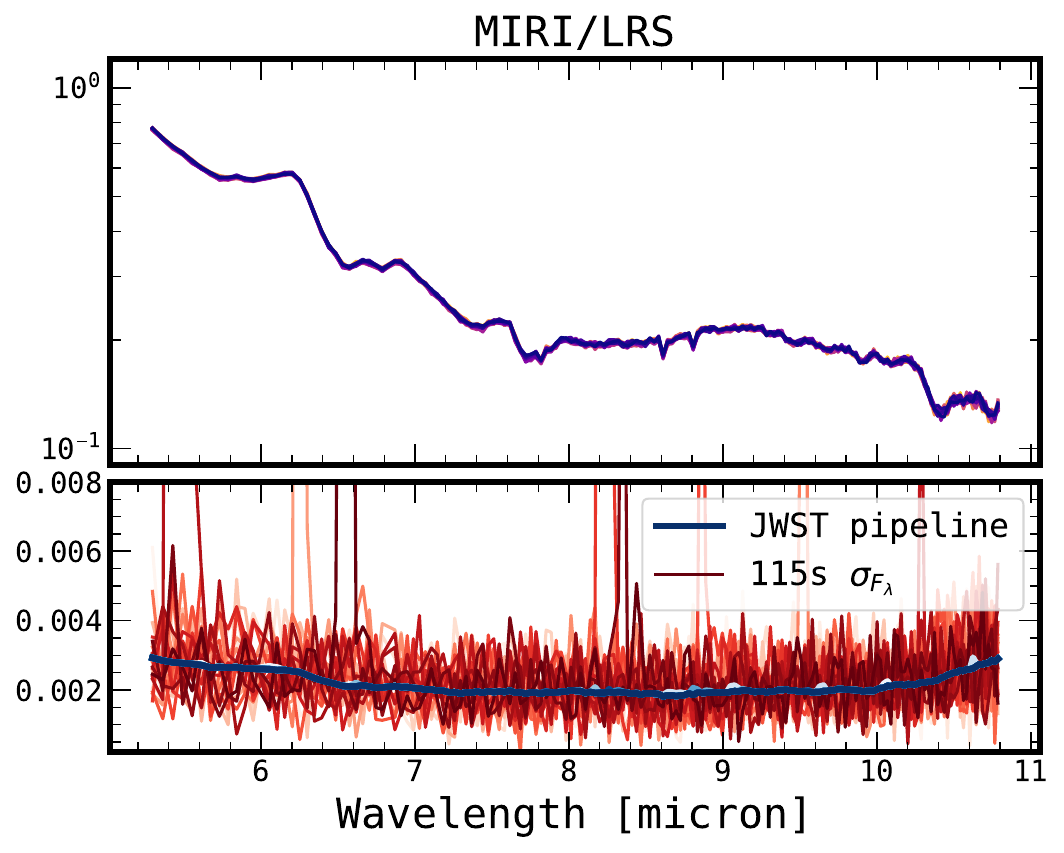}\\
    \hspace{0.95cm}\includegraphics[width=0.91\linewidth]{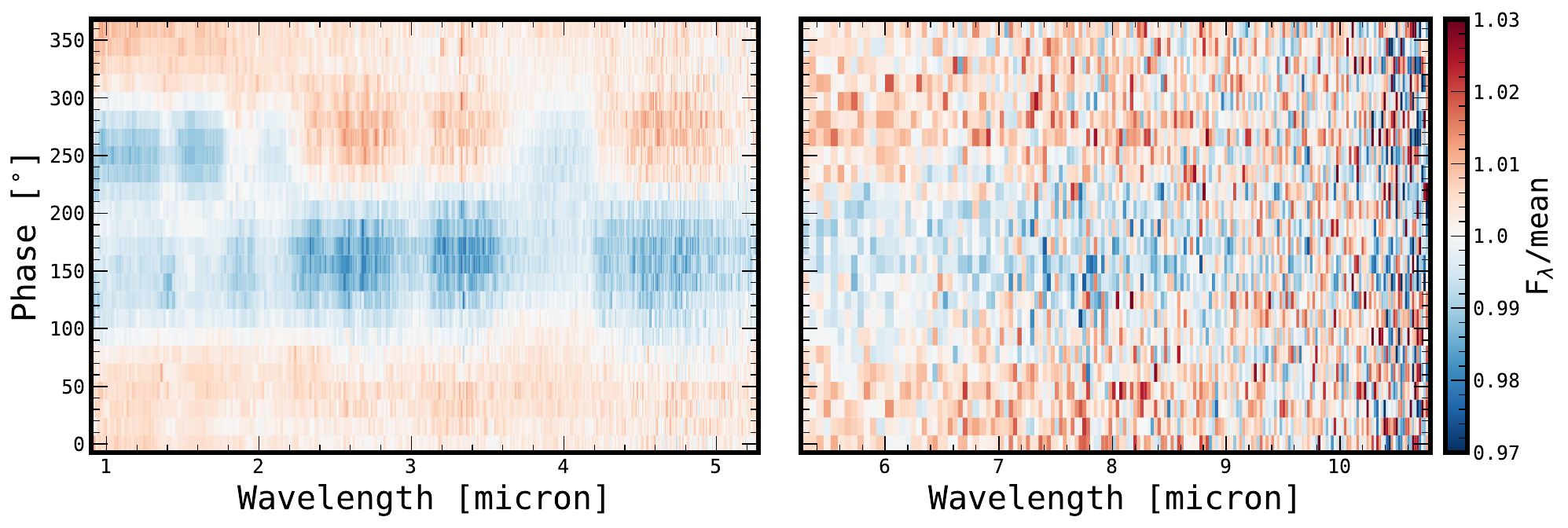}
    \caption{NIRSpec/PRISM (left) and MIRI/LRS (right)  observations of SIMP-0136. \textbf{Top:} Emission spectra in 15$^{\circ}$ rotation bins.  Each binned NIRSpec spectrum is averaged over 90~s of integration, while each LRS spectrum is averaged over 38~s integrations to ensure that the overall signal to noise of the input spectra for the retrievals is proportional to the emitted flux. \textbf{Centre:}  Uncertainties calculated by the JWST pipeline for each spectrum in blue. Standard deviation of the 90s (115s for the LRS) intervals  in red, demonstrating that the pipeline uncertainties accurately reflect the underlying statistical distribution of the noise. \textbf{Bottom:} Binned variability maps, also known as dynamic spectra. These variability maps are aligned in phase, as the MIRI observations were taken after the NIRSpec observations and are binned to highlight the wavelength dependence of the variability. The 24 phase bins used in these maps correspond to the binned spectra used as inputs for the retrievals. }
    \label{fig:binned_spectra_nirspec}
\end{figure*}

We obtained time-resolved spectra of SIMP-0136 using NIRSpec/PRISM and the MIRI/LRS as part of GO 3548 (PI: Vos).
These were originally presented in \citetalias{mccarthy_simp_2025}, and the observing setup is summarised in Table \ref{tab:observations}.
A total of 5726 spectra were obtained with NIRSpec/PRISM \citep{jakobsen_nirspec_2022} in bright object time-series (BOTS) mode. These observations were made with a cadence of 1.8\,s spanning a total of 3.4 hours, or 1.4 rotations \citep[P=2.4 hours,][]{yang_extrasolarstorms_2016}.
These spectra have a spectral resolving power ranging from $R=30-400$, and cover 0.6-5.3 $\upmu$m, and are shown in Fig.\ref{fig:binned_spectra_nirspec}.
The MIRI/LRS \citep{kendrew_lrs_2015} observations were taken subsequently and 575 spectra were obtained with a cadence of 19.2 s, over a total of 3.55 hours of observations. 
The LRS observations have a spectral resolving power varying from $R=40-160$ across the wavelength range from 5-14 $\upmu$m  (also presented in Fig. \ref{fig:binned_spectra_nirspec}).
Similar to Fig. 4 of \citetalias{mccarthy_simp_2025} we phase-fold the LRS spectrum to align with the NIRSpec observations to produce a complete spectrum that varies with the rotation of SIMP-0136.

\subsection{Data reduction}
The \textit{JWST} NIRSpec/PRISM data were processed using the standard stage 1 of the JWST pipeline \citep[v1.16.1,][]{Bushouse_JWST_Calibration_Pipeline_2025} for the bright-object time series (BOTS) mode and a customised version of the stage 2 (\texttt{calwebb\_detector2}) pipeline, which was optimised for time-series spectral extraction and the mitigation of systematic noise sources. The reduction used CRDS version 12.0.6 with the \texttt{jwst\_1298.pmap} context file. The output of our modified pipeline was the flux-calibrated, time-series spectra. 

Unlike \cite{mccarthy_simp_2025}, who used the default stage 2 reduction, we modified stage 2 to process each integration individually, resulting in better correction of systematics.
By default, the BOTS reduction does not apply the \texttt{nsclean} step to correct for $1/f$ noise, which manifests as low-frequency spatial structure in the detector readout \citep{rauscher_nsclean_2024}. Since this correction was applied to fixed-slit observations, we modified the \texttt{EXP\_TYPE} metadata from \texttt{NRS\_BOTS} to \texttt{NRS\_FIXEDSLIT} and applied \texttt{nsclean} to each integration.
The standard stage 2 BOTS reduction assumes a fixed background and detector properties and applies single correction factors to each. 
To avoid these biases, we applied flat-field and photometric calibrations to each spectrum individually.
By default, the pipeline assumes one fixed extraction aperture and spectral trace across all integrations, which does not account for time-dependent shifts due to optical distortions, telescope jitter, or variations in prism dispersion. 
These shifts can introduce flux losses and wavelength misalignment if a static extraction aperture is used, and so our modified pipeline automatically determines a spectral trace for each integration. 
We fit a first-order polynomial to the trace positions across the detector columns for each individual exposure, improving both wavelength calibration and flux accuracy.
This corrects for systematics introduced by time variable detector drifts, pointing variations, and changes in pixel sensitivity.
We found that these improvements resolved systematic flux variations between integrations, independent of wavelength. These are likely caused by systematic detector effects such as small gain or bias drifts, charge trapping, or variations in readout electronics. The default pipeline's application of a static flat-field correction, as well as a uniform spectral trace, can introduce artificial time-dependent offsets if pixel sensitivity fluctuates over the course of the observation.
With the spectra extracted, we removed low signal-to-noise ($S/N$) data from wavelengths shorter than 0.9 $\upmu$m and longer than 5.3 $\upmu$m. 
Three wavelength bins near 1.58 $\upmu$m were contaminated by a hot pixel, and were also removed. 
We binned the time-series spectra to highlight the astrophysical variation of SIMP-0136 over its rotation. 
The bins were centred at every 15$^{\circ}$ of rotation, and combined 25 spectra on either side of these bin centres, for a total of 90 s of integration time into each bin.
These binned integrations both improve the $S/N$ by a factor of $\sqrt{50}$ compared to a single exposure, but are also sufficiently short to prevent the blending of the spectrum as SIMP-0136 will only rotate by 3.75$^{\circ}$ in 90 s.

The MIRI/LRS observations were reduced using v1.16.1 of the \textit{JWST} pipeline using the settings described in \citetalias{mccarthy_simp_2025}.
As with the PRISM data, we removed low $S/N$ data at wavelengths shorter than 5.3 $\upmu$m and longer than 10.8 $\upmu$m. 
This range also ensures that there is no overlap with the NIRSpec data and it extends out to cover the 10 $\upmu$m ammonia absorption feature.
Additional wavelength bins near 6.0 $\upmu$m  and 6.68 $\upmu$m were also removed due to poor hot-pixel correction.
While we excluded data below 5.3 $\upmu$m, the LRS spectrum was offset by an average of 2\% below the PRISM spectrum between 5.1 and 5.3 $\upmu$m. This is compatible to within the PRISM uncertainties, but the three overlapping LRS data points range between fully compatible to offset by 10$\sigma$. 
The data were binned similarly to the PRISM data; however, only two integrations were combined for the LRS, for a total of 38 s of integration time. This produces a $S/N$ across the spectrum that is proportional to the emitted flux and matches the $S/N$ at 5.3 $\upmu$m in the binned PRISM spectrum. These bins were aligned such that they were in phase with the NIRSpec bins.
The mean-normalised variability maps for both NIRSpec and the LRS are presented in Fig. \ref{fig:binned_spectra_nirspec}.

\begin{table}
\caption{SIMP-0136 properties and photometry.}
\vspace{-1.5em}
\begin{footnotesize}
\begin{center}
\begin{tabular}{l|ll}
\toprule
\textbf{Property}       &  \textbf{SIMP-0136}  & \textbf{Reference} \\
\midrule
RA                      & 01:36:57                &  C03       \\
Dec                     & 09:33:47                &  C03       \\
$\mu\alpha$ cos $\delta$ (mas yr$^{-1}$)&  $1238.982\pm1.189$   & C03, G18    \\
$\mu\delta$ (mas yr$^{-1}$) & $-17.353 \pm 0.841$ & S13, G18   \\
RV (km s$^{-1}$)               &   $12.3 \pm 0.8$ & V17        \\
Spectral Type  (Opt/IR) &      T2/T2.5            & B06, A06 \\
Parallax (mas)    &     $163.45\pm0.46$           &  GDR3      \\
Distance (pc) & $6.12\pm0.02$ & GDR3      \\
Rotation Period (hr)    & $2.414\pm0.078$         & Y16        \\
$v\sin i$ (km s$^{-1}$)  &$49.8_{-0.5}^{+0.5}  $   & H21        \\ 
$i$ ($^{\circ}$) & $80_{-12}^{+10}\,^{\circ}$ & V17\\
\midrule
\multicolumn{3}{l}{Fundamental Parameters (\texttt{SEDkit})}\\
\midrule
$\log\, $L$_{\rm bol}$ (L$_{\odot}$)   & $-4.641\pm0.003$  & This work    \\
T$_{\mathrm{eff}}$ (K)    & $1136\pm22$              &  ---   \\
Radius $(R_{\mathrm{Jup}})$ & $1.20\pm0.05$  &  ---   \\
Mass  $(M_{\mathrm{Jup}})$  & $15\pm3$             &   ---  \\
$\log g$  (cm s$^{-2}$)     & $4.4 \pm 0.1$              &   ---  \\
\midrule
\multicolumn{3}{l}{Fundamental Parameters (\texttt{pRT})}\\
\midrule
$\log\, $L$_{\rm bol}$ (L$_{\odot}$)   & $-4.715\pm0.002$  & ---  \\
T$_{\mathrm{eff}}$ (K)    & $1245\pm1.4$             &   ---  \\
Radius $(R_{\mathrm{Jup}})$ & $0.933 \pm 0.005$  &  ---   \\
Mass  $(M_{\mathrm{Jup}})$  & $10.38\pm0.41$             &   ---  \\
$\log g$  (cm s$^{-2}$)     & $4.49 \pm 0.02$              &  ---   \\
$\left[\mathrm{M/H}\right]$ & $0.18 \pm 0.01$&--- \\
C/O & $0.65 \pm 0.01$& ---\\
\bottomrule
\end{tabular}
\end{center}
\end{footnotesize}
\vspace{-0.5em}
 \begin{tablenotes}
    \item \textbf{Notes:} Uncertainties on fundamental parameters include astrophysical variability. References: C03: \citet{cutri_vizier_2003}; G18: \citet{gaia_collaboration_gaia_2018};  S13: \citet{smart_nparsec_2013}; V17: \citet{vos_viewing_2017}; Y16: \citet{yang_extrasolarstorms_2016}; Z21: \citet{zhang_hawaiiV_2021}; P16: \citet{pineda_ha_2016}; B06: \citet{burgasser_method_2006}; A06: \citet{artigau_discovery_2006}; H21: \citet{hsu_kinematics_2021}.
\end{tablenotes}
\label{tab:props}
\end{table}

\subsection{Uncertainty estimation}
The high time resolution of the observations allowed us to validate the uncertainties produced by the \textit{JWST} pipeline.
As shown in Fig \ref{fig:binned_spectra_nirspec}, we grouped the time-series spectra into bins of 90 s for NIRSpec and 115 s for the LRS, with the same 15$^{\circ}$ cadence as the binned spectra, taking the standard deviation of the spectra within these bins. 
This timescale is short enough to ensure only small astrophysical changes in the spectra and, thus,  the dominant variation between the samples is random noise.
We compared these measurements of the random noise to the uncertainties estimated by the \textit{JWST} pipeline and found that this estimate closely matches the measured noise as shown in the centre panels of Fig. \ref{fig:binned_spectra_nirspec}.

In contrast to known under-estimation of uncertainties with the MIRI/MRS \citep[e.g.][]{miles_jwst_2023,patapis_geometric_2024,matthews_hcnc2h2_2025}, the uncertainties associated with NIRSpec/PRISM and MIRI/LRS time-series data accurately reflect the underlying noise distribution.
As the noise is normally distributed, we can use standard error analysis to combine $N$ integrations, reducing the uncertainty by a factor of $1/\sqrt{N}$.
In addition to the statistical uncertainties, both NIRSpec/PRISM and the LRS have systematic uncertainties in the absolute flux calibration. The PRISM spectrum is expected to have an absolute flux calibration accurate to within 5\%, while the LRS spectrum calibration can vary between 3-8\% \citep{jdox_2016, gordon_jwstcalplan_2022}. We did not account for these systematic biases in our uncertainty estimates, although we note that offsets in the absolute flux will lead to biased measurements, particularly in terms of the luminosity and effective temperature.

\section{Atmospheric modelling}\label{sec:methods}
\begin{table}
    \caption{Priors used for the fiducial retrieval setup and median retrieved parameter values across the full rotation of SIMP-0136.}
    \label{tab:priors}
    \centering
    \begin{small}
    \begin{tabular}{l|ll}
    \toprule
    \textbf{Parameter} & \textbf{Prior} & \textbf{Median}\\
    \midrule
    T$_{\rm bot}$ [K]& $\mathcal{U}\left(2000, 12000\right)$ & $5474 \pm 928$\\
    $\nabla T_{\rm 1}$ & $\mathcal{N}\left(0.25, 0.1\right)$ & $0.26 \pm 0.07$\\
    $\nabla T_{\rm 2}$ & $\mathcal{N}\left(0.25, 0.1\right)$ & $0.24 \pm 0.07$\\
    $\nabla T_{\rm 3}$ & $\mathcal{N}\left(0.26, 0.1\right)$ & $0.19 \pm 0.01$\\
    $\nabla T_{\rm 4}$ & $\mathcal{N}\left(0.2, 0.1\right)$ & $0.207 \pm 0.002$\\
    $\nabla T_{\rm 5}$ & $\mathcal{N}\left(0.12, 0.1\right)$ &  $0.150 \pm 0.004$\\
    $\nabla T_{\rm 6}$ & $\mathcal{N}\left(0.07, 0.1\right)$ & $-0.11 \pm 0.02$\\
    $\nabla T_{\rm 7}$ & $\mathcal{N}\left(0.0, 0.5\right)$ & $0.04 \pm 0.03$\\
    $\nabla T_{\rm 8}$ & $\mathcal{N}\left(0.0, 0.5\right)$ & $-0.10 \pm 0.04$\\
    $\nabla T_{\rm 9}$ & $\mathcal{N}\left(0.0,0.05\right)$ & $0.03 \pm 0.04$\\
    $\nabla T_{\rm 10}$ & $\mathcal{N}\left(0.0,0.05\right)$ & $0.01 \pm 0.04$\\
    $\sigma_{\rm LN}$ & $\mathcal{U}\left(1.005,3\right)$ & $1.1 \pm 0.2$\\
    $\log P_{c}$(Mg$_{2}$SiO$_{4}$) & $\mathcal{U}\left(-2.0, 3.0\right)$ & $1.27 \pm 0.08$\\
    $\log P_{c}$(Fe) & $\mathcal{U}\left(-2.0, 3.0\right)$ & $0.96 \pm 0.06$\\
    $\log$(Mg$_{2}$SiO$_{4}$) & $\mathcal{U}\left(-6.0, -0.3\right)$ & $-3.0 \pm 1.2$\\
    $\log$(Fe) & $\mathcal{U}\left(-12.0, -0.3\right)$ & $-8.5 \pm 3$\\
    $f_{\rm sed}$(Mg$_{2}$SiO$_{4}$) & $\mathcal{U}\left(0, 10\right)$ & $8 \pm 1$\\
    $f_{\rm sed}$(Fe) & $\mathcal{U}\left(0, 10\right)$ & $4 \pm 3$\\
    $\log r_{c}$(Mg$_{2}$SiO$_{4}$) [cm]& $\mathcal{U}\left(-7, 3\right)$ & $-3.86 \pm 0.06$\\
    $\log r_{c}$(Fe) [cm]& $\mathcal{U}\left(-7, 3\right)$ & $0.6 \pm 3$\\
    $f_{\rm cloud}$ & $\mathcal{U}\left(0, 1\right)$ & $0.81 \pm 0.04$\\
    log H$_{2}$O & $\mathcal{U}\left(-12, -0.3\right)$ & $-2.717 \pm 0.002$\\
    log CO$_{2}$ & $\mathcal{U}\left(-12, -0.3\right)$ & $-5.647 \pm 0.005$\\
    log C$_{2}$H$_{2}$ & $\mathcal{U}\left(-12, -0.3\right)$ & $-9 \pm 1$\\
    log HCN & $\mathcal{U}\left(-12, -0.3\right)$ & $-5.73 \pm 0.06$\\
    log NH$_{3}$ & $\mathcal{U}\left(-12, -0.3\right)$ & $-5.26 \pm 0.03$\\
    log H$_{2}$S & $\mathcal{U}\left(-12, -0.3\right)$ & $-2.97 \pm 0.01$\\
    log FeH & $\mathcal{U}\left(-12, -0.3\right)$ & $-8.62 \pm 0.02$\\
    log Na & $\mathcal{U}\left(-12, -0.3\right)$ & $-4.48 \pm 0.05$\\
    log K & $\mathcal{U}\left(-12, -0.3\right)$ & $-5.67 \pm 0.07$\\
    $\log(P_{\rm CH_{4}})$ & $\mathcal{U}\left(-6, 3\right)$ &  $0.9 \pm 2$\\
    $\log$ CH$_{4}$(P$_{0}$) & $\mathcal{U}\left(-12, -0.3\right)$ & $-4.6 \pm 3$\\
    $\log$ CH$_{4}$(P$_{1}$) & $\mathcal{U}\left(-12, -0.3\right)$ & $-3.5 \pm 0.2$\\
    $\log$ CH$_{4}$(P$_{2}$) & $\mathcal{U}\left(-12, -0.3\right)$ & $-11 \pm 4$\\
    $\log$ CO(P$_{0}$) & $\mathcal{U}\left(-12, -0.3\right)$ & $-3 \pm 1$\\
    $\log$ CO(P$_{1}$) & $\mathcal{U}\left(-12, -0.3\right)$ & $-2.1 \pm 0.3$\\
    $\log$ CO(P$_{2}$) & $\mathcal{U}\left(-12, -0.3\right)$ & $-1.6 \pm 3$\\
    \bottomrule
    \end{tabular}
    \end{small}
    \begin{tablenotes}
    \item\textbf{Notes:} Uncertainties are taken from the standard deviation of the median retrieved values across all phases. $\mathcal{N}(\mu, \sigma)$ indicates a Gaussian distribution centred at $\mu$ with standard deviation $\sigma$. $\mathcal{U}(a, b)$ indicates a uniform distribution with bounds of $(a, b)$.
    \end{tablenotes}
\end{table}

To determine the mechanisms underpinning the observed variability, we must be able to accurately infer the atmospheric state as a function of time. 
We first compared the SIMP-0136 spectrum to self-consistent, radiative-convective equilibrium forward models to fit the spectrum, the results of which are discussed in Appendix \ref{app:gridfits}.
These models did not fit the spectrum sufficiently well to resolve the observed level of variability, motivating the use of a data-driven retrieval approach to infer the atmospheric properties. 
In addition, we estimated the fundamental parameters of SIMP-0136 using \texttt{SEDkit} \citep{filippazzo_fundamental_2015}, the results of which are included in Table \ref{tab:props}.
These measurements were used to provide a baseline against which to compare the results of the atmospheric retrievals. 

Overall, \ptrad (\prt) is a widely used, open source tool for computing emission and transmission spectra \citep{molliere_petitradtrans_2019,blain_prt_2024}.
It combines a flexible, 1D parametric forward model with a Bayesian framework \citep{nasedkin_atmospheric_2024}, which has been used to characterise the atmospheres of directly imaged exoplanets \citep[e.g.][]{molliere_retrieving_2020,nasedkin_hr8799_2024,landman_betapicb_2024,hoch_silicates_2025} and brown dwarfs \citep[e.g.][]{deregt_supjupI_2024,zhang_supjupIII_2024,deregt_supjupVII_2025} via atmospheric retrievals of their emission spectra.
These retrievals are enabled by the use of the correlated-k method for radiative transfer \citep{goody_correlated-k_1989,lacis_description_1991}, allowing efficient computation of the emission spectra up to a spectral resolving power of 1000, and the Feautrier method for calculating scattering in the atmosphere in the presence of clouds \citep{feautrier_rt_1964}. 
We used \prt version 3.2.0 to perform the retrievals in this study, using \texttt{exo-k} \citep{leconte_spectral_2021} to bin the correlated-k tables to R=400 for model computation.
As in previous studies, we used an adaptive pressure grid, using a total of 134 pressure levels between 10$^{3}$ and $10^{-6}$ bar, with a 10$\times$ finer grid spacing at the location where the clouds are found to condense. 
To validate the effectiveness of \prt in identifying different potential mechanisms for variability, we produced a set of forward models in Appendix \ref{app:forward}. 
This allowed us to demonstrate that different processes, such as changes in the temperature structure, cloud coverage, and chemical profiles  impact the spectrum in uniquely identifiable ways, reproducing the self-consistent modelling analysis of \cite{morley_spectral_2014}.

The \prt retrieval module combines the forward model with \texttt{Multinest} \citep{feroz_multimodal_2008,feroz_importance_2019} implemented in \texttt{pyMultinest} \citep{buchner_x-ray_2014} to sample the parameter space, estimate the model evidence, and infer posterior parameter distributions.
Unless otherwise noted, we use the constant efficiency mode of \texttt{Multinest} with a sampling efficiency of 5\% and 400 live points to reduce the computation time of the retrievals to make retrieving time-resolved spectra feasible.
When combined with importance nested sampling, this reduces the accuracy of the evidence estimate, making model comparison unreliable \citep{feroz_importance_2019}, although the parameter estimation remains minimally affected. 
A full listing of the priors used in the fiducial retrieval setup is available in Table \ref{tab:priors}.

Each model was convolved with a Gaussian kernel calculated from the wavelength-dependent instrumental spectral resolving power, before being binned to the wavelength bins of the data.
The spectral resolution functions for NIRSpec/PRISM and MIRI/LRS are obtained from the JWST CRDS system \citep{greenfield_crds_2016}.
The log-likelihood is then calculated to measure the goodness-of-fit between the data, $\vec{D}$, model, $\vec{F}$, and covariance matrix, $\vec{C}$, expressed as
\begin{equation}\label{eqn:loglike}
    -2\log\mathcal{L} = \left(\mathbf{D}-\mathbf{F}\right)^{T}\mathbf{C}^{-1}\left(\mathbf{D}-\mathbf{F}\right) + \log\left(2\pi\det\left(\mathbf{C}\right)\right).
\end{equation}
For the SIMP-0136 data, the covariance matrix is diagonal and consists only of the uncertainties associated with each wavelength bin.
Given the broad wavelength coverage and large numbers of pixels sampled, correlations due to inter-pixel cross talk will not impact the overall slope of the spectrum, and will likely only impact a few neighbouring wavelength channels.
Likewise, as the SIMP-0136 spectra were obtained using slit spectroscopy with nearly straight traces spread over many pixels, relatively few interpolations were made during the data reduction that would impart additional correlations into the data.
The assumption of independent Gaussian uncertainties is therefore a reasonable approximation for this dataset, although future analyses could include fitting for a covariance matrix.
We did not use any error inflation in the retrievals presented in this work, although the impacts of including error inflation are discussed in Section \ref{sec:errorinflation}.
We also tested the addition of an offset between the PRISM and LRS data. 
While a small offset was found, it did not significantly impact the retrieved parameters.

\subsection{Temperature profile}
Our fiducial temperature parameterisation is based on that of \cite{zhang_elemental_2023} (hereafter, \citetalias{zhang_elemental_2023}).
This approach has become widely adopted for sub-stellar atmospheric retrievals; for example, in its application in \cite{zhang_99770_2024} or \cite{nasedkin_atmospheric_2024}.
In the \citetalias{zhang_elemental_2023} setup, the retrieved parameters are the $i^{th}$ temperature gradients ($\left. d\log T/d\log P\right|_{i}$) at a set of log-spaced pressure nodes, with the temperature profile calculated by integrating the temperature gradient and interpolating onto the full pressure grid.
The temperature at the bottom of the atmosphere (\Tbot) was also retrieved.
Thus, for $i$ increasing from the bottom of the atmosphere, we have 
\begin{align}\label{eqn:grad}
    &T_{0} = T_{\rm Bot},\\
    &T_{i+1} = \exp\left(\log T_{i} + \left(\log P_{i+1} - \log P_{i}\right)\left(\frac{d\log T}{d\log P}\right)_{i}\right).
\end{align}
Throughout this work, we use the notation 
\begin{equation}
    \nabla T_{i}\equiv\frac{d\log T}{d\log P}_{i}.
\end{equation}
In contrast to the original implementation of this profile, we retrieved the temperature gradient from the bottom to the top of the atmosphere, without fixing an isothermal region at pressures lower than 10$^{-3}$ bar.

In \citetalias{zhang_elemental_2023}, the priors on the temperature gradient parameters were determined by measuring the temperature gradients across a set of radiative-convective equilibrium models.
By using relatively narrow Gaussian priors centred at these values, the retrieval was constrained to physically plausible solutions.
We explored the impact of varying the prior width on the retrieved temperature profile, finding no significant variation and confirming that the posterior distributions are not determined by the prior widths at pressures between 10 bar and 0.1 mbar. 
To allow the fits to be driven primarily by the data, we used a modestly wider set of priors, as listed in Table \ref{tab:priors}.

To validate our choice of temperature profile, we performed test retrievals using the profiles of \cite{molliere_retrieving_2020} and \cite{guillot_radiative_2010}. 
The more flexible Molli\`{e}re profile verified that the temperature structure did not depend on the choice of parametrisation, finding qualitatively similar results. 
However, the Guillot profile, which was only allowed to monotonically decrease with decreasing pressure, was insufficiently flexible to fit the data, producing reduced $\chi^{2}$ values 10 to 100 times worse than with the \citetalias{zhang_elemental_2023} profile.

\subsection{Chemistry}\label{sec:retrieval_chem}
Through a small set of test retrievals, we found that a free-chemistry approach produced significantly better fits than one relying on equilibrium chemistry, with disequilibrium abundances for \water, CO, and \methane, as was used in \cite{molliere_retrieving_2020}.
In particular, the equilibrium approach was not able to reproduce the CO$_{2}$, H$_{2}$S, and PH$_{3}$ abundances.
We therefore freely retrieve abundance profiles for 
\water \citep{ExoMol_H2O}, 
CO \citep{rothman_hitemp_2010},
\cotwo \citep{ExoMol_CO2},
\methane \citep{exomol_ch4MM_2024},
C$_{2}$H$_{2}$ \citep{ExoMol_C2H2},
HCN \citep{ExoMol_HCN},
FeH \citep{wende_feh_2010},
H$_{2}$S \citep{ExoMol_H2S},
\ammonia \citep{ExoMol_NH3},
PH$_{3}$ \citep{exomol_ph3},
Na \citep{allard_new_2019},
and K \citep{allard_k-h_2016}.
The mass fraction abundances of each of these molecular species were retrieved as free parameters, and the remaining atmosphere was filled with a mixture of 76\% H$_{2}$ and 23\% He by mass, similar to the composition of Jupiter \citep{atreya_composition_2003,benjaffel_helium_2015}.
In addition to the impact of the molecular species on the spectra from their line absorption and emission, we also account for Rayleigh scattering from  H$_{2}$ and He, as well as the collisionally induced absorption for the  H$_{2}$-H$_{2}$ and H$_{2}$-He interactions.

Previous studies have shown that retrievals are sensitive to changes in chemical abundances with pressure. For instance FeH in L-dwarfs \citep{rowland_feh_2023,deregt_supjupVII_2025} and \water in Y-dwarfs \citep{kuhnle_depletion_2024}.
\cite{rowland_feh_2023} also suggested that non-uniform \methane abundances will be important for interpreting the emission spectra of L-T transition objects, such as SIMP-0136.
In the presence of upper atmosphere heating from aurora or other phenomena, this effect will likely be magnified as \methane becomes disfavoured in equilibrium compared to CO.
Therefore, we also developed a parametrisation for variable abundance profile throughout the atmosphere. 
The abundance at the top and bottom of the atmosphere are freely retrieved parameters.
Between these boundaries, pressure-temperature pairs are also retrieved. 
This setup allows the retrieval of the location where changes in abundance occur, as well as the abundance at that position.
The location of the pressure node is determined by retrieving the difference in log pressure between the node and the previous node, measured from the top of the atmosphere down.
Between these nodes, the abundance profile can be interpolated using a linear spline, as illustrated in Fig. \ref{fig:model-non-constant-abund-profile}.
For the present study, we used a single pressure-temperature pair as free parameters for the \methane abundance to measure potential changes induced by the thermal structure of the atmosphere and to better fit the depths of \methane absorption features in the spectrum.

\begin{figure*}[t]
    \centering
    \includegraphics[width=0.85\linewidth]{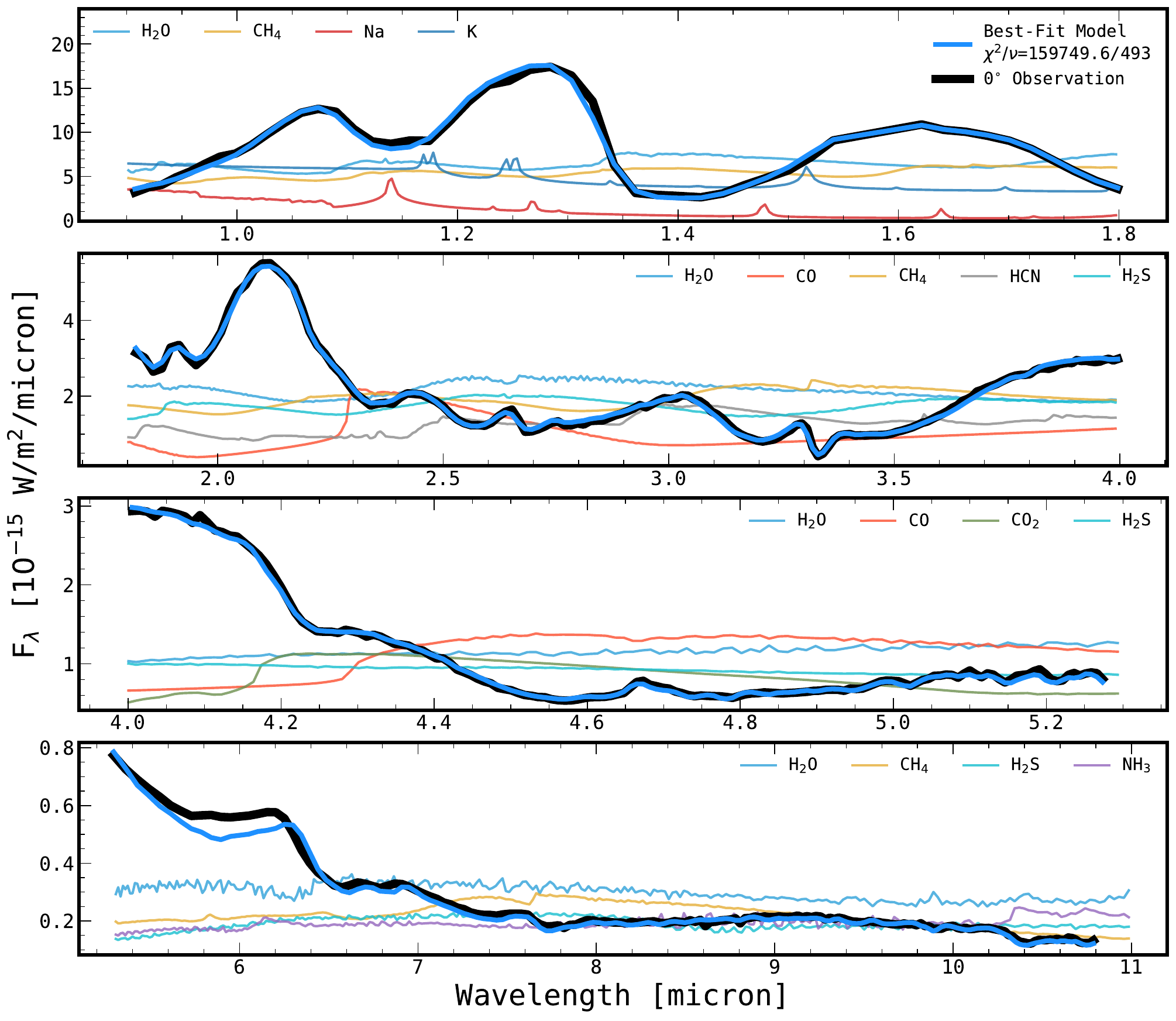}
    \caption{Best-fit model from the fiducial retrieval to the observed spectrum at 0$^{\circ}$ phase, whose uncertainties are smaller than the displayed line width. This model assumed a fixed mass and radius of 10.381 M$_{\rm Jup}$ and 0.9329 R$_{\rm Jup}$ respectively. Also shown for visual reference are the log opacities at the 1 bar level of the primary absorbers in each wavelength range. The \methane abundance was allowed to vary as a function of altitude. The reduced $\chi^{2}$ of the fit is poor at 295, due to the photometric precision of \textit{JWST}. However, the model clearly matches the key absorption features of \water, \methane, CO, and \cotwo, as well as additional species such as H$_{2}$S and NH$_{3}$.}
    \label{fig:retrieval-fit}
\end{figure*}

\subsection{Clouds}
Clouds are a critical component of the atmosphere, contributing a source of continuum opacity and reddening.
We used a cloud model inspired by \cite{vos_patchy_2023}; similarly to their retrievals of SIMP-0136, our fiducial model uses an amorphous Mg$_{2}$SiO$_{4}$ \citep{servoin_infraredmg2si04_1973} cloud together with a crystalline Fe cloud \citep{henning_dust_1996}.
The cloud opacities are calculated using a distribution of hollow spheres approach \citep{toon_algorithms_1981,min_modelingoptical_2005}, similarly to \cite{molliere_retrieving_2020}.
We also explored varying the cloud composition and number of cloud layers.
We tested replacing the iron cloud layer with crystalline KCl \citep{palik_handbook_1985} and Na$_{2}$S clouds \citep{morley_neglected_2012}, which are expected to condense at photosphere pressures in T-dwarfs, while the iron cloud should condense too deep to be observable.
As \cite{mccarthy_multiple_2024} found multiple layers of silicate clouds, we also tested the inclusion of patchy Mg$_{2}$SiO$_{4}$ and MgSiO$_{3}$ \citep{jaeger_silicates_1998}, as well as both silicate clouds and a full-coverage Na$_{2}$S cloud.
As the clouds in SIMP-0136 are expected to occur near the base of the photosphere or deeper, we did not expect the results to depend strongly on the details of the cloud composition or particle geometry. Thus, we leave a more detailed study of the cloud properties to a future analysis.

We retrieved the mean particle radius of each layer, together with the width of a log-normal distribution about the mean.
The cloud base pressure for each cloud was freely retrieved, along with the mass fraction at the base of each cloud layer.
We find that this mass fraction decreases with decreasing pressure as
$X_{i}(P) = X_{i,0}\left(P/P_{0}\right)^{f_{\rm sed}},$
where $f_{\rm sed}$ is used as a parameter to determine the rate of the abundance decrease.
As in \cite{vos_patchy_2023}, the cloud coverage fraction of the silicate cloud, $f_{\rm cloud}$ was also allowed to vary while the non-silicate cloud is assumed to enshroud the entire object.
From test retrievals, we found that the patchiness of the silicate cloud and the mass fraction were moderately degenerate; occasionally, the retrieval would find a solution where the cloud coverage approached 1, while the cloud mass fraction approached 0.
On average, the silicate cloud mass fraction at the cloud base was found to be between 10$^{-3}$ and $10^{-4}$ by mass. 
As we did not expect variations in the cloud abundance by any significant order of magnitude, we set the lower limit of the mass fraction prior of the silicate cloud to $10^{-6}$, thus avoiding the degenerate solution.

\section{Results}\label{sec:results}

\subsection{Setting mass and radius}

\begin{table*}
\caption{Measuring mass and radius of SIMP-0136 under different prior assumptions.}
    \label{tab:mass}
    \vspace{-1em}
    \begin{center}
    \begin{tabular}{r|llll}
    \toprule
    \textbf{Parameter} & $\mathbf{0^{\bm\circ}}$ & $\mathbf{90^{\bm\circ}}$& $\mathbf{180^{\bm\circ}}$& $\mathbf{270^{\bm\circ}}$\\
    \midrule
     & \multicolumn{4}{c}{Uniform Mass Prior}\\
     \midrule
       Mass [M$_{\rm Jup}$] & 3.813 $\pm$ 0.002 &  3.744 $\pm$ 0.002 & 3.973 $\pm$ 0.002 & 3.419 $\pm$ 0.001 \\
       Radius [R$_{\rm Jup}$]& 0.9339 $\pm~3\times10^{-5}$ &  0.9370 $\pm~6\times10^{-5}$&  0.9339 $\pm~2\times10^{-5}$ & 0.9278 $\pm~4\times10^{-5}$\\
       $\log g$ [cgs] & 4.05 &  4.043 &  4.072 &  4.012\\
       $\chi^{2}/\nu$ & 306.4 & 307.0 & 307.2 & 318.2\\
    \midrule
     & \multicolumn{4}{c}{Evolutionary Model Mass Prior}\\
     \midrule
       Mass [M$_{\rm Jup}$] & 7.734 $\pm$ 0.003 &  10.146 $\pm$ 0.006 &  10.137 $\pm$ 0.007 & 10.8589 $\pm$ 0.003 \\
       Radius [R$_{\rm Jup}$] & 0.9439 $\pm~3\times10^{-5}$ &  0.9363 $\pm~6\times10^{-5}$ &  0.9281 $\pm~6\times10^{-5}$ & 0.9385 $\pm~4\times10^{-5}$\\
       $\log g$ [cgs] & 4.35 &  4.47 &  4.484 &  4.504\\
       $\chi^{2}/\nu$ & 289.6 & 305.6 & 318.2 & 302.3\\
       \bottomrule
    \end{tabular}
    \end{center}
    \vspace{-0.5em}
    \begin{tablenotes}
        \item \textbf{Notes:} For the full set of phase-resolved retrievals we fix the mass and radius to the mean of the evolutionary prior values, excluding the outlying retrieval at 0$^{\circ}$ phase. While these measurements are inconsistent with the \texttt{SEDkit} values, the resulting surface gravity is compatible.
    \end{tablenotes}

\end{table*}

To observe the variability of atmospheric parameters, it is necessary to first constrain the properties of SIMP-0136 that cannot vary as a function of phase; namely, its mass and radius.
We ran a small set of retrievals at each of 0$^{\circ}$, 90$^{\circ}$, 180$^{\circ}$, and 270$^{\circ}$ phase, where mass and radius were included as free parameters.
This experiment was repeated using both a uniform prior distribution on the mass and radius, and with a Gaussian prior based on evolutionary expectations.
For the uniform case, the radius prior was set to $\mathcal{U}(a=0.7, b=3.0)$ R$_{\rm Jup}$ and the mass to $\mathcal{U}(3, 18)$ M$_{\rm Jup}$, while for the evolutionary prior case, the priors were set based on the measurements of \cite{gagne_simp_2017} $\mathcal{N}(\mu=1.1, \sigma=0.1)$ R$_{\rm Jup}$ and $\mathcal{N}(12, 1)$ M$_{\rm Jup}$. 
We also performed an additional set of retrievals using the \texttt{SEDkit} measurements as priors on the mass and radius.
Notably, this enforced a very strong prior on the radius of $\mathcal{N}(\mu=1.2, \sigma=0.05)$ R$_{\rm Jup}$. 
This resulted in a radius more consistent with the evolutionary models, but resulted in unphysical cloud and temperature structure parameters; thus, we discarded this set of models. 
Overall, this is related to the well-known `small radius problem', discussed in detail in \cite{balmer_aflep_2025}. 
There are known inconsistencies between the radii measured using atmospheric and evolutionary models, possibly due to an incomplete treatment of the clouds.
As such, for the remainder of this work, we treat the mass and radius as nuisance parameters for obtaining the surface gravity, which has a more direct impact on the emission spectrum. 

Both the uniform and \cite{gagne_simp_2017} prior setups produced qualitatively similar atmospheric states.
The resulting masses and radii at each phase are presented in table \ref{tab:mass}.
The evolutionary prior retrievals produce marginally better fits for 3/4 phases, as measured by the reduced $\chi^{2}$.
This set of retrievals inferred masses closer to expectations from evolutionary models, along with measured $\log g$ values consistent with literature measurements.
Thus, we were able to take the average of this set of retrievals as our fixed mass and radius, excluding the outlying measurement of 7.7 M$_{\rm Jup}$.
The mass was therefore fixed at 10.381 M$_{\rm Jup}$ and the radius at 0.9329 R$_{\rm Jup}$, resulting in a log surface gravity of 4.490. 
The surface gravity was kept consistent with the \texttt{SEDkit} results, while the mass and radius were both set to be lower, which we attribute to  the small radius problem.

\subsection{Atmospheric properties of SIMP-0136}\label{sec:single-retrieval}
Although we performed an ensemble of retrievals to understand the observed spectroscopic variability, it is insightful to consider the bulk planet properties obtained from a single retrieval. 
We discuss the retrievals of the 0$^{\circ}$ observations as a representative example, with the best-fit model shown in Figure \ref{fig:retrieval-fit}.
From the emission contribution function, shown in Fig. \ref{fig:contribution_function}, we see that the emission spectrum probes four decades in pressure, from 10 bar where the spectrum is cut off in the NIR by clouds, to methane features at 3.3 $\upmu$m and ${\sim}$7 $\upmu$m at pressures as low as 10$^{-3}$ bar.
This implies that the retrievals are sensitive to the thermal structure and chemical abundances across this pressure range, while there is no information imparted onto the observed spectrum from both deeper pressures and at very high altitudes.

\subsubsection{Stratospheric heating}
\begin{figure}
    \centering
    \includegraphics[width=\linewidth]{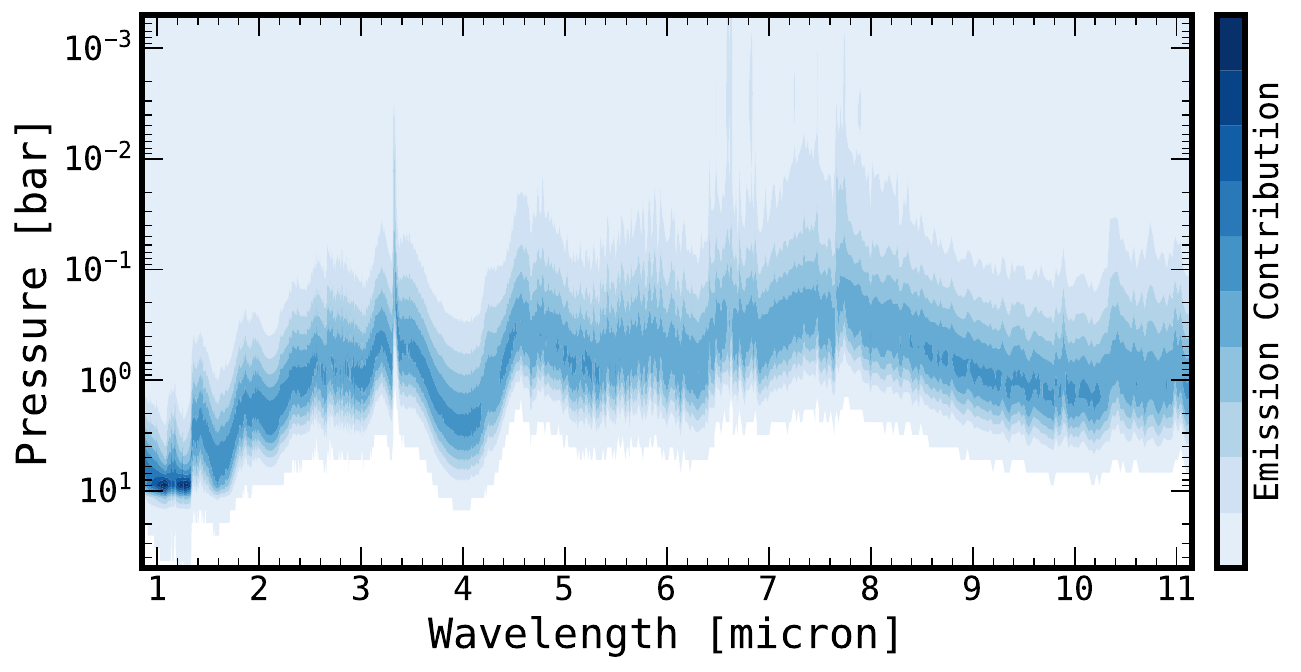}
    \caption{Emission contribution function for SIMP-0136 at 0$^{\circ}$ phase. The emission contribution is shown in square-root colour scale to highlight subtle variations. Significant contributions are present across four decades in pressure, bounded by a cloud layer near 10 bar and by methane absorption up to 10$^{-3}$ bar.}
    \label{fig:contribution_function}
\end{figure}

\begin{figure}
    \centering
    \includegraphics[width=\linewidth]{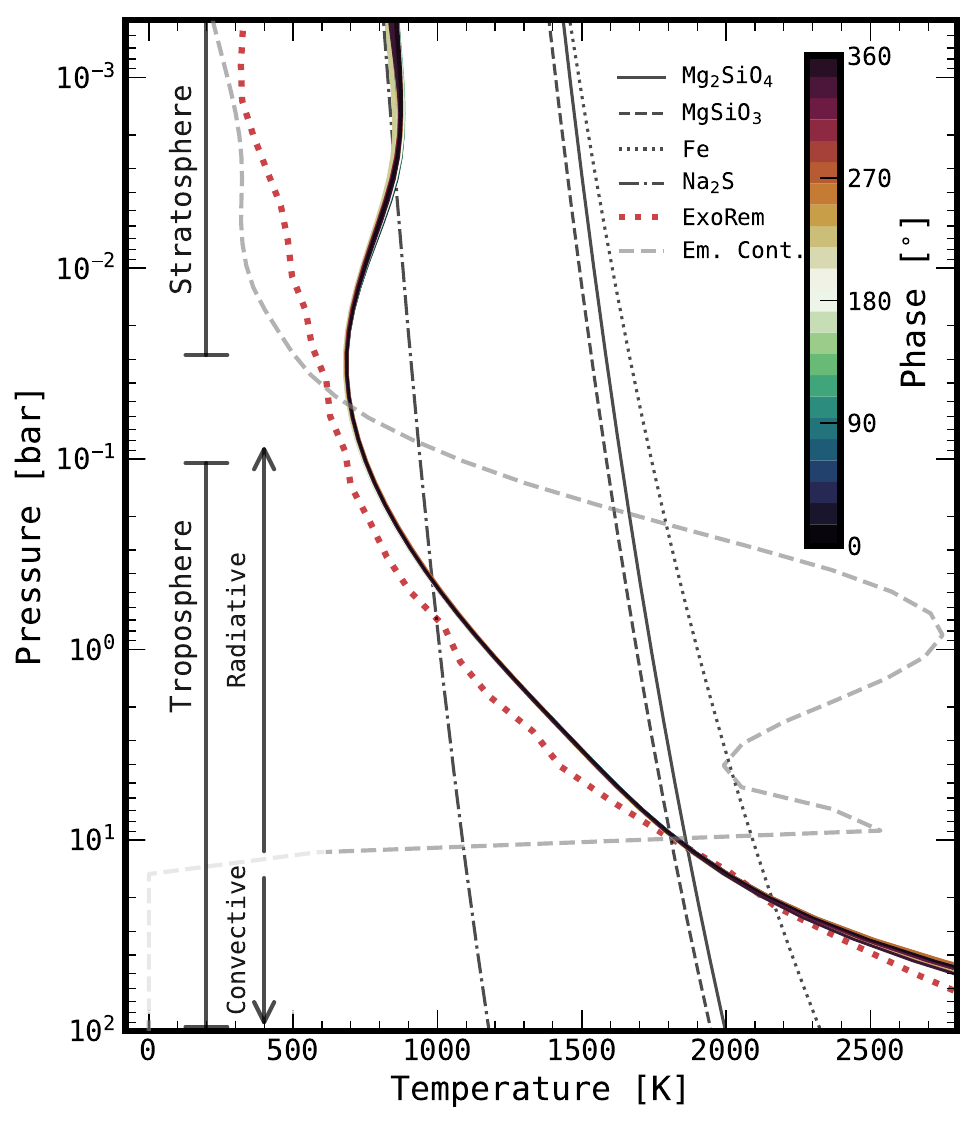}
    \caption{Retrieved temperature profiles for each phase bin. Temperature profiles are given as 90\% confidence intervals. In grey is the emission contribution function, averaged over wavelength. We show the square root of the contribution to highlight that there is small, but significant emission from the upper atmosphere, coincident with the location of the temperature inversion. In black lines are condensation curves for enstatite, forsterite, and iron, showing that the retrieved cloud base pressure for the silicate cloud is coincident with the expected condensation location. We also show  the temperature profile from a self-consistent \texttt{ExoRem} model ($\log g = 4.5$, [M/H]$=0$, C/O$=0.55$, T$_{\rm eff}=1000$~K) as a red dotted line.}
    \label{fig:pt-profile}
\end{figure}

The retrieved temperature profile is presented in Fig \ref{fig:pt-profile}. In the troposphere, at pressures greater than 0.1 bar, the temperature profile follows an adiabat, similar to the temperature profiles predicted by radiative-convective equilibrium forward models.
There is no evidence of an isothermal region, such as those produced by fingering convection \citep{tremblin_fingering_2015,tremblin_cloudless_2016}, which would act to redden the spectrum without requiring clouds.
Between $10^{-1}$ bar and $10^{-3}$ bar the temperature gradient inverts, and begins increasing with increasing altitude, reaching a maximum near $2\times 10^{-3}$ bar.
The transition from an adiabatic profile in the troposphere to an inverted temperature profile in the stratosphere at 0.1 bar is qualitatively similar to the structure of Jupiter's atmosphere, where the tropopause is also located at 0.1 bar \citep{gupta_jupiterpt_2022}.
This is clearly in contrast with the self-consistent forward models, which are usually monotonically decreasing with altitude in the absence of external irradiation.
Compared to a temperature profile that becomes isothermal at the point of the inversion, the hotspot at 0$^{\circ}$ phase peaks at 265 K warmer at $3\times10^{-3}$ bar.
From the forward models of Section \ref{sec:temp-forward-model}, and from the emission contribution function, shown in Fig. \ref{fig:contribution_function}, we found that this region is probed by the 3.3 $\upmu$m and 7.7 $\upmu$m \methane features.
This is the strongest source of opacity in the atmosphere, and probes the lowest pressures.
We performed two additional retrievals to asses  the importance the \methane features in shaping the inversion. These retrievals were run while masking out the 3.1 - 3.5 $\upmu$m feature, and masking out both 3.1 - 3.5 $\upmu$m and 7 - 8 $\upmu$m. 
In both cases, the strength of the inversion is greatly reduced, with the temperature profile weakly oscillating around an isothermal solution at pressures lower than $10^{-2}$ bar.
Without any strong opacity source we could not reliably infer the temperature structure at pressures lower than 1 mbar.

We calculated the derivative of the retrieved temperature profile, $d\log T/d\log P$, and compared this to the adiabatic temperature gradient to determine where the atmosphere is convectively unstable, following the Schwarzchild criterion.
The adiabatic temperature gradient was calculated for an atmosphere with a metallicity of 0.2 and a C/O ratio of 0.55. It was found using the interpolated \texttt{easyCHEM} chemical equilibrium grid \citep{molliere_observing_2017}.
Both temperature gradients are shown in Appendix \ref{app:retrievals} in Fig. \ref{fig:temp-gradient}.
At 17 bar, the retrieved temperature gradient is consistent with an adiabatic gradient, but is constrained by the prior distribution. 
At pressures lower than 5 bar the atmosphere is stable against convection, and is therefore radiative.
Based on the contribution function, this transition coincides with the top of the silicate cloud layer.

At 0$^{\circ}$ phase, we measured the effective temperature to be $1246.82\pm0.02$ K by integrating a low resolution model between 0.5 $\upmu$m and 250 $\upmu$m. 
Given the fixed radius, this is simply a function of the bolometric luminosity of SIMP-0136 at 0$^{\circ}$ phase, which is measured to high precision due to the precise spectrophotometry and wavelength coverage of the combined NIRSpec/PRISM and MIRI/LRS observations.
Comparing the full wavelength range of the model to the wavelength range of the data, we found that the \textit{JWST} observations measured 95.7\% of the total emitted flux.
We compared the $T_{\rm eff}$ found by integrating the low resolution model to that obtained by directly integrating the observed spectrum. 
There was significant systematic discrepancy between the methods: as several wavelength bins are excluded from the data due to hot pixels, the flux is somewhat underestimated, and the effective temperature was found to be 1197 K after applying the bolometric correction factor.
This is around 100 K warmer than expectations from evolutionary models or SED fitting (see Appendix \ref{app:gridfits}), but is also about 80 K colder than found by \cite{vos_patchy_2023}, which we attribute to their smaller retrieved radius.
While the mean T$_{\rm eff}$ varied between methods, the relative variation was found to be consistent, varying by 5 K between the coldest phase and the warmest.


\begin{figure}
    \centering
    \includegraphics[width=\linewidth]{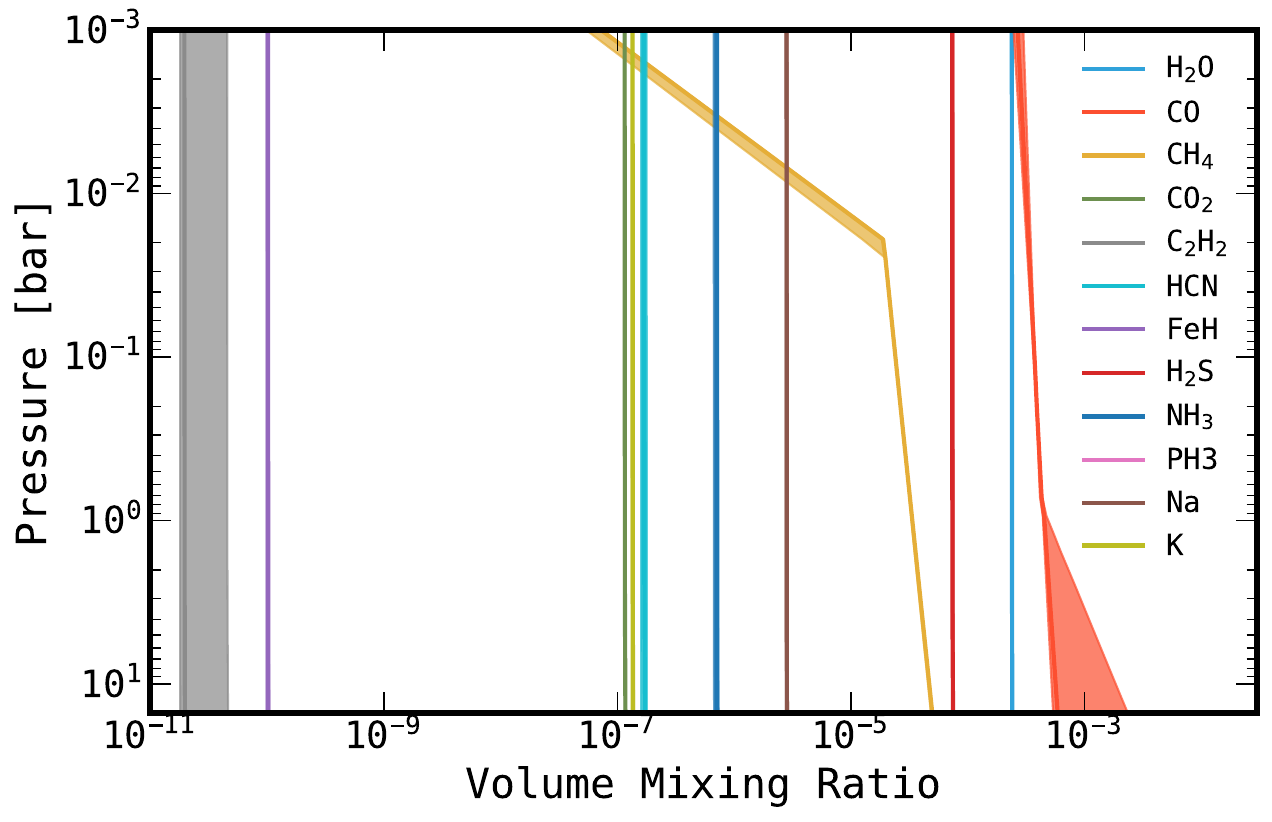}
    \caption{Volume mixing ratio profiles of gas-phase species for all phases. The shaded regions indicate 68\% confidence intervals, taken from the set of median abundances measured at each phase. The measured precision for a single phase is typically an order of magnitude smaller than the variation between phases.}
    \label{fig:abundance_profiles}
\end{figure}

\subsubsection{Atmospheric chemistry}
We were able to measure the abundance of 10 of the 12 chemical species included in the free retrieval, placing upper limits on the abundance of acetylene (C$_{2}$H$_{2}$) and phosphine (PH$_{3}$), with combined abundance profiles from each phase for each species shown in Fig. \ref{fig:abundance_profiles}.
Full posteriors are available in the appendix in Fig. \ref{fig:vmr_hists}.
Due to the computational cost of retrievals when not using \texttt{Multinest's} constant efficiency mode, we were unable to reliably measure the Bayesian evidence and calculate detection significances for each species.
Of the species with constrained abundances, \water, \methane, CO, \cotwo, and \ammonia have clearly visible absorption features present in the spectrum, as seen by comparing the opacities to the spectrum in Fig. \ref{fig:retrieval-fit}. In particular,
\htwos is consistently a strong source of opacity in the NIR, with its opacity reaching a maxima between 2.5 $\upmu$m and 3 $\upmu$m, as well as between 3.5 $\upmu$m and 4.1 $\upmu$m, although it lacks uniquely identifiable absorption lines in the spectrum.
As the retrievals fit these features well, with only small variations in the abundances with phase, the abundances of these species are reliably measured.
HCN, FeH, Na, and K all contribute to the continuum opacity. 
The low spectral resolution of NIRSpec/PRISM prevents the identification of the alkali lines in the near-IR, and so the abundances of Na and K are largely constrained by their impact on the NIR slope, which is degenerate with the impact of clouds in this region.
FeH acts similarly, and its low abundance indicates that it does not strongly impact the spectrum.
Such a low abundance is consistent with all of the iron condensing out deep within the atmosphere.

The retrieved abundances of these species are mostly consistent with an equilibrium chemistry model with [M/H]=0.2 and C/O=0.55, suggesting that the retrievals have inferred plausible abundances for each of these species. Both the retrieved abundances and the equilibrium profiles are shown in the appendices in Fig. \ref{fig:vmr_hists}.
The \water abundance agrees well with the equilibrium abundance at a pressure of 2 bar.
Then, \methane and CO are consistent near 2 bar, where they appear to quench.
\htwos is strongly enriched compared to the equilibrium prediction.
Na is very consistent up to 1 bar.
\ammonia is somewhat under abundant, but intersects with the equilibrium profile around 0.05 bar.
K and HCN are both consistent up to 1 bar.
Both FeH and C$_{2}$H$_{2}$ drop off to 0 at around 2 bar in equilibrium, consistent with the upper limits set by the retrievals;
PH$_{3}$ is severely under abundant compared to equilibrium expectations, although this is consistent with the missing phosphine problem widely observed in brown dwarfs \cite{visscher_sulfurphos_2006,rowland_protosolar_2024}. 

To test the assumption of chemical equilibrium, we ran a series of retrievals allowing different levels of flexibility in the \methane and CO abundance profiles. 
As described in Section \ref{sec:retrieval_chem}, we compared using a vertically constant abundance profile, a profile with a single freely retrieved pressure-abundance pair, and a profile with three retrieved pressure-abundance pairs.
Using vertically constant abundance profiles, we found that the retrievals were unable to capture both the 1.2 $\upmu$m and 3.3 $\upmu$m \methane absorption features, and the goodness-of-fit was poor, with a reduced $\chi^{2}$ of ${\sim}$360.
Adding a node to the \methane abundance profile provided a significantly better fit to the data,  with $\Delta\chi^{2}/\nu\approx80$ and a physically plausible temperature profile.
While allowing additional flexibility in the \methane profile led to significantly better fits, adding additional nodes to the CO profile only marginally improved the reduced $\chi^{2}$, implying that a vertically constant abundance is sufficient for CO.
With a single node, we find that (as shown in Fig. \ref{fig:abundance_profiles}) the \methane abundance decreases sharply above $2\times10^{-2}$ bar.
Lastly, when including three nodes for each of \methane and CO, we find that the profiles match that of the single-node retrievals. 
Even with sufficient flexibility, the abundances do not reproduce the equilibrium chemistry profiles, instead suggesting that both \methane and CO are in chemical disequilibrium.
This exercise was repeated, allowing \water to vary as well.
Adding the additional flexibility to fit the water abundance profile resulted in a worse reduced $\chi^{2}/\nu$, suggesting that the inclusion of these additional parameters is not justified by the data.
Regardless of the degree of flexibility used, the overall interpretation of the atmosphere remains qualitatively similar, with only minor variation in the inferred atmospheric metallicity.

As both \methane and CO appear to be in chemical disequilibrium, we can apply the approach of \cite{zahnle_methane_2014}  to derive the vertical mixing strength, \kzz, required to quench both species. We calculated the \kzz from the quench point of \methane and CO.
\methane intersects the equilibrium profile at 17 bar, near the cloud base, while CO intersects the equilibrium profile at about 2 bar.
Calculating \kzz using the temperature at the quench point, around 1500 K at 2 bar or 1850 K at 17 bar, then we find a $\log K_{zz}$ between $6-9$ log cm$^{2}$s$^{-1}$, which is relatively consistent with the GCM predictions of a $\log K_{zz}\sim 9$ at the temperatures of SIMP-0136 \citep{lee_dynamicallII_2024}. 
However, the value of K$_{ZZ}$ depends strongly on the temperature, which varies throughout the atmosphere. If instead we use the effective temperature instead of the temperature at the quench point, we find a range of $\log K_{zz}=3.3 - 4.6$ in units of log cm$^{2}$s$^{-1}$, depending on the choice of whether the methane or CO quench pressure is used. 
This relatively low value is similar to the mixing strength found for HR 8799 b \citep[$2.9^{+0.6}_{-0.7}$ log cm$^{2}$s$^{-1}$,][]{nasedkin_hr8799_2024}, which has a similar spectral type to SIMP-0136 \citep{bonnefoy_characterization_2014}.
This orders-of-magnitude variation in the mixing strength suggests that non-vertically-constant K$_{zz}$ parameterisations are necessary to interpret the chemical state of the atmosphere.

\begin{figure}[t]
    \centering
    \includegraphics[width=\linewidth]{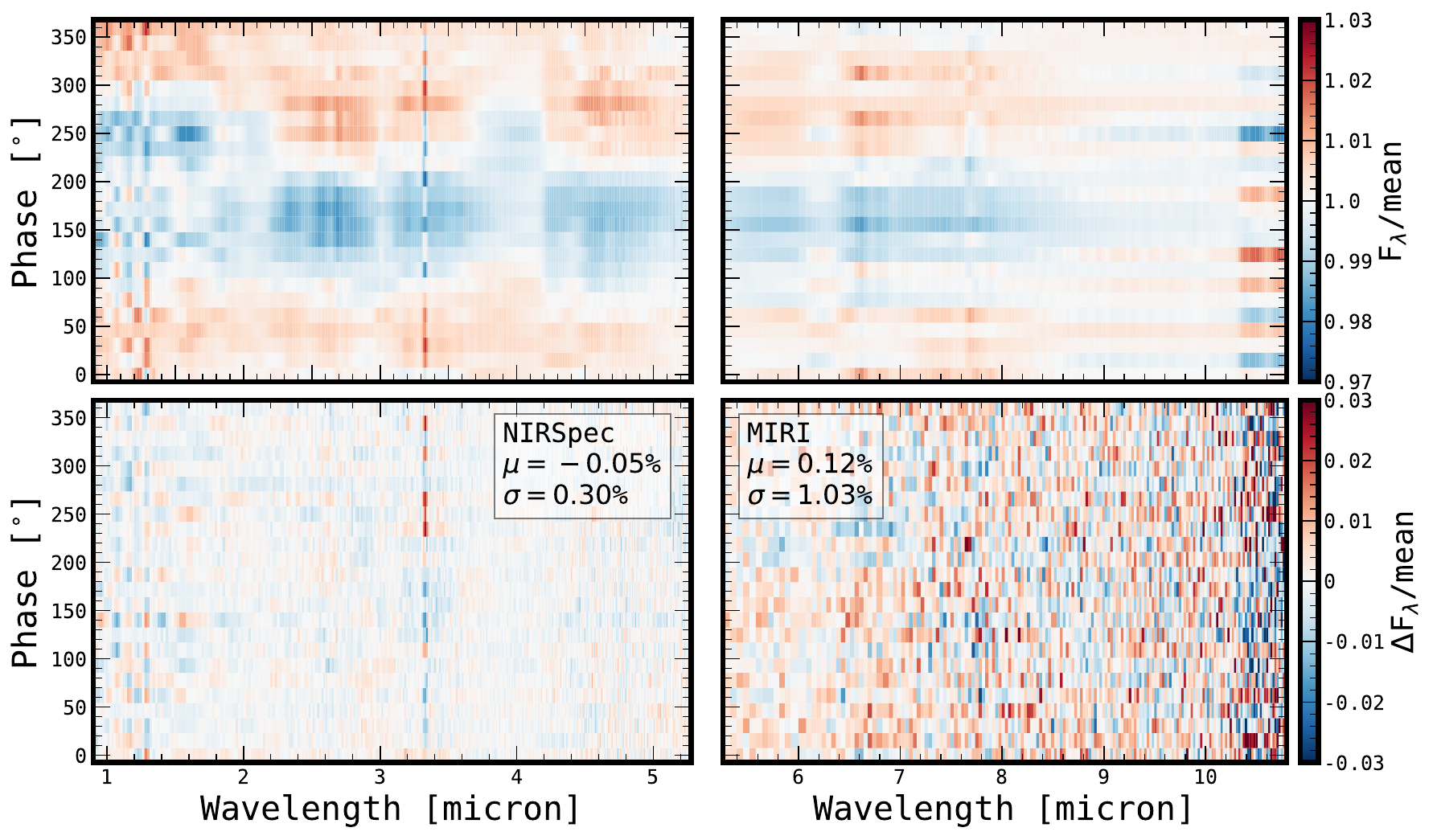}
    \caption{\textbf{Top:} Phase-resolved best-fit spectra divided by the mean retrieved spectrum for NIRSpec/PRISM (left) and the MIRI/LRS (right), taken from the fiducial retrieval. \textbf{Bottom:}  Difference between the observed and retrieved variability maps. The remaining difference is normally distributed, centred at 0, and with a standard deviation of 0.3\% for NIRSpec and 1\% for MIRI.
    }
    \label{fig:retrieved-variability-map}
\end{figure}

\subsubsection{Patchy silicate clouds}
While there is no distinct silicate feature identified in the MIR, as seen in the case of late-L dwarfs \citep[e.g.][]{suarez_emergence_2022, suarez_ultracool_2023_grains}, silicate clouds are found to contribute strongly to the shape of the emission spectrum.
At 0$^{\circ}$, the forsterite cloud base is located at $13.10 \pm 0.03$ bar, forming the base of the photosphere and strongly absorbing at wavelengths shorter than 2 $\upmu$m.
The condensation curves shown in Fig. \ref{fig:pt-profile}, taken from \cite{visscher_atmospheric_2010}, show that the silicate clouds are expected to condense between about 8 bar and 15 bar given the inferred temperature structure.
At this depth, the silicate clouds do not significantly contribute their own MIR absorption features, as the MIR photosphere lies above the cloud layer.
At the cloud base, the log mass fraction of Mg$_{2}$SiO$_{4}$ is found to be $-4.33 \pm 0.01$. 
The $f_{\rm sed}$ value, which determines the rate at which the silicate abundance decreases with altitude, was found to be $f_{\rm sed, \mathrm{Mg}_{2}\mathrm{SiO}_{4}}=9.90 \pm 0.05$, consistent with the upper limit of the prior, and creating a very compact cloud layer.
Micron-sized particles were found to best fit the observed cloud properties, with a narrow log-normal distribution ($\log r_{c, \mathrm{Mg}_{2}\mathrm{SiO}_{4}}=-3.994 \pm 0.003, \sigma_{\rm LN} = 1.19 \pm 0.01$. 
The silicate clouds were found to cover $87.14 \pm 0.08$\% of the atmosphere at this phase.

In addition to the forsterite clouds, an additional cloud is necessary to fit the observations; several different compositions were tested.
In our fiducial model, we included a crystalline iron cloud, which was found to be optically thin and condensed at a slightly lower pressure (i.e. higher altitude) than the forsterite cloud.
Based on the condensation curves, iron would be expected to condense at pressures greater than 30 bar, significantly deeper than the silicate clouds.
When the iron cloud was replaced with an Na$_{2}$S cloud, the condensation location was found to remain constant, without significant changes in the goodness-of-fit.
As with the iron clouds however, the location of the Na$_{2}$S cloud did not correspond with its expected condensation pressure.
A KCl cloud was also tested, and resulted in a marginally worse fit than either the iron or Na$_{2}$S clouds.
The addition of an additional patchy enstatite cloud also had no significant impact on the goodness-of-fit of the spectrum.
When combined, this suggests that in addition to a patchy silicate cloud, an additional source of grey opacity near the base of the photosphere near 10 bar is required to fit the data.
This cloud may be compositionally different from the silicate cloud, or it may be due to the cloud parametrisation failing to accurately model the silicate clouds.

\begin{figure}
    \centering
    \textbf{Fixed Temperature Profile}\\
    \includegraphics[width=\linewidth]{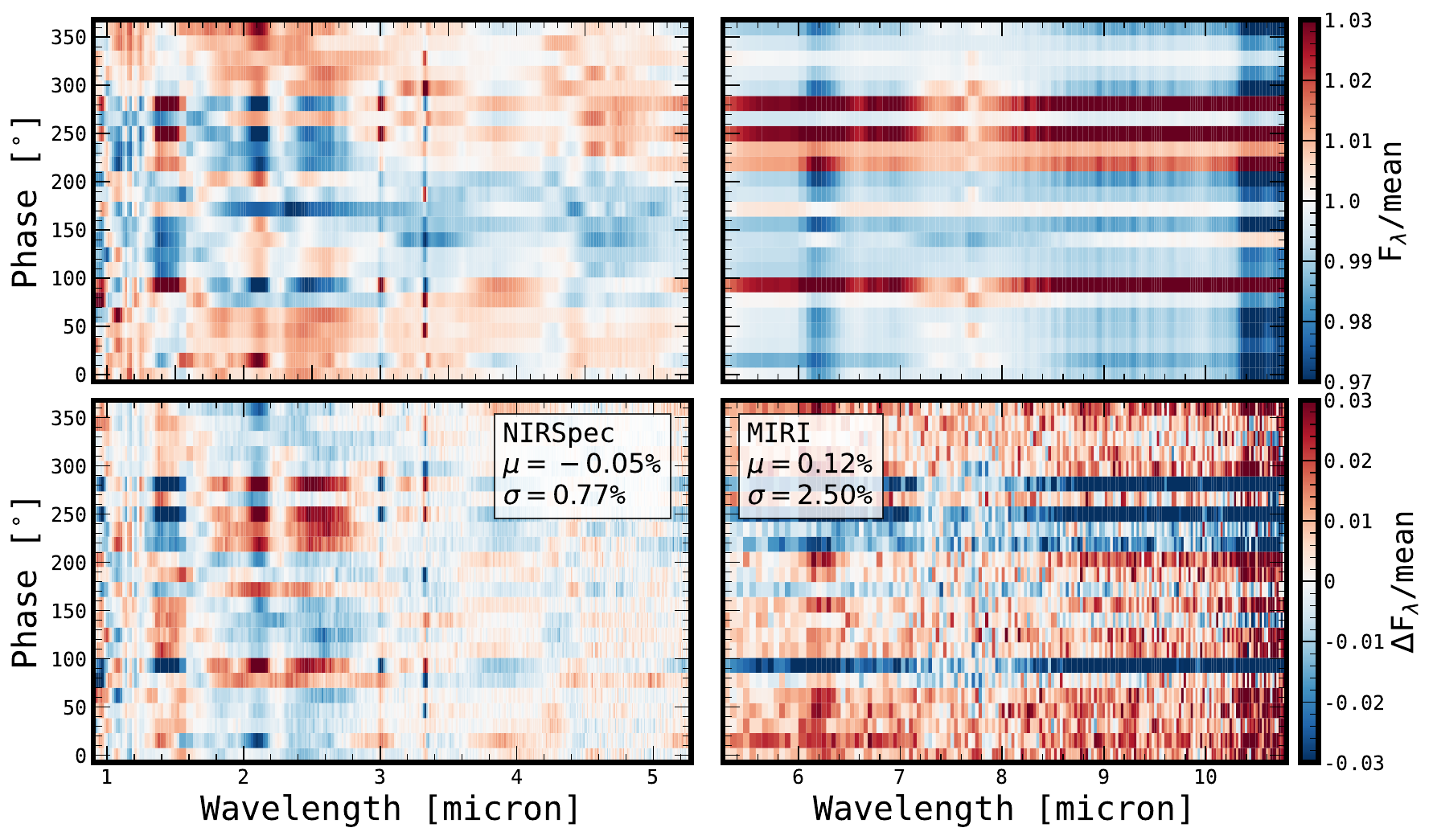}
    \textbf{Fixed Cloud Parameters}\\
    \includegraphics[width=\linewidth]{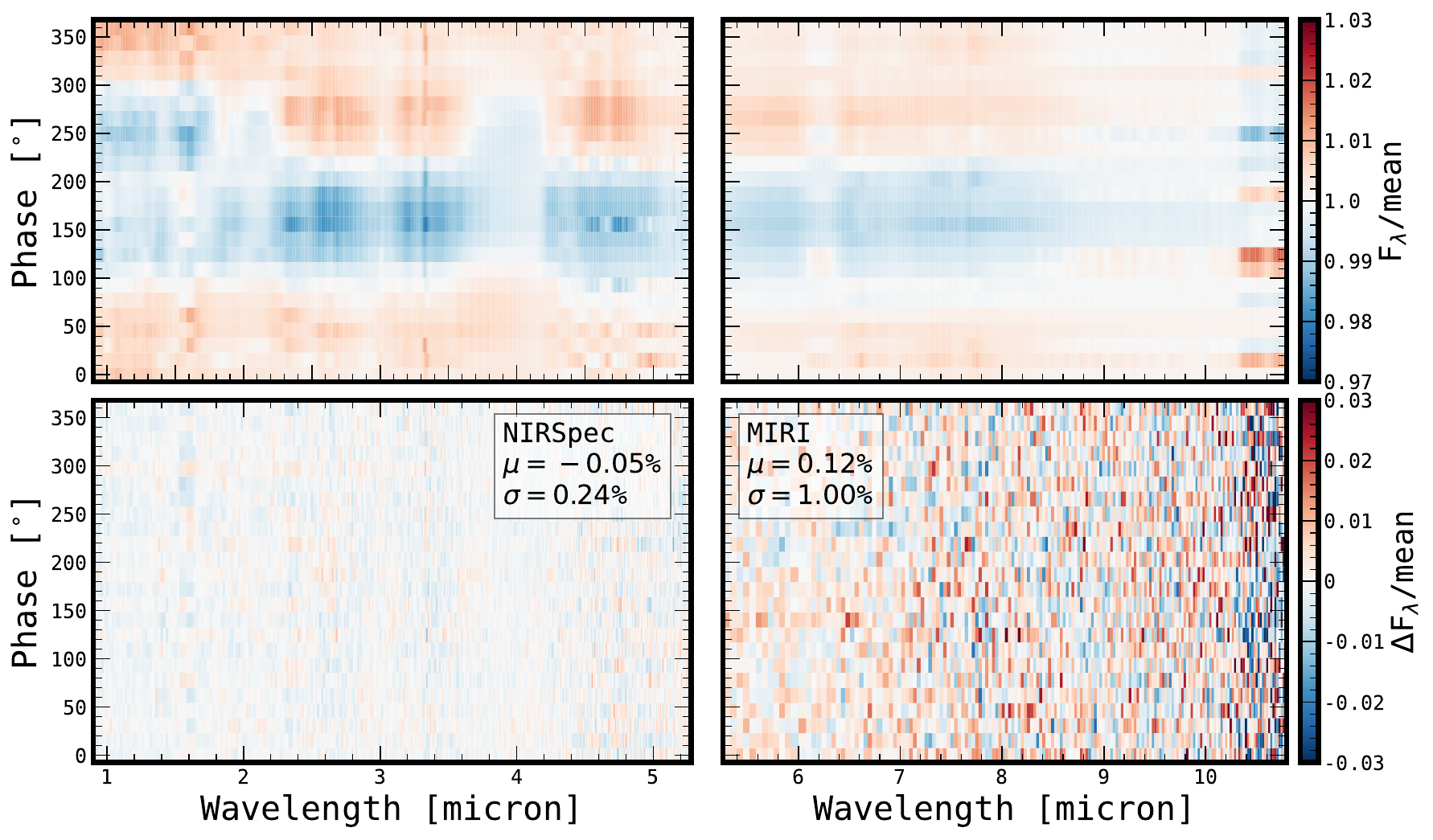}
    \caption{Same as Fig. \ref{fig:retrieved-variability-map}, but with sets of parameters kept constant across phase. \textbf{Top:} Variability map keeping the temperature profile constant, showing that the retrievals are unable to reproduce the observed variability. \textbf{Bottom:} Variability map keeping the cloud parameters constant, which better reproduces the observed variability than the full retrieval.}
    \label{fig:fixed-param-vmaps}
\end{figure}

\subsection{Time-resolved atmospheric retrievals}
We ran independent retrievals on each of the 24 phase-binned spectra of SIMP-0136, using the same setup as described in Section \ref{sec:single-retrieval}.
We refer to this setup as the fiducial, or baseline setup for the retrievals.
Across both the NIRSpec and MIRI spectra the models are consistent to within 15\% of the observed flux at every wavelength bin;  the RMS deviation between the NIRSpec data and the models is $2.60\pm0.08$\%, while for MIRI it is $3.8\pm0.2$\%.
Across all phases, the largest discrepancy occurs between 5.86 $\upmu$m and 6.3 $\upmu$m, where the models find stronger water absorption than is present in the data, and the onset of the water absorption feature at 6.4 $\upmu$m is slightly redshifted in the model compared to the data.
For each retrieval, the residuals between the models and the data were dominated by systematic effects which were consistent across phase, as shown in Fig. \ref{fig:residuals_all} in Appendix \ref{app:retrievals}. 
These discrepancies may be caused by missing physical processes in the models, inaccurate line-lists for the molecular species, or unknown issues in the observations or data processing.
However, these systematic effects are effectively removed when dividing by the mean spectrum to examine the variability trends, resulting in  typical discrepancies between the mean-normalised models and data of less than 0.3\% across the NIRSpec range and <1\% across the MIRI data, as shown in the lower panels of Fig. \ref{fig:retrieved-variability-map}.

The retrievals were able to accurately fit the spectra, and were able to reproduce the morphology of the observed spectroscopic variability, as shown in Fig. \ref{fig:retrieved-variability-map}, which can be compared to the observed variability maps of Fig. \ref{fig:binned_spectra_nirspec}.
For example, the brightness between 2 $\upmu$m and 3.5 $\upmu$m decreases between 100$^{\circ}$ and 200$^{\circ}$ of rotation, while shorter wavelengths display a higher frequency phase variation.
Fig. \ref{fig:retrieved-variability-map} also shows the difference between the observed and modelled variability maps.
In general, the residuals are normally distributed, and centred at 0, suggesting that there are few remaining systematic biases in the models.
With a standard deviation of 0.3\%, the residuals between the data and the models in the NIRSpec range are much smaller than the observed variability of ~3\%, which implies that the models are accurately capturing the varying atmospheric state.

\subsubsection{Fixed parameter retrieval tests}
To determine the impact of specific mechanisms driving the variability, we performed a series of tests, fixing different sets of parameters to the median retrieved values and observing if the retrievals retained sufficient flexibility to reproduce the observations.

For the first test, we fixed the temperature profile to the median profile across all phases, leaving the rest of the setup the same as in the baseline model.
As shown in the top panels of Fig. \ref{fig:fixed-param-vmaps} this setup is unable to reproduce the observed variability maps. 
The overall fits are poor and the structure of the variability does match that of Fig. \ref{fig:binned_spectra_nirspec}. 
It is clear that allowing the temperature profile to vary is critical to reproducing the variability, and that the variability cannot be explained solely through changes in the cloud coverage and chemistry.

In the second test, we fixed the cloud parameters to the median retrieved values from the baseline retrieval. In contrast to fixing the temperature profile, this setup is able to reproduce the observed variability better than the baseline retrieval, as shown in the bottom panels of Fig. \ref{fig:fixed-param-vmaps}.
This reinforces the findings of the fixed-temperature test, implying that changes in the thermal structure are driving the observed variability. While patchy clouds are necessary to fit the spectrum, variations in the cloud coverage or opacity are not required to explain the spectroscopic variability.
This model provides a better match to the observed variability map, with a standard deviation in the residuals of 0.24\% for PRISM (compared to 0.3\% for the fiducial model) and 1\% for the LRS (compared to 1.03\%). 
Crucially, the remaining residuals are smooth, without the visible features between 1 to 1.5 $\upmu$m and at 3.3 $\upmu$m, suggesting that the fixed cloud model provides a better measure of the \methane abundance than the fiducial model.
As such, we continue to use this model throughout this section to observe parameter variations as a function of phase, and to produce the figures in this work unless otherwise noted.
For completeness, the phase variation of each parameter is presented in Appendix \ref{app:retrievals} in Fig. \ref{fig:param_variation}.

\subsubsection{Statistical tests}
\begin{figure}
    \centering
    \includegraphics[width=\linewidth]{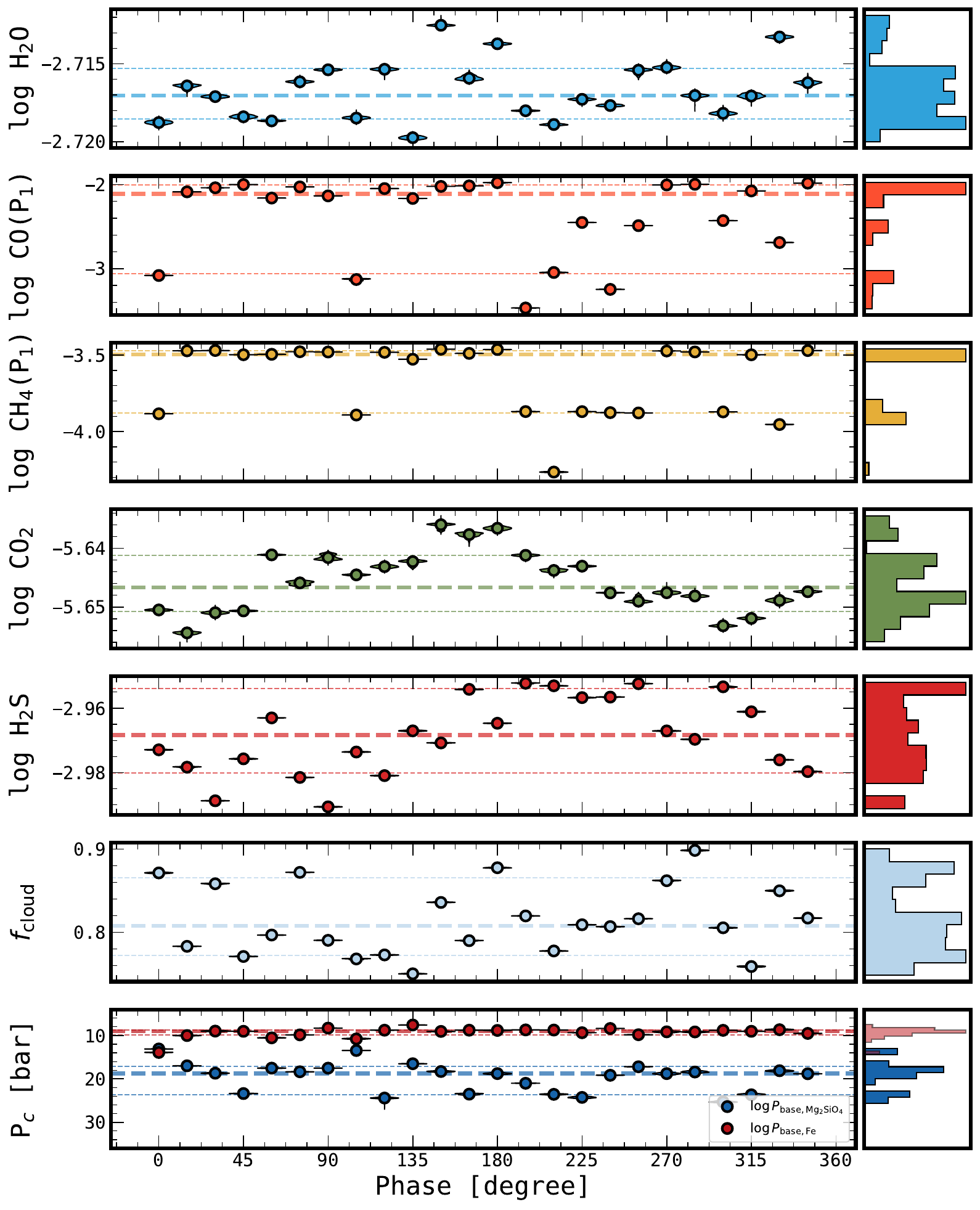}
    \caption{Measurements of key atmospheric parameters as a function of phase, taken from the fiducial retrieval. Also indicated are the median and $\pm1\sigma$ confidence intervals for each set of measurements.}
    \label{fig:parameter-variation}
\end{figure}

The retrieved posterior distributions, both in terms of median values and posterior widths are relatively consistent over the course of the observations, with only a few outliers.
In Fig. \ref{fig:parameter-variation}, we show the inferred posterior distributions at each phase for a set of key atmospheric parameters.
The inferred posterior width is significantly smaller than the observed variation in the parameters.
 To determine whether the measured variation in the atmospheric parameters is physical, or simply due to scatter in the measurements, we perform runs tests on each set of parameter measurements \citep{wald_runs_1940}.
We apply the \texttt{runstest\_1samp} function implemented in the \texttt{statsmodels} package, correcting for small sample statistics \citep{seabold2010statsmodels}.
This test determines the probability $1-p$ that a given series of runs (sequences of consecutive observations either greater than or less than the mean value) are drawn from a normal distribution. 
The sequence is considered to be distinct from random draws if $p<0.05$.

Across the ensemble of retrievals, \cotwo is the only parameter that is reliably found to follow a trend with $p_{\rm CO_{2}}=0.002$. For most of the retrieval sets, \htwos is also found to be non-random, with $p_{\rm H_{2}S}\approx0.002$; however, in the fiducial case, this is reduced to $p_{\rm H_{2}S}\approx0.14$.The fiducial model is the only setup where $p_{\rm H_{2}S}$ is not significant. 
Both \cotwo and \htwos follow similar trends in each set of retrievals, increasing until around 180$^{\circ}$ before decreasing towards 360$^{\circ}$.
The effective temperature is also found to be non-random, with $p_{\rm T_{eff}}=2\times10^{-5}$.

In addition to the non-parametric runs test, we also compare each parameter variability curve to the clustered light curves of \citetalias{mccarthy_simp_2025}, and to each other.
Using the clustering analysis of \cite{biller_weather_2024}, the authors identified nine unique light-curve patterns in the NIRSpec data and a further two in the LRS data.
These clusters were found to correspond to distinct pressures of the atmosphere. 
The combination of pressure and wavelength was linked to major opacity sources when compared to a Sonora Diamondback radiative-convective equilibrium forward model \citep{morley_diamondback_2024}. 
In the NIRSpec data, cluster one was assigned to the lowest
pressure, 0.12 bar, while cluster nine was assigned to the deepest pressure at 10.44 bar.
The light curves curve clusters could broadly be grouped into two categories.
Light curves one through five show approximately sinusoidal variation with phase, while clusters six through nine have a double-peaked shape.
This suggests that these groups of light curves are sensitive to different mechanisms driving the variability, with different physical processes causing the different characteristic shapes.

\begin{figure}
    \centering
    \includegraphics[width=\linewidth]{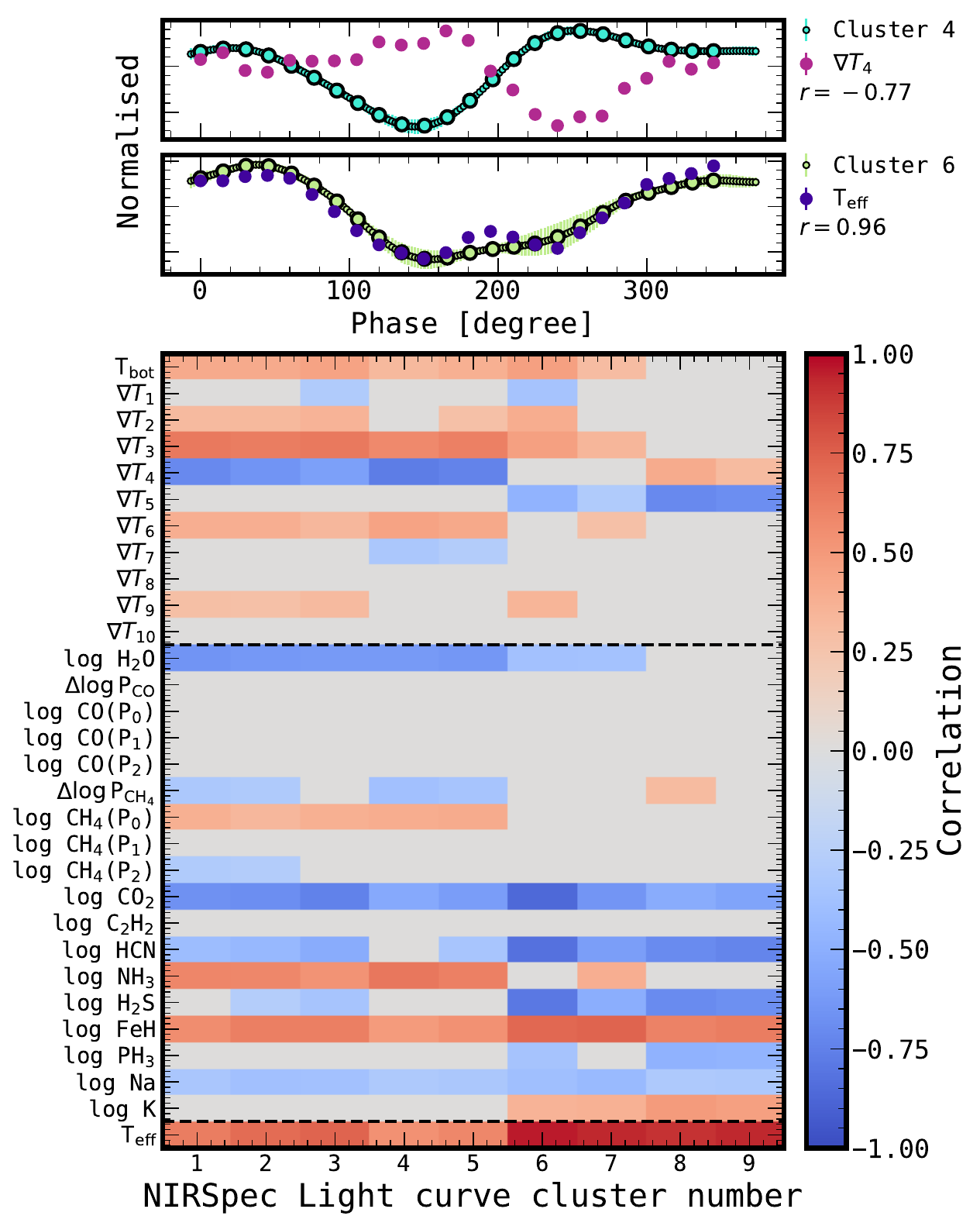}
    \caption{\textbf{Top:} Comparison of normalised effective temperature and the $\nabla T_{4}$ parameter with the clustered light curves of \citetalias{mccarthy_simp_2025}. The effective temperature is strongly correlated with the light curve, while $\nabla T_{4}$ is anticorrelated.
    \textbf{Bottom}: Matrix of Pearson correlation coefficients between atmospheric parameters and clustered light curves from \citetalias{mccarthy_simp_2025}. 
    Non-significant correlations ($|r|<0.271$) are not shown.}
    \label{fig:lightcurve-correlation}
\end{figure}

The correlation between the retrieved parameters as a function of phase and the clustered light curves is illustrated in the top panels of Fig. \ref{fig:lightcurve-correlation}.
For each pair we calculate the Pearson correlation coefficient, $r$. 
An $r$ value of -1 indicates that the pair are fully anti-correlated, while a value of 1 indicates full correlation.
Given the 24 measurements for each parameter and setting a threshold of 90\% confidence to determine significance, we find using a 2-sided t-test that the threshold for significant correlation is $|r|>0.271$.
The full set of $r$ values between the parameters and light curves is shown in Fig. \ref{fig:lightcurve-correlation}.
Most parameters are uncorrelated with the light curves.
The strongest correlation is with the effective temperature, which is strongly correlated with the light curves from 0.66 bar and deeper in the atmosphere.
\cotwo is strongly anti-correlated with the light curves, with the strongest anti-correlation with light curve 6.
\citetalias{mccarthy_simp_2025} found that light curve 6 probes 0.66 bar in pressure, and was also co-located with the 4.2 $\upmu$m \cotwo feature.

\begin{figure}
    \centering
    \includegraphics[width=\linewidth]{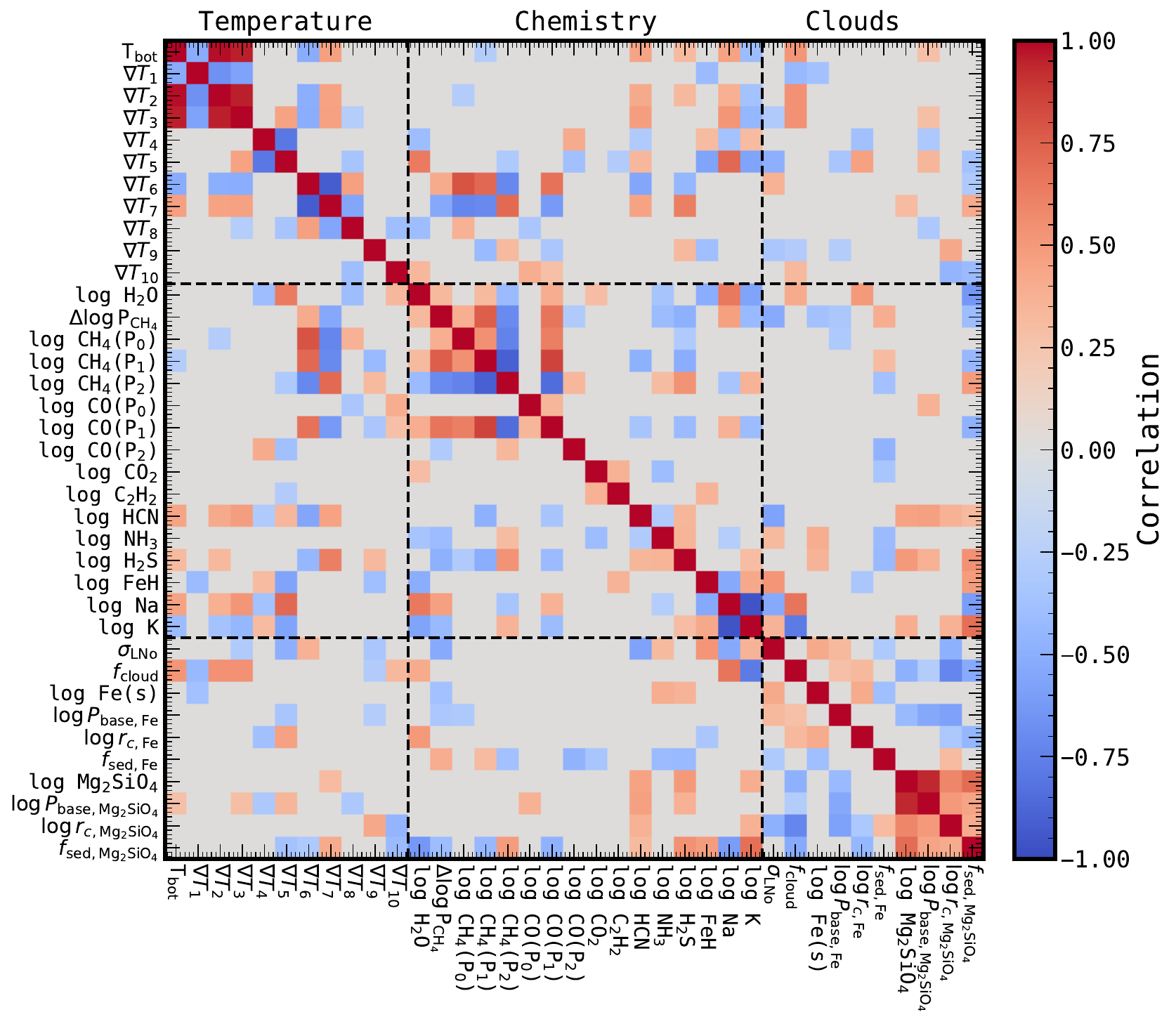}
    \caption{Pearson correlation coefficients between the phase-variation of each pair of retrieved parameters in the fiducial retrieval. Using a two-sided t-test with 24 samples, the minimum statistically significant correlation at a 90\% confidence level is 0.271, thus setting the limits on which correlations are shown. While some correlations are natural degeneracies of the model, such as the correlation between $\Delta\log$P$_{\rm CH_{4}}$ and the \methane abundance at that pressure node, other pairs are likely astrophysical in nature.
    }
    \label{fig:parameter-correlation}
\end{figure}

We also compare the correlations between the atmospheric parameters, shown in Fig. \ref{fig:parameter-correlation}.
Many of the parameter correlations are due to degeneracies in the model. 
For example, the parametrisation for the \methane abundance profile can set the pressure node at either the top or bottom of the photosphere to achieve the same overall abundance gradient.
Likewise, the silicate cloud mass fraction and the cloud coverage fraction are anti-correlated, as a decrease in the cloud coverage can be compensated for by an increase in the cloud mass fraction.

However, some correlations seem to be astrophysical in nature.
The strongest correlation between chemical parameters and temperature profile parameters is the correlation between the methane abundance parameters and the temperature gradient at the location of the temperature inversion, $\nabla T_{6}$. 
As methane is an effective thermometer for the atmosphere of SIMP 0136, this correlation demonstrates that the observed temperature inversion is likely found because of the measured methane abundance.


\subsubsection{Thermal structure evolution}
\begin{figure}
    \centering
    \includegraphics[width=\linewidth]{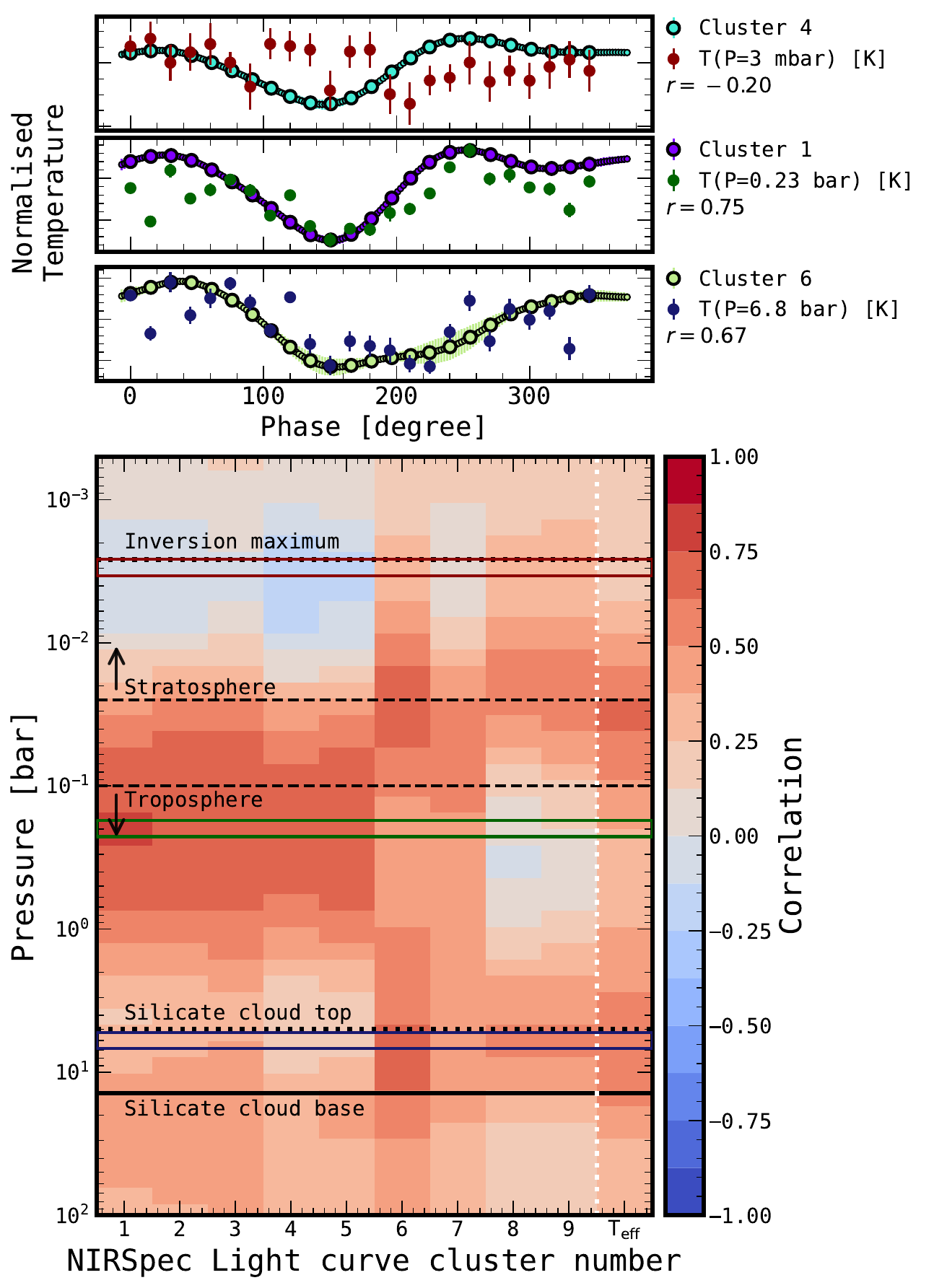}
    \caption{Correlation between the phase-variation of the temperature profile and the clustered light curves from \citetalias{mccarthy_simp_2025}. \textbf{Top:} Example temperature-variation curves drawn from three representative pressure levels in the atmosphere (highlighted boxes) compared to correlated or anti-correlated light curve clusters. \textbf{Bottom}:  Correlation coefficient between the temperature-variation at that pressure and the clustered light curves for each pressure level, as well as the correlation to the effective temperature in the right-most column. The pressure levels for the temperature curves in the top panels are highlighted with coloured boxes. Horizontal lines and annotations indicate the location of atmospheric features.}
    \label{fig:temp-lightcurve-correlation}
\end{figure}

The temperature profile in the troposphere (between 0.1 bar and 10 bar) is retrieved with high precision ($\pm$0.5 K per temperature point) at every phase, shown in Fig. \ref{fig:pt-profile}.
In this region the temperature changes by ${\pm}3$~K between phases per pressure level.
This level of temperature variation is consistent with the ${\sim}$1~K temperature variations predicted by \cite{showman_atmospheric_2013} for the top of the convective region in a rapidly rotating object. For SIMP-0136 we find the radiative-convective boundary to occur at 17 bar, as shown in Fig. \ref{fig:temp-gradient}.
At pressures between 17 bar and 0.1 bar the temperature profile is moderately sub-adiabatic, and is broadly consistent with the \texttt{ExoRem} temperature profile of an 1100 K object in radiative-convective equilibrium.
Below the base of the photosphere at 17 bar the temperature profile is essentially unconstrained, with the width of $\nabla \log T_{1}$ and $\nabla \log T_{2}$ posteriors determined by the prior distributions.

Every retrieved temperature profile shows evidence of a thermal inversion in the stratosphere (Fig. \ref{fig:pt-profile}), beginning at around 40 mbar, and reaching a maximum near 3 mbar.
The amplitude of the inversion varies more strongly with phase than the deep atmosphere, with a standard deviation of 40 K between phases at the peak of the inversion.
The amplitude of the inversion peak does not vary smoothly with phase, but is generally the strongest around $200^{\circ}$ and weakest near $345^{\circ}$, although the actual largest amplitude hot spot occurs at $0^{\circ}$. 
The location of the inversion coincides with the top of the photosphere, as measured by the $\tau=1$ surface in the 3.3 $\upmu$m \methane feature.
As in the case of the single retrieval, there is no contribution at pressures lower than 1 mbar, and therefore this region is also unconstrained. 
At the top of the atmosphere, the uncertainties reach about 100 times larger than the uncertainties in the adiabatic region, with the standard deviation between phases reaching 70 K.

\begin{figure}
    \centering
    \includegraphics[width=0.85\linewidth]{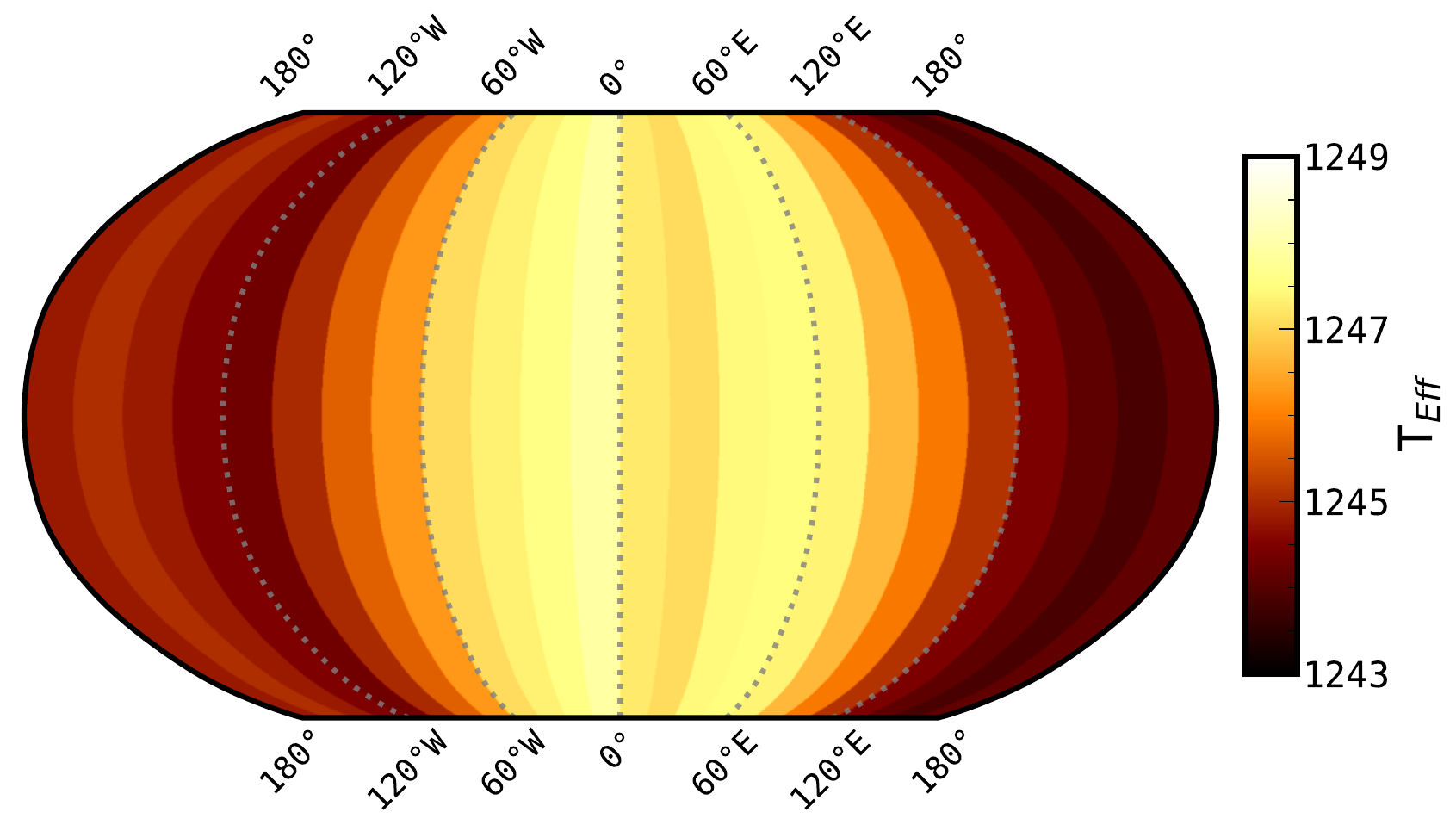}
    \caption{Effective temperature for each observed phase.  Over one rotation T$_{\rm eff}$ varies by 5 K. }
    \label{fig:teffmap}
\end{figure}

Using the same statistical tests as applied to the inferred atmospheric parameters, we test the measured temperature at each pressure for randomness using the runs test, and correlate the temperature variations with the clustered light curves of \citetalias{mccarthy_simp_2025}, which is shown in Fig. \ref{fig:temp-lightcurve-correlation}.
None of the individual temperature variation curves were found to be statistically different from random draws.
However, the effective temperature is found to be significantly different from random, with $p_{\rm T_{eff}}=2\times10^{-5}$.
The effective temperature varies smoothly over the surface of SIMP-0136, with hemisphere-averaged temperature variations of about 5 K as shown in Fig. \ref{fig:teffmap}.

While the correlations between the temperature profile and the clustered light curves at any individual pressure level are generally weak, there are clear trends between different atmospheric regions.
The PT profile is correlated with light curves 1 through 5 at all pressures below the stratosphere, with statistically significant anti-correlations in the inversion and near the top of the troposphere, as shown in Fig. \ref{fig:temp-lightcurve-correlation}.
This implies that at least some of the observed variability in light curves one through five can be attributed to changes in the strength of the inversion, and its resulting impact on the opacities in this region of the atmosphere.
There are transitions in the behaviour of the correlations both between the troposphere and stratosphere, and in the region of the silicate clouds.
We observe a general correlation between the tropospheric temperature profile and the effective temperature at both a few hundred mbar and deep in the atmosphere at a few bar. As the bulk of the observed emission is from the troposphere, we attribute the primary variation in the luminosity with the change in temperature in this reason, consistent with the ad hoc approach of \cite{tremblin_rotational_2020}.
These changes may be due to cloud radiative feedback \cite{tan_atmospheric_2021} driving temperature fluctuations, or by the turbulent atmosphere distributing energy inhomogeneously \citep{showman_atmospheric_2013}.

Conversely, the stratospheric inversion is anti-correlated with the light curves associated with molecular absorption features, and is only weakly correlated with the effective temperature. 
This is likely due to the strength of the inversion being probed by the methane absorption at 3.3 $\upmu$m and around 7 $\upmu$m, acting as a thermometer for the upper atmosphere.
While we cannot attribute a causal link between the changing stratospheric temperature and the methane absorption, we propose that mechanism may be changing auroral strength, which we discuss further in Section \ref{sec:disc:aurora}.

\subsubsection{Chemical variation with phase}
Of all the measured species, only CO$_{2}$ and H$_{2}$S are found to vary smoothly as a function of phase.
Both of these species are anti-correlated with the effective temperature, with $r_{\rm CO_{2}}=-0.69$ and $r_{\rm H_{2}S}=-0.41$.
However, in absolute terms both species only vary by small amounts, with \cotwo varying by about $\pm2$\% about its median value, while the precision on each measurement is about 0.06\%.
H$_{2}$S varies by $\pm4$\% about the median value, with a precision of 0.03\% on each measurement.
This level of variation is comparable to the scatter seen in other species: \water varies by about $\pm1$\%, albeit without any clear trend identified as a function of phase.
Without such a trend, we conclude that the abundances of most species, particularly \water, \methane, and CO are homogeneous over the surface of SIMP-0136, to within the scatter of the measurements, or about 1\%. 
In contrast, the abundances of CO$_{2}$ and H$_{2}$S are likely determined by the temperature of the atmosphere, increasing in abundance as the temperature falls. This pattern, particularly the sharp transition in the H$_{2}$S abundance in the fixed cloud retrieval (Fig. \ref{fig:param_variation}), is qualitatively similar to the correlation between the chemical abundances and gas temperature predicted in \cite{lee_dynamicallII_2024}. This suggests that the changes in the chemical abundances may be due to a storm rotating into view, bringing localised cloud, chemistry, temperature variations. 

\begin{figure}
    \centering
    \includegraphics[width=0.95\linewidth]{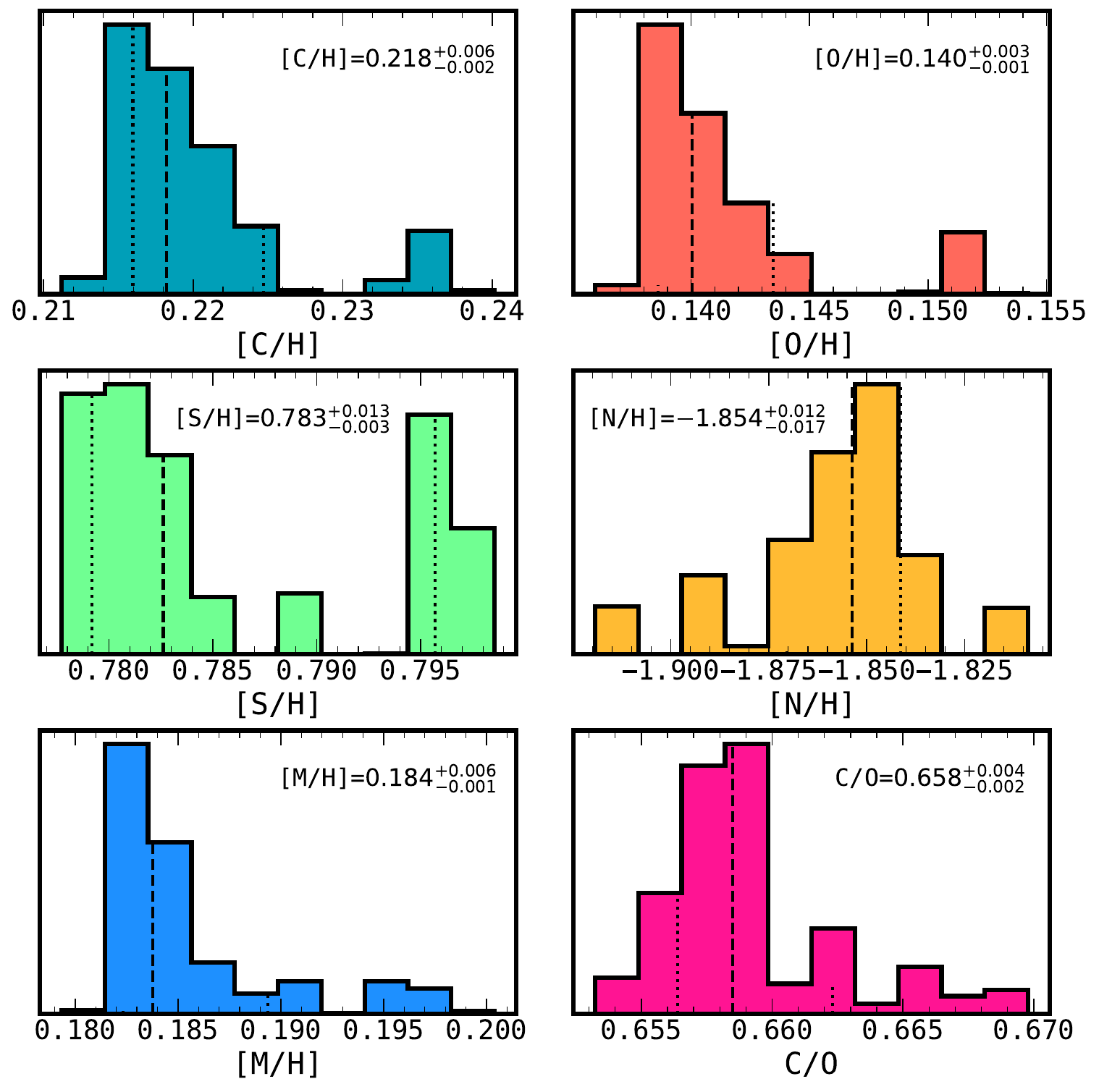}
    \caption{Elemental abundance ratios derived from retrieved molecular abundances combined from all phases.}
    \label{fig:evmr_hists}
\end{figure}

By measuring the abundances of the primary carriers of carbon, oxygen, sulphur, and nitrogen, we are able to calculate the atmospheric metallicity, as well as the elemental abundance ratios relative to the solar value \citep{lodders_abundances_2009}.
As the primary molecular species do not show trends with time, these elemental ratios can be measured by combining the abundance measurements at each phase.
We find that the metallicity is slightly enriched compared to the solar value, with $\left[\mathrm{M/H}\right] = 0.184_{-0.001}^{+0.006}$.
The $\left[\mathrm{C/H}\right]$ ratio is slightly more enhanced than the oxygen ratio, with
$\left[\mathrm{C/H}\right] = 0.218_{-0.002}^{+0.006}$ and
$\left[\mathrm{O/H}\right] = 0.140_{-0.001}^{+0.003}$.
$\left[\mathrm{S/H}\right]$ as measured by the H$_{2}$S abundance is strongly enhanced compared to solar, with $\left[\mathrm{S/H}\right] = 0.783_{-0.003}^{+0.013}$, while the nitrogen abundance is very strongly depleted
$\left[\mathrm{N/H}\right] = -1.85_{-0.02}^{+0.01}$. However, the nitrogen abundance is measured only through \ammonia and HCN, and does not account for N$_{2}$, which should be the dominant carrier at the temperatures of SIMP-0136 \citep{zahnle_methane_2014}.
From these elemental abundances, we can also calculate additional elemental abundance ratios. 
We find that the gas phase C/O ratio in the troposphere is modestly super-solar, with C/O = $0.658_{-0.002}^{+0.004}$, although this does not account for oxygen sequestration in the silicate clouds, which would reduce the bulk C/O ratio by around 20\% \citep{lodders_solarsystem_2003,calamari_predicting_2024}.
Accounting for this, the C/O ratio is $0.54\pm0.01$, consistent with the solar value.
Both values are significantly lower than the ratio found by \cite{vos_patchy_2023}, where C/O$=0.79\pm0.02$.
Similarly, we find the atmospheric S/O=$0.12\pm0.00$ and C/S=$5.60\pm0.15$.
The elemental abundance ratios are shown in Fig. \ref{fig:evmr_hists}.
\cite{biazzo_lowmass_metallicities_2012} measured metallicities for low-mass stars in the Carina-Near Association, of which SIMP-0136 is a member. 
While they do not measure elemental abundance ratios for volatile species, the typical metallicities for these objects is [Fe/H]=$0.08\pm0.05$, suggesting that SIMP-0136 is only modestly enriched in metals relative to objects with a similar origin.
Compared with its siblings, our metallicity measurement is more compatible with the expectations for a low-mass brown dwarf than previous metallicity measurements for SIMP-0136 \citep{line_uniform_2017}, as in \cite{vos_patchy_2023} who found a metal-poor atmosphere with $\left[\mathrm{M/H}\right]=-0.32_{-0.05}^{+0.05}$.

\subsubsection{No change in patchy clouds  with phase}
From the runs tests there is no evidence to support systematic variation of the cloud coverage as a function of phase $p_{f_{\rm cloud}}>0.05$, reinforcing that variations in the clouds do not drive the observed variability for this early T-dwarf.
Likewise, the cloud coverage is not found to be correlated with any of the clustered light curves.
The cloud parameters are highly (anti)correlated with each other,
and are weakly correlated with some temperature and chemical parameters.
While the patchy clouds are required to reproduce the observed emission spectrum, changes in the cloud coverage do not appear to drive the observed spectroscopic variability.
Finally, Fig. \ref{fig:fixed-param-vmaps} shows that the fixed-cloud set of retrievals provided a better fit to the variability map than the fully free retrievals, highlighting that changes in the cloud parameters are not necessary to reproduce the observed variability.
This is consistent with the picture of \cite{freytag_gravitywave_2010}, where the cloud patchiness is driven by gravity waves, resulting in patchiness on small spatial scales. 

\section{Discussion}\label{sec:discussion}
\subsection{Error inflation}\label{sec:errorinflation}
In contrast to previous studies \citep[e.g.][]{vos_patchy_2023}, the \textit{JWST} data in this work were obtained with much higher $S/N$, and developments in atmospheric models allow us to more accurately fit the broad wavelength spectrum.
However, the same $S/N$ that allows us to measure changes in the spectra over time also lays bare the limitations of our 1D parametric models: typical reduced $\chi^{2}$ values of retrievals in this study are ${\sim}300$, with systematic residuals of the order of 5\% between the models and data.
The typical solution to this is to add an error-inflation term to the retrieval, and increase the uncertainties until a reduced $\chi^{2}$ of 1 is achieved. 
However, we found this would increase the uncertainties, and by extension the widths of the posterior distributions, to the point where we were no longer sensitive to the observed level of variability.

Error inflation is a standard technique for inferring uncertainties on the data when they are difficult to directly measure, as is the case for the MIRI/MRS \citep[e.g.][]{kuhnle_depletion_2024,barrado_15nh3_2024,matthews_hcnc2h2_2025}, as well as for accounting for the uncertainties in the models.
By comparing the stochastic variations in the observed spectrum to the uncertainties calculated by the JWST pipeline in Fig. \ref{fig:binned_spectra_nirspec}, it is evident that the uncertainties are well understood.
However, in Fig \ref{fig:residuals_all} there are strong systematic residuals present between the data and the best fit models, indicating that the models poorly describe the observations.
This is consistent with the large reduced $\chi^{2}$ values, which demonstrate poor goodness-of-fit.

To compensate for these model uncertainties, we performed an additional set of retrievals using error inflation based on the approach of \cite{line_uniform_2015}, where
\begin{equation}
    \sigma_{\rm Total} = \sqrt{\sigma^{2} + 10^{b}}.
\end{equation}
Here, the total uncertainty $\sigma_{\rm Total}$ is found by adding an additional inflation term to the measured uncertainties on a per-instrument basis, where $b$ is a freely retrieved parameter.
We found that the uncertainties were substantially inflated for both NIRSpec and MIRI to compensate for the systematic deviation in the models, with $b_{\rm PRISM}=-31.41\pm0.02$ and $b_{\rm LRS}=-33.765\pm0.03$, and a resulting $\chi^{2}/\nu$ of ${\sim}1$ for each retrieval performed.
The typical uncertainties $\sigma$ on the NIRSpec PRISM data are of the order of $10^{-16}$ - $10^{-17}$ W/m$^{2}$/$\upmu$m. This implies that the total uncertainty is dominated by the 10$^b$ term, as $\sigma^2 + 10^{b}\approx 10^{b}$, resulting in a total uncertainty between 3-30 times greater than the measured uncertainties.

However, the inflation of the uncertainties on the posterior parameter distributions due to the increased measurement uncertainties resulted in an inability to distinguish variations as a function of phase. 
Key parameters such as the cloud coverage fraction and the cloud base pressure were unconstrained.
The median values of these parameters were also less physically consistent than in the non-inflated retrievals: for example, in the fiducial retrievals the silicate clouds are found to condense near 10 bar, at the pressure expected from the condensation curves.
In the inflated case, the silicate clouds condense at 0.01 bar, far above where they would be expected in an object with the effective temperature of SIMP-0136. 

Based on the inability to reproduce the observed variability, measure variations in atmospheric parameters, and the physical implausibility of the results of the inflated retrievals, we only consider the non-inflated case in presenting results for SIMP-0136. 
Although there remain significant systematic discrepancies between the models and the data, the magnitude of these variations is typically small, and they do not impact the ability to reproduce the observed variability. 
Future model development incorporating 3D atmospheric dynamics, improved line lists, and better cloud parameterisations is likely necessary to further reduce these discrepancies.

\subsection{Auroral heating in upper atmosphere}\label{sec:disc:aurora}
\begin{figure}
    \centering
    \includegraphics[width=\linewidth]{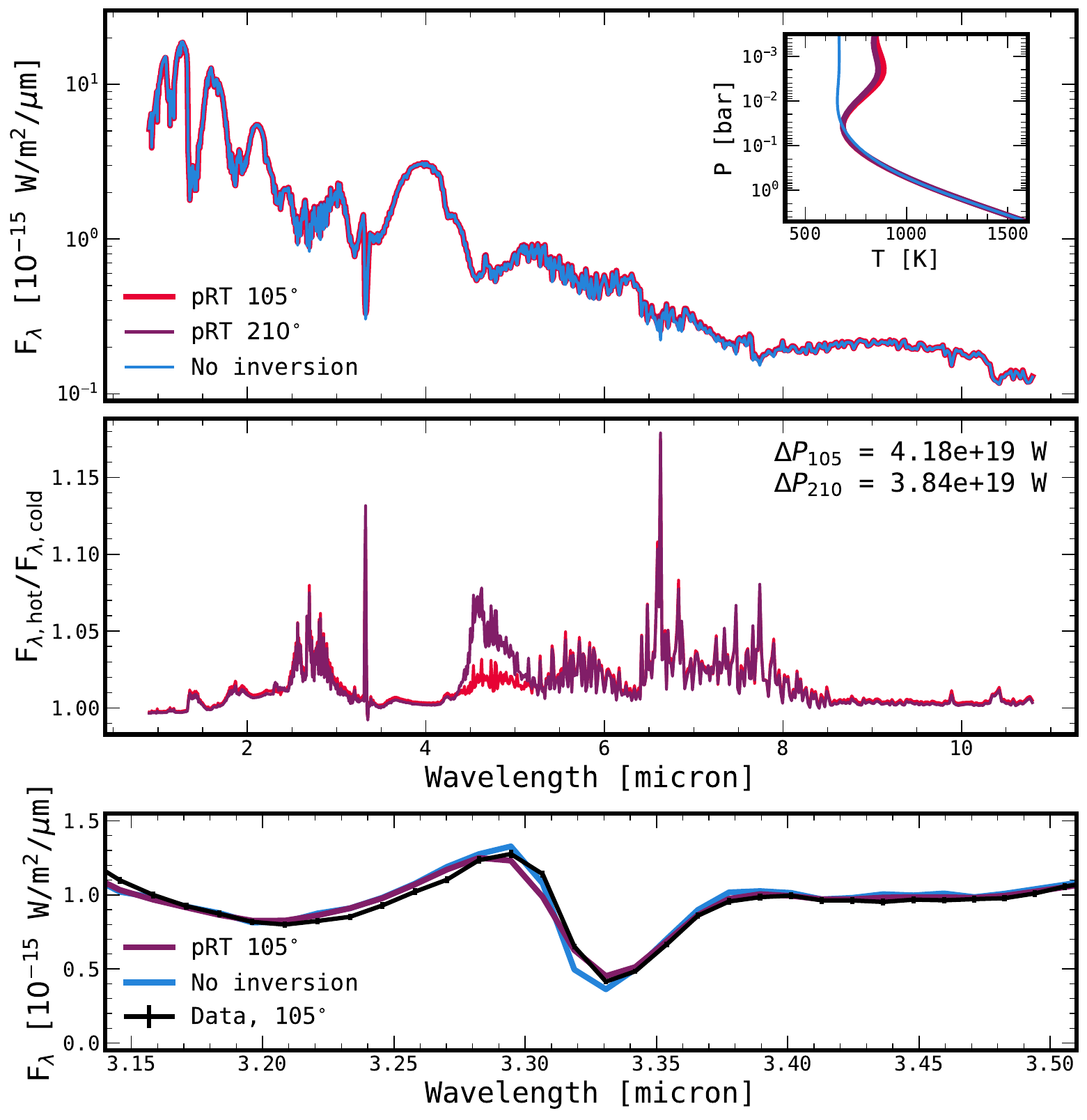}
    \caption{Comparison between retrieved best-fit spectrum and models without an upper atmosphere inversion to determine the contribution from the hot spot, similar to Fig. \ref{fig:hot_spot_model}. \textbf{Top:} Comparison of each of the spectra, with the inset panel showing the differences in the temperature profile. \textbf{Centre:} Ratio between the two retrieved spectra with the maximum ($105^{\circ}$) and minimum ($195^{\circ}$) inversion strengths, as well as a model without an inversion. \textbf{Bottom:} Comparison of the best-fit model at $105^{\circ}$ and the non-inverted model to the observed spectrum. The data are shown with $10\sigma$ uncertainties for visibility.}
    \label{fig:atmosphere_heating_spectra}
\end{figure}
SIMP-0136 is known to emit pulsed radio signals \citep{kao_aurora_2016,kao_aurora_2018}, which is attributed to the large scale currents generated by electrons flowing along magnetic field lines.
These electrons interact with the atmosphere, releasing energy and forming aurorae.
While our retrievals find heating in the stratosphere of SIMP-0136, it is challenging to directly attribute this to heating from auroral processes: there is still no direct evidence for UV or IR auroral emission in SIMP-0136.
Characteristic signatures of aurorae such as H$_{3}$+ \citep{gibbs_h3plus_2022,pineda_h3plus_2024} emission have not been observed, although higher resolution NIRSpec/G395H observations obtained with GO 5814 (PI: Fitzgerald) may place meaningful constraints on this emission.
Likewise, there no observations of UV emission from any brown dwarf aurora available at present \citep{saur_uv_2021}.
Without such direct measurements, we can explore whether auroral activity could reasonably explain the observed temperature inversion and place this in context with auroral observations from within the solar system.

The stratospheric heating appears to be present at all phases, and while we cannot resolve latitudinal differences, must cover a significant fraction of the observed hemisphere.
The maximum difference in the stratospheric temperature from an isothermal model is found to be about 265 K at 3 mbar, which is much stronger than the  $37\pm4$ K heating at 1 mbar in Jupiter \citep{rodriguez_temperature_2024}, but comparable to the 300 K auroral heating at 1-10 mbar on WISE J1935 \citep{faherty_emission_2024}.
However, the mechanism through which the atmosphere is heated by an aurora is unclear.
In Jupiter, the hotspot in the outer reaches (at pressures as low as $10^{-6}$ bar) of the atmosphere is thought to be heated by electron precipitation \citep{sinclair_aurora_2018}.
The mechanism for heating Jupiter's second hotspot, located at 1 mbar, is poorly understood \citep{rodriguez_temperature_2024}. 
Multiple mechanisms have been proposed, including adiabatic heating from local downwelling initiated by electron precipitation, and radiative heating of stratospheric aerosols produced by precipitation \citep{sinclair_photochemical_2017}, with the former explanation preferred in the literature \citep{sinclair_midir_2023,rodriguez_temperature_2024}. 
During strong auroral events, driven by interactions of the solar wind with plasma emitted by Io, Jupiter has experienced global stratospheric heating due to its aurora \citep{migliorini_ch4emission_2023,odon_globalheating_2021}.
Likewise, Neptune's aurora is not confined to its poles, and acts to significantly heat its atmosphere \cite{melin_neptune_2025}.
It is therefore plausible for a local aurora on SIMP-0136 to affect the global temperature structure.

From radio observations, \cite{kao_aurora_2018} inferred the presence of a strong magnetic field, of the order of 3000 G.
This is far stronger than the magnetic fields that drive the aurora on Jupiter, whose magnetic field strength is about 4~G.
Such strong magnetic fields will accelerate electrons, which will follow the field lines before precipitating in the stratosphere.
The depth of the atmosphere at which the electrons precipitate depends on the kinetic energy of the electron beam \citep{pineda_h3plus_2024}.
To precipitate between 1 and 10 mbar, electron energies between 100 and 1000 keV are required.
In Jupiter, the electrons typically carry 100 keV of energy \citep{gerard_mappingelectron_2014,allegrini_energy_2020}, so it is plausible that the higher magnetic field strength of SIMP-0136 is capable of accelerating the electrons to high enough energies to precipitate at the pressure of the observed hotspot.
Given the relatively high pressure at which a 1000 keV electron beam deposits its energy, \cite{pineda_h3plus_2024} also discuss how the reaction timescales of H$_{3}$+ with water are sufficient to prevent significant H$_{3}+$ emission, which would explain the lack of observed IR auroral emission in T-dwarfs.

We can estimate the required power deposited in the stratosphere to maintain the thermal inversion by comparing the difference in emission between a model with the inversion to a model with an isothermal stratosphere, shown in Fig. \ref{fig:atmosphere_heating_spectra}.
For this comparison we use the best fit retrieved spectrum at two different phases, and compare them to identical models with the temperature gradients above 0.1 bar set to 0.
The two phases plotted were chosen based on the variability map in Fig. \ref{fig:retrieved-variability-map} which highlights the phases where there appears to be enhanced emission in the 3.3 $\upmu$m and 7.7 $\upmu$m methane absorption features.
The structure of the ratio between the retrieved best-fit models and those with the isothermal stratospheres clearly resembles the forward models of Fig. \ref{fig:hot_spot_model}.
By taking the difference between the models, we can calculate the difference in the bolometric luminosity, and thus the emitted power.
By conservation of energy, this emitted power must be equal to the power input into the stratosphere to maintain the thermal inversion.
We find that for between the minimum inversion strength and the maximum, this power varies between $3.8-4.2\times10^{19}$ W.
This is significantly stronger than the IR auroral emission of Jupiter, which emits between $1-5\times10^{13}$ W \citep{gerard_mappingelectron_2014}.
However, in relative terms, Jupiter's aurora accounts for about 0.002\% to 0.01\% of Jupiter's luminosity, while for SIMP-0136 this is about 0.5\% of the total emitted power.
By scaling Jupiter's auroral UV emission of 10$^{12}$ W, \cite{saur_uv_2021} finds that brown dwarfs can emit $10^{19}$ W in the UV. 
Their scaling relations imply that a similar increase would be observed in the IR, which is consistent with our measurement of the additional auroral IR emission.

We can compare the \cite{saur_uv_2021} UV luminosity prediction to an estimate of the energy deposited in SIMP-0136's atmosphere from electron precipitation.
Juno/JEDI measurements find that for characteristic electron energies of 100 keV, the auroral electron beam deposits 100 mW/m$^{2}$ \citep{clark_precipitating_2018}.
The authors show that for 1000 keV electron energies, an aurora would deposit around 10$^{4}$ mW/m$^{2}$, assuming similar electron densities surrounding the object.
Optimistically assuming this is deposited evenly over the surface of SIMP-0136, we calculate a total power input of $5\times10^{17}$ W, or about 1\% of the observed excess emitted power.
Likewise, integrating the ionisation rate (${\sim10^{8}}$ cm$^{-3}$s$^{-1}$) of \cite{pineda_h3plus_2024} for a 1000 keV beam uniformly depositing its energy in a 25 km thick shell at 1 R$_{\rm Jup}$ produces an input power of $3\times10^{17}$ W. 
This inconsistency suggests that either additional heating mechanisms are necessary to maintain the stratospheric inversion, or that the mechanisms of auroral energy deposition in brown dwarfs are physically distinct from those in Jupiter.

While electron precipitation is thought to be the primary mechanism for auroral heating of the upper atmospheres of brown dwarfs and giant planets \citep[e.g.][]{sandel_euv_1982,melin_cassini_2011,mueller_magnetosphere_2012,pineda_h3plus_2024}, our rough estimates of the energy deposition from electron precipitation into the atmosphere of SIMP-0136 may be insufficient to fully explain the magnitude of the stratospheric inversion.
Joule heating is another mechanism that may play a key role, as the strong electric currents from the auroral electron flux will heat the atmosphere.
From Cassini and HST observations of Saturn, \cite{cowley_simple_2004} found that the Joule heating rate may be several orders of magnitude larger than from the electron precipitation.
In brown dwarfs, the magnetic fields are far stronger than in Saturn, and the higher temperature creates a higher ionisation fraction than in the cold atmospheres of the Solar System planets, and so the resulting electric currents will also be correspondingly larger in magnitude.
Dynamical effects such as gravity wave \citep{young_gravitywave_1997} have been proposed as an additional mechanism for heating the upper atmospheres of brown dwarfs \citep{freytag_gravitywave_2010}, and such waves have been observed to transport energy throughout the atmospheres of Jupiter and Saturn \citep{brown_evidence_2022,ingersoll_overturning_2021}.

\subsection{Dynamical regime}
A key prediction of GCMs is that the dynamical state of brown dwarf atmospheres is dependent on both the temperature, which controls the radiative timescale \citep[e.g.][]{showman_atmospheric_2013}, and the rotation rate \citep[e.g.][]{tan_atmospheric_2021}.
Changing these parameters changes the qualitative behaviour of the atmospheric structure, determining whether the presence and scale of bands and vortices \citep{hammond_shallowwater_2023}.
The measurement of the surface gravity, radius, effective temperature and the rotation rate, allows us to compute the thermal Rossby Number, Ro$_{\rm T}$ and the Non-dimensional Radiative Time-scale, $\hat{\tau}$, following equations 7 and 8 of \cite{hammond_shallowwater_2023}, and using the same values for the variables of $\gamma$ and $\kappa$. 
We find dimensional quantities of $\tau_{rad}$ to be between 10000 and 16000 s, assuming the bulk of the flux is emitted from the 1 bar level, and depending on whether  retrieved surface gravity and effective temperature are taken from the retrievals or \texttt{SEDkit} analysis. 
With a rotation rate of $\Omega = 7.27\times10^{-4}$ s$^{-1}$, this gives a non-dimensional radiative timescale of $\hat{\tau}$ between 15-23.
We compute an equilibrium geopotential of 10$^{5}$-10$^{6}$ m$^{2}$ s$^{-2}$, which gives a thermal Rossby number of Ro$_{\rm T}$ between $10^{-3}-10^{-4}$, placing SIMP-0136 in a regime where radiation forcing dominates and jets are not expected. 
Instead, the atmosphere is expected to be largely homogeneous only small-scale spatial variation, although the predicted variability amplitude of $\sim$0.1\% from the shallow water models in \cite{hammond_shallowwater_2023}   is inconsistent with the observed variability amplitude in SIMP-0136.
This inconsistency motivates the need to for GCMs to model the dynamic state of the atmosphere across this parameter space, to better capture the observed variability compared to the simplified shallow water models.

\subsection{Time-dependent retrieval parametrisations}
In this study, we performed a series of nearly independent atmospheric retrievals on time-series spectra of SIMP-0136.
However, this is only an initial step towards full three-dimensional atmospheric characterisation of this object.
To achieve this, there are several underlying assumptions that limit our approach.
First, we assume that the spectrum we observe at each time is independent.
In reality, we are always observing a full hemisphere of the object, and so the emission from neighbouring longitudes is correlated.
This is easily captured in 3D GCM models by integrating the emission over a hemisphere, but 3D radiative transfer is still too computationally expensive to be applied to atmospheric retrievals.
New approaches, such as GPU-optimised radiative transfer codes \citep[e.g. ExoJAX, gCMCRT,][]{kawahara_autodifferentiable_2022,lee_gcmcrt_2022}, or machine-learning accelerated approaches \citep[e.g. FlopPITy,][]{martinez_floppity_2024} may enable sufficiently fast 3D calculations for future retrieval frameworks.
At the same time, new sampling approaches, using gradient-descent based methods with autodifferentiable models \citep{kawahara_autodifferentiable_2022}, or machine-learning posterior estimation \citep{vasist_neural_2023,lueber_faster_2025} will reduce the number of model computations required to perform parameter estimation and model comparison, allowing for the use of more complex models.

Even with sufficiently fast model computation, it remains to be seen what practical approaches exist for parameterising time- and spatially-dependent atmospheric phenomena such as winds, bands, storms, and aurora.
Several approaches have been proposed for hot Jupiters, where the dynamics can be described to first order by just the equatorial jet \citep{blecic_implication_2017,dobbsdixon_gcmomotivated_2022,macdonald_trident_2022}.
However, for brown dwarfs the dynamics are  subtler, without displaying any strong diurnal contrasts.
From our independent retrievals, we see that several parameters, such as the effective temperature, and abundances for \cotwo and \htwos vary nearly sinusoidally. 
Such parameters $\theta$ could be modelled as 
\begin{equation}
    \theta(t) = \theta_{0}f\left(\phi(t) + \delta\right)
,\end{equation}
for phase $\phi(t)$ and initial phase offset $\delta$. Both $\theta_{0}$ and $\delta$ could be freely retrieved. 
In the simplest form, the function $f$ could be sinusoidal in nature, although more complex functional forms could be used in general.
Such an approach could be extended to higher order variations by including additional Fourier components to approximate an arbitrary periodic function, at the cost of a large number of additional free parameters. 
Using such parameterisations, models could be calculated at every binned timestep $t$, and jointly fit to the full set of time-resolved spectra, computing a single likelihood. 
To speed up computation, the models could be computed at relatively coarse time steps and interpolated onto the intermediate timesteps.
From our parameter measurements for SIMP-0136, we observe that most parameters would not need to be variable, at least at the precision available with current observations. 
The choice of parametrisation of the time-variability is difficult, and further study of atmospheric variability using both independent retrievals and through efforts to approximate 3D GCMs will be necessary to determine appropriate prescriptions.

\subsection{Future observations}
Even with precise \textit{JWST} observations, further study of SIMP-0136 is necessary to fully unravel its atmospheric structure and the processes driving the upper atmosphere heating.
High resolution observations of the 3.3 $\upmu$m feature could probe an extreme dynamic range in pressure, allowing for better characterisation of the temperature structure of the upper atmosphere.
Likewise, moderate resolution observations using NIRSpec/G395H and the MIRI/MRS would allow for observations of the full set of methane lines and the detection of H$_{3}$+, improving our ability to constrain the thermal structure, abundance profiles, and observe tell-tale signs of auroral emission.
Deep observations with HST could test to see if SIMP-0136 is a strong UV emitter, which would conclusively demonstrate the presence of an aurora.
Finally, long baseline time-series observations will allow for the observation of the same longitude across time.  This would enable us to better resolve the surface structure of SIMP-0136 and determine weather patterns that vary over timescales beyond the rotational period.

\section{Conclusions}\label{sec:conclusions}
Broad wavelength observations of SIMP-0136 from the NIR to the radio make it a benchmark for understanding the processes driving variability near the L-T transition. 
Its complex atmosphere varies with time, experiencing an evolving thermal structure, as well as stratospheric heating which we attribute to aurora.
Using our novel time-resolved retrieval method applied to high-precision data from \textit{JWST}, we were able to place the most precise constraints to date on its thermal structure, chemistry, and cloudiness.
For the first time, we have been able to quantify how the atmospheric state evolves over time.

\begin{enumerate}
\item Using \texttt{petitRADTRANS} atmospheric retrievals applied to \textit{JWST} NIRSpec/PRISM and MIRI/LRS time-series observations, we were able to precisely constrain the properties of SIMP-0136. Using \texttt{SEDKit}, we found the mass to be $15 \pm 3 $M$_{\rm Jup}$ and the radius to be $1.20 \pm 0.05 $R$_{\rm Jup}$. The retrievals suffered from the `small-radius' problem and we therefore treated the mass and radius as nuisance parameters, fixing them to 10.1 M$_{\rm Jup}$ and 0.93 R$_{\rm Jup}$, respectively. This produces a  surface gravity ($\log g = 4.49 \pm 0.01$) result that is consistent with the \texttt{SEDKit} estimate and provided a good fit to the observations. The effective temperature was found by integrating the models calculated by the retrievals, and was found to be $1245 \pm 1.4$ K, varying from 1243 K at the coldest to 1248 K at the warmest. 

\item We found that SIMP-0136 has a stratospheric inversion, peaking at $3\times10^{-3}$ bar, and with a maximum temperature approximately 300 K warmer than if the atmosphere remained isothermal above the base of the stratosphere near 0.03 bar. We hypothesise that this inversion is due to auroral heating driven by electron precipitation. SIMP-0136 must either have a very strong aurora to sufficiently heat the stratosphere or additional heating sources are required to maintain the inversion. More detailed modelling, including ion precipitation, Joule heating, and related magnetic effects is likely required to fully understand the heating mechanisms involved. 

\item  The bulk atmospheric metallicity is found to be constant as a function of phase. Using precisely measured chemical abundances, we found that the atmospheric metallicity is $0.18 \pm 0.01$.This is only slightly enriched compared to low mass stars in Carina Near association \citep{biazzo_lowmass_metallicities_2012}. The atmospheric C/O ratio was found to be $0.65 \pm 0.01$; however, when a 20\% oxygen sequestration into silicate clouds is accounted for, the bulk C/O ratio is $0.54 \pm 0.01$, consistent with the solar value.

\item In addition to the chemical abundance results, we demonstrated that our study is sensitive to vertical chemical gradients in the photosphere. We demonstrated that an altitude-varying chemical profile is necessary to accurately fit the \methane absorption features. 
Methane and carbon monoxide are in chemical disequilibrium. The methane abundance decreases at the location of the temperature inversions, while both \methane and CO are quenched between ${\sim}$10 bar and the base of the inversion.

\item We performed time-resolved atmospheric retrievals to investigate changes in atmospheric properties over time. The primary driver of the variability of SIMP-0136 is the changing thermal structure as a function of phase, similar to the temperature variations predicted by \cite{showman_atmospheric_2013}. Changes in the troposphere drive an overall change in the effective temperature of about 5 K during one rotation, while changes in the strength of the stratospheric inversion are tied to variations associated with changes at wavelengths dominated by \methane absorption. While patchy clouds were necessary to fit the spectrum at wavelengths shorter than 2 $\upmu$m, we did not observe statistically significant variations in the cloud properties or coverage fraction as a function of phase.

\item We found that most atmospheric properties are consistent with being constant across the surface of SIMP-0136. However, we found that the abundances of \cotwo and \htwos vary as a function of phase and were correlated with the effective temperature, suggesting that we may be observing chemical changes driven by dynamics and storms.
\end{enumerate}

SIMP-0136 is uniquely well-suited to studying the dynamic auroral, thermal, and chemical processes that drive variations in brown dwarf atmospheres. 
Further observations will reveal not only how these processes shape the atmosphere, but the timescales over which they vary.
Surprisingly, we find that changes in the cloud coverage are not necessary to explain the observed variability, although this is often pointed to as the primary mechanism.
Studies of a larger sample objects on both sides of the L-T transition will be necessary to determine where cloud variability becomes a relevant mechanism.
Lastly, model development is necessary to fully exploit the high-precision \textit{JWST} data, but also to reveal the physical processes that cause the measured changes in the thermal structure and auroral heating strength.

\begin{acknowledgements}
The authors would like to thank the anonymous reviewer for their thorough and  helpful feedback. 
We would also like to thank Elspeth Lee and Paul Molli\`{e}re for their insightful discussions about this project.
E. N., J. M. V., and C. O’T acknowledge support from a Royal Society - Research Ireland University Research Fellowship (URF/1/221932, RF/ERE/221108). M. L. and M. S. acknowledge support from Trinity College Dublin via the Trinity Research Doctoral Awards. This work is based on observations made with the NASA/ESA/CSA James Webb Space Telescope. The data were obtained from the Mikulski Archive for Space Telescopes at the Space Telescope Science Institute, which is operated by the Association of Universities for Research in Astronomy, Inc., under NASA contract NAS 5-03127 for JWST. These observations are associated with program GO:3548, and can be found at \href{https://doi.org/10.17909/pfnd-md36}{10.17909/pfnd-md36}. This work was supported in part by JWST-GO-03548.004-A.
Computations were performed on the HPC system Viper at the Max Planck Computing and Data Facility.
B. B. acknowledges support from UK Research and Innovation Science and Technology Facilities Council [ST/X001091/1].
A. M. M. acknowledges support from the National Science Foundation Graduate Research Fellowship Program under Grant No. DGE-1840990.

Approved for unlimited public release (United States Air Force, Public Affairs \# USAFA-DF-2025-476). The views expressed in this article are those of the authors and do not necessarily reflect the official policy or position of the United States Air Force Academy, the Air Force, the Department of Defense, or the U.S. Government.

Software used: \texttt{petitRADTRANS, species, jwst-pipeline, pyMultiNest, phot\_utils, Python, numpy, matplotlib, astropy}.
\end{acknowledgements}

\bibliographystyle{yahapj}
\bibliography{bib}

\begin{appendix}

\section{Benchmarking SIMP-0136}\label{app:gridfits}
\begin{table}
\caption{SIMP-0136 photometry.}
\label{tab:photometry}
\vspace{-1.5em}
\begin{footnotesize}
\begin{center}
\begin{tabular}{l|l}
\toprule
\textbf{Filter}       &  \textbf{Magnitude} \\
\midrule
\multicolumn{2}{l}{Synthetic absolute photometry}\\ 
\midrule
$J_{\mathrm{MKO}}$  (mag) & 14.385 $\pm$ 0.006 \\
$H_{\mathrm{MKO}}$  (mag)&  13.899 $\pm$ 0.006 \\
$K_{\mathrm{MKO}}$  (mag)& 13.665 $\pm$ 0.006 \\
$L'_{\mathrm{MKO}}$  (mag) &  12.026 $\pm$ 0.006 \\
$L_{\mathrm{MKO}}$  (mag) & 12.845 $\pm$ 0.006 \\
$M_{\mathrm{MKO}}$  (mag)& 12.225 $\pm$ 0.006 \\
2MASS $J$ (mag)& 14.529 $\pm$ 0.006 \\
2MASS $H$ (mag)& 13.809 $\pm$ 0.006 \\
2MASS $Ks$ (mag) &  13.630 $\pm$ 0.006 \\
NIRCam $F115W$ (mag)&  14.861 $\pm$ 0.006 \\
NIRCam $F140M$ (mag)&  15.376 $\pm$ 0.006 \\
NIRCam $F200W $(mag)&  13.977 $\pm$ 0.006 \\
NIRCam $F277W $(mag)&  13.698 $\pm$ 0.006 \\
NIRCam $F335M $(mag)&  13.390 $\pm$ 0.006 \\
NIRCam $F356W$ (mag)&  12.488 $\pm$ 0.006 \\
NIRCam $F444W$ (mag)&  11.870 $\pm$ 0.006 \\
MIRI $F770W$ (mag)&  11.519 $\pm$ 0.006 \\
MIRI $F1000W$ (mag)&  10.679 $\pm$ 0.006 \\
MIRI $F1065C$ (mag)&  10.665 $\pm$ 0.007 \\
\midrule
\multicolumn{2}{l}{Variability amplitudes}\\
\midrule
$J_{\mathrm{MKO}}$  ($\Delta$mag) & 0.021 $\pm$ 0.006\\
$H_{\mathrm{MKO}}$  ($\Delta$mag)&  0.020 $\pm$ 0.006\\
$K_{\mathrm{MKO}}$  ($\Delta$mag)& 0.013 $\pm$ 0.006\\
$L'_{\mathrm{MKO}}$  ($\Delta$mag) &  0.013 $\pm$ 0.006\\
$L_{\mathrm{MKO}}$  ($\Delta$mag) & 0.018 $\pm$ 0.006\\
$M_{\mathrm{MKO}}$  ($\Delta$mag)& 0.019 $\pm$ 0.006\\
2MASS $J$ ($\Delta$mag)& 0.021 $\pm$ 0.006\\
2MASS $H$ ($\Delta$mag)& 0.020 $\pm$ 0.006\\
2MASS $Ks$ ($\Delta$mag) &  0.013 $\pm$ 0.006\\
NIRCam $F115W$ ($\Delta$mag)&  0.022 $\pm$ 0.006\\
NIRCam $F140M$ ($\Delta$mag)&  0.017 $\pm$ 0.006\\
NIRCam $F200W $($\Delta$mag)&  0.013 $\pm$ 0.006\\
NIRCam $F277W $($\Delta$mag)&  0.021 $\pm$ 0.006\\
NIRCam $F335M $($\Delta$mag)&  0.025 $\pm$ 0.006\\
NIRCam $F356W$ ($\Delta$mag)&  0.015 $\pm$ 0.006\\
NIRCam $F444W$ ($\Delta$mag)&  0.013 $\pm$ 0.006\\
MIRI $F770W$ ($\Delta$mag)&  0.016 $\pm$ 0.006\\
MIRI $F1000W$ ($\Delta$mag)&  0.008 $\pm$ 0.006\\
MIRI $F1065C$ ($\Delta$mag)&  0.016 $\pm$ 0.007\\
\bottomrule
\end{tabular}
\end{center}
\end{footnotesize}
\vspace{-0.5em}
\begin{tablenotes}
    \item \textbf{Notes:} Synthetic absolute photometry is derived from binned spectra using \texttt{species} \citep{stolker_miracles_2020_species}.
\end{tablenotes}
\end{table}
To infer the bulk properties of SIMP-0136, we used \texttt{SEDkit}, which reproduces the methods of \cite{filippazzo_fundamental_2015}. We found L$_{\rm bol} = -4.641 \pm 0.003$,
T$_{\rm eff}$ = $1136 K \pm 22$ K,
Radius = 1.20 R$_{\rm Jup} \pm 0.05 $R$_{\rm Jup}$,
Mass = 15 M$_{\rm Jup} \pm 3 $M$_{\rm Jup}$ and
log~$g = 4.4 \pm 0.1$.
These measurements are largely consistent with existing literature values, \citep[e.g.][]{vos_patchy_2023}, although the improved precision and spectral coverage of the \textit{JWST} observations reduce the uncertainty on the measurements.
We also used the rejection sampling method of \cite{dupuy_masses_2023,miles_jwst_2023} to compare the properties of SIMP-0136 to the evolutionary models of \cite{saumon_evolution_2008}. 
These results were consistent with the \texttt{SEDkit} measurements, finding 
T$_{\rm eff}$ = $1136 K \pm 22$ K,
radius = 1.15 R$_{\rm Jup} \pm 0.05 $R$_{\rm Jup}$,
Mass = 15.8 M$_{\rm Jup} \pm 2.3 $M$_{\rm Jup}$, and
log~$g = 4.47 \pm 0.1$.
Similarly to VHS 1256 b \citep{miles_jwst_2023}, the posterior distributions were slightly bimodal, as SIMP-0136 is both young and has a mass near the 13 M$_{\rm Jup}$ deuterium burning limit.

We performed a small set of fits using self-consistent radiative-convective equilibrium grids using \texttt{species} \citep{stolker_miracles_2020_species} to determine a baseline for the planet properties.
We used the low resolution \texttt{ExoRem} models \citep{charnay_self-consistent_2018,blain_exorem_2021}, the cloud-free \texttt{ATMO} models from \cite{petrus_x-shyne_2023}, and the cloudy Sonora Diamondback models \citep{morley_diamondback_2024}.
The fits of each of these are included in appendix \ref{app:gridfits}. 
None of the self-consistent grids found good fits to the JWST spectrum, with reduced $\chi^{2}$ values much greater than 1000.
There was also significant variation in the inferred properties, with T$_{\rm eff}$ varying between 1000 K and 1325 K, $\log g$ between 3.5 and 5, and [M/H] between -0.2 and 0.2. 
While these results are broadly consistent with the literature, they lack the accuracy or goodness-of-fit to characterise the observed variability in the spectrum.
This motivates the need for a retrieval analysis of SIMP-0136, whereby the precise spectra obtained with JWST can be used to infer the atmospheric parameters, which may disagree with forward model assumptions.

We additionally provided synthetic photometry for a range of filters, as well as the variability amplitude in each filter in Table \ref{tab:photometry}.

\begin{figure}
    \centering
    \includegraphics[width=\linewidth]{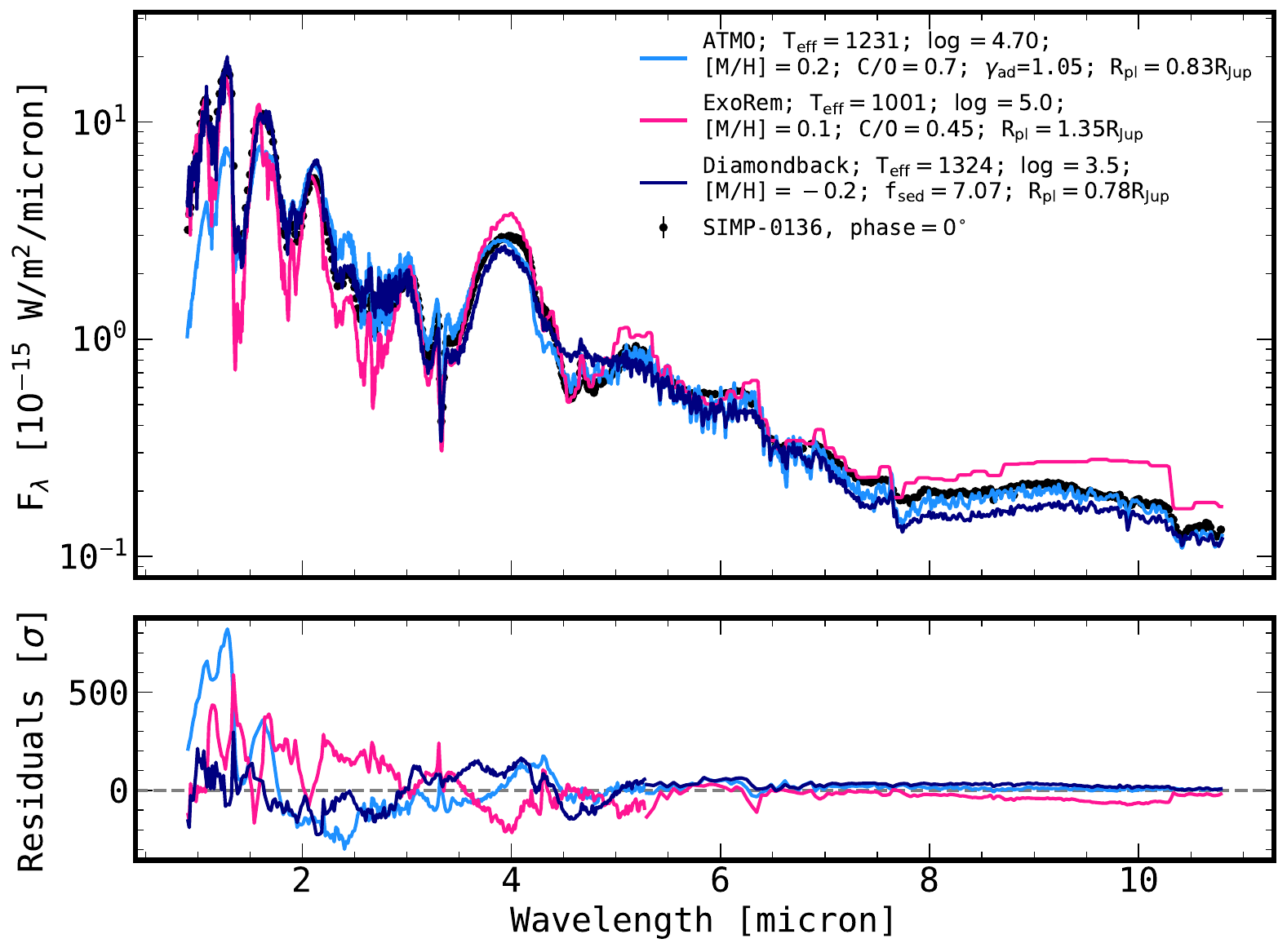}
    \caption{Fits of three self-consistent models to the SIMP-0136 spectrum at $0^{\circ}$ phase. Statistical goodness-of-fits are poor, with Diamondback having the lowest reduced $\chi^{2}$ at 6327. Inferred parameters also significantly vary between each of the fits.}
    \label{fig:grid-fits}
\end{figure}

\section{Forward modelling of variability processes}\label{app:forward}
The primary drivers of variability in SIMP-0136 have been attributed to varying cloud coverage \citep[e.g.][]{artigau_photometric_2009,vos_patchy_2023,mccarthy_multiple_2024}, auroral processes driving heating in the upper atmosphere \citep[e.g.][]{pineda_panchromatic_2017,richey_correlation_2020}, changing carbon chemistry \citep[e.g.][]{tremblin_cloudless_2016,tremblin_thermocompositional_2019,tremblin_rotational_2020,biller_weather_2024, mccarthy_simp_2025}. 
We used \prt to model how each of these processes can impact the spectrum of SIMP-0136, keeping all other parameters constant to isolate the independent effects.
Across these three processes, we found that each of them imparts distinct changes in the spectrum, providing confidence that retrievals were not only be able to distinguish between the impacts, but could be used to characterise the extent to which each of these processes contributes to the observed spectroscopic variability.
When compared to the typical uncertainties on each wavelength bin of 1\%, we demonstrated that relatively modest changes in parameters resulted in easily detectable changes in the observed spectrum.

\subsection{Upper atmosphere heating}\label{sec:temp-forward-model}
\begin{figure}
    \centering
    \includegraphics[width=\linewidth]{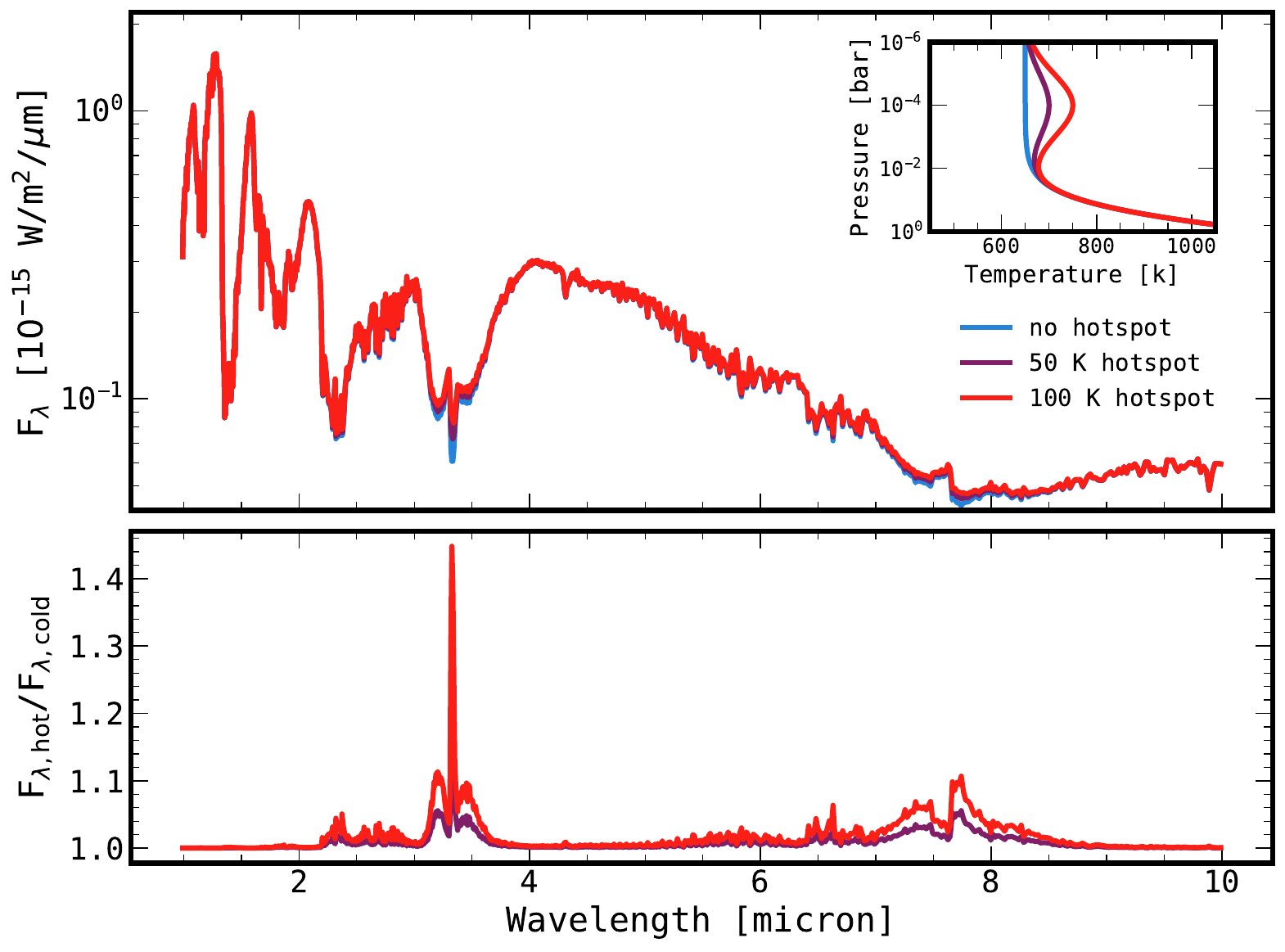}
    \caption{Variation in the spectrum due to an upper atmosphere temperature perturbation, centred at 10$^{-4}$ bar. Due to the changes in the temperature structure, both the equilibrium chemical abundances and the opacities structure change, inducing changes in the outgoing emitted spectrum. The largest impact of this upper atmosphere warming is on the 3.3 $\upmu$m absorption feature, due to its relatively high abundance in the upper atmosphere and large cross section at 3.3 $\upmu$m, which causes the atmosphere to become optically thick at lower pressures. }
    \label{fig:hot_spot_model}
\end{figure}

SIMP-0136 is known to be auroral from radio observations \citep{kao_aurora_2016,kao_aurora_2018}.
Auroral processes can induce heating of the upper regions of the atmosphere due to the ionisation of atoms by high-energy electrons, leading to an increase in the temperature, as has been measured on Jupiter \citep[e.g.][]{gerard_mappingelectron_2014,odon_globalheating_2021,sinclair_midir_2023,rodriguez_temperature_2024}, Saturn \citep{gezari_newaurorasaturn_1989}, Neptune \citep{melin_neptune_2025} and on Y-dwarf WISE-J1935 \citep{faherty_emission_2024}.
 To test the impact of the upper atmosphere thermal structure on the observed spectrum, we generated an emission spectrum using parameters derived from test retrievals of SIMP-0136.
We used an equilibrium chemistry model \citep[\texttt{easyCHEM},][]{molliere_observing_2017}, with solar metallicity and C/O (0.55, \cite{lodders_abundances_2009}.
This was repeated using vertically constant abundances  to ensure variations in the spectrum were attributable to changes in the relative absorption and emission in each layer, as opposed to changes in the chemical abundances.
As a baseline, we use an Eddington temperature profile \citep{eddington_effect_1930}, where 
\begin{equation}\label{eqn:eddington}
T^{4}(P)=\frac{3T_{\rm int}^{4}}{4}\left(\frac{2}{3} + \frac{\kappa_{\rm IR}P}{g}\right),
\end{equation}
setting $T_{\rm int}=773$ K, $\kappa_{\rm IR}=0.18$ and $\log g=4.49$ based on a test retrieval of SIMP-0136 using the Guillot profile.
We applied a Gaussian perturbation to the temperature profile, centred at 10$^{-4}$ bar and with a width of 1 bar. 
The peak of the perturbation was set to 50 K and 100 K from the nominal temperature profile.
This is a similar setup to \cite{morley_spectral_2014}, who identified such upper atmosphere heating as a source of variability in brown dwarfs.

The temperature profiles, together with the spectra computed for each model are shown in Fig. \ref{fig:hot_spot_model}.
The increase in the upper atmosphere temperature primarily impacts the 3.3 $\upmu$m methane feature.
At the temperatures of SIMP-0136, this is the strongest source of opacity in the atmosphere, and thus probes the lowest pressures, and is the only molecule which probes the region affected by the temperature perturbation.
With sufficient increase in temperature (around 400 K at the effective temperature of SIMP-0136), this methane feature transitions from an absorption feature to an emission feature, as observed in a Y-dwarf by \cite{faherty_emission_2024}.
Given the clear absorption feature in the SIMP-0136 spectrum, we therefore expect to be sensitive to the temperature structure at pressures as low as $10^{-3}$ bar.
This temperature sensitivity is limited by the low resolution of NIRSpec/PRISM, and higher resolution observations would probe a larger dynamic range in pressure.

These results were similar between the equilibrium models and for the constant abundance models, which saw an even larger change in the 3.3 $\upmu$m methane feature, as well as larger changes in the 2.3 $\upmu$m CO feature.
We also explored varying the location of the inversion, as well as how the inversion impacts objects with different effective temperatures.
At higher pressures, the inversion acts similar to an increase in the effective temperature, broadly increasing the flux emitted throughout the spectrum.
This results in a closer match to the flux ratios of \cite{morley_spectral_2014}, who placed the hotspot at pressures of 0.1 bar or deeper.
At lower pressures, the impact becomes more localised to the strongest opacity sources, primarily the 3.3 $\upmu$m and 7.7 $\upmu$m \methane features.
At lower effective temperatures, the dominant source of opacity is ammonia, and so the strongest impact of an inversion is on the 10 $\upmu$m absorption feature.
At higher temperatures, as CO becomes dominant over \methane the relative impact on the methane features decreases, while the impact on the CO features increases.

\subsection{Carbon chemistry}\label{sec:chem-forward-model}
\begin{figure}
    \centering
        \includegraphics[width=\linewidth]{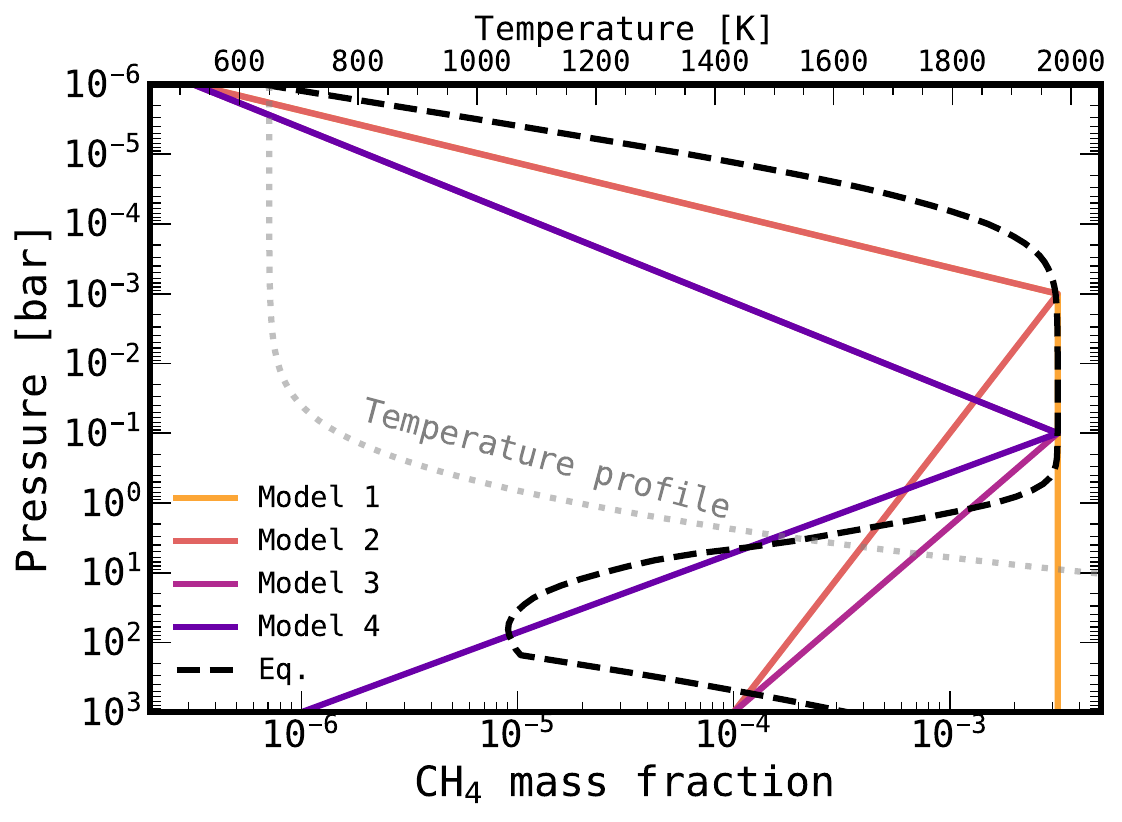}
        \includegraphics[width=\linewidth]{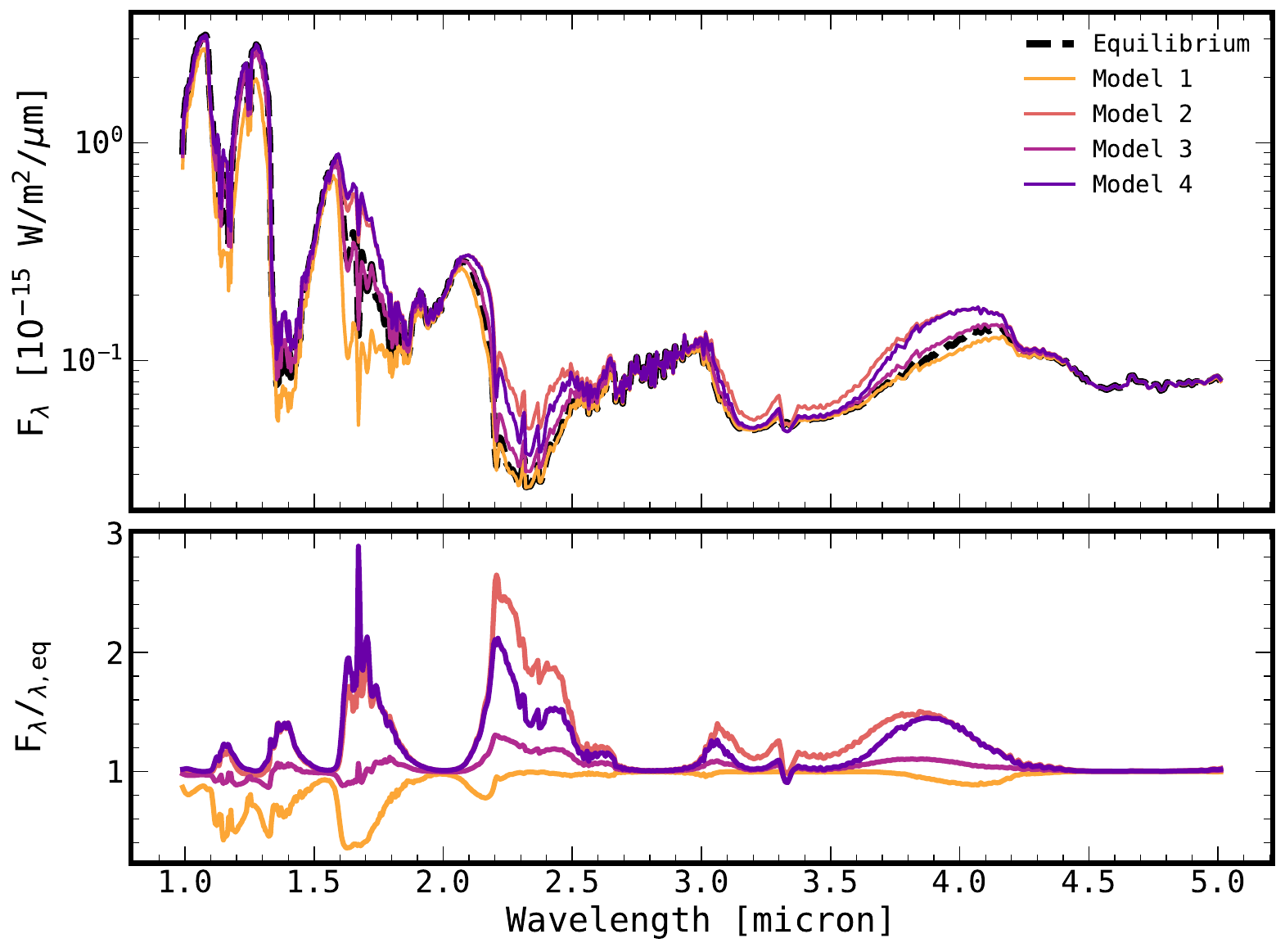}
    \caption{\textbf{Top:} Set of mass-fraction \methane profiles, demonstrating the linear spline parameterisations. The black dashed line shows the equilibrium profile for \methane for [M/H]=0 and C/O = 0.55, for the temperature profile illustrated by the grey dotted line.
    \textbf{Bottom:} Equilibrium chemistry models have large variations in abundances throughout the atmosphere. In this model, we fixed all of the abundances to the median retrieved values of Table \ref{tab:priors}, except for that of methane. The different colours correspond to the same abundance profiles as in the top panel. Even though these parameterisations only affect the methane abundance, the impact on the spectrum is clearly distinct from the variations caused by the hotspot. }
    \label{fig:model-non-constant-abund-profile}
\end{figure}

Changes in the carbon chemistry have been identified as a potential mechanism for the observed variability in the absorption features of carbon-bearing molecules, and thus precise characterisation of the abundances of these species is necessary to explore their evolution over time \citep{tremblin_cloudless_2016,tremblin_thermocompositional_2019,tremblin_rotational_2020}.
Equilibrium chemistry calculations show that varying the temperature at a given pressure will change the chemical composition.
While many L-T transition objects are thought to experience strong vertical mixing \citep{mukherjee_probing_2022}, which homogenises the chemical abundances, this has not yet been tested using a data-driven approach.
Furthermore, the vertical mixing strength is expected to decrease with altitude \citep{tan_atmospheric_2021,mukherjee_probing_2022}, allowing the chemical abundances to equilibrate, albeit at very low temperatures; then, the reaction rates become low enough to remain out of equilibrium.
The resulting abundances in the upper atmosphere would therefore be sensitive to upper atmosphere heating.

 To determine the sensitivity to changes in the abundance profile, we calculated an emission spectrum for each of the profiles of, keeping all species other than \methane vertically constant.
For the equilibrium case, we only allow the \methane abundance to vary under equilibrium assumptions ([M/H]=0, C/O=0.55), again keeping all other abundances vertically constant.
The abundances at the top and bottom of the atmosphere, as well as at the pressure node were chosen to approximate the equilibrium case.
Model 1 is vertically constant, with an abundance equal to the equilibrium abundance in the upper regions of the photosphere (between 1-10$^{-3}$ bar).
Model 2 reaches the same abundance at 10$^{-3}$ bar, and approximates the slope in the very upper atmosphere.
Model 3 reaches the same abundance deeper in the atmosphere, near 1 bar, and a comparison with Model 4 allows us to determine the sensitivity to the chemical gradient in the photosphere.
Each of these profiles is shown in the top panel of Fig. \ref{fig:model-non-constant-abund-profile}. 

The resulting spectra are shown in the bottom panel of Fig. \ref{fig:model-non-constant-abund-profile}.
The relative strengths of the 1.6 $\upmu$m and 2.2 $\upmu$m \methane features make each case clearly identifiable relative to each other and to the equilibrium case.
The structure of the variation is clearly distinct from the variations induced by changing the thermal structure in the upper atmosphere, and primarily changes the strength of the 3.3 $\upmu$m \methane feature.
We expect this sensitivity to chemical gradients to extend to other species as well.
In practice, the methane abundance could not decrease without a sink for the additional carbon atoms in another carbon bearing species, and from equilibrium chemistry we would expect the CO abundance to increase in proportion to the decrease in methane.

\subsection{Patchy clouds}
\begin{figure}
    \centering
    \includegraphics[width=\linewidth]{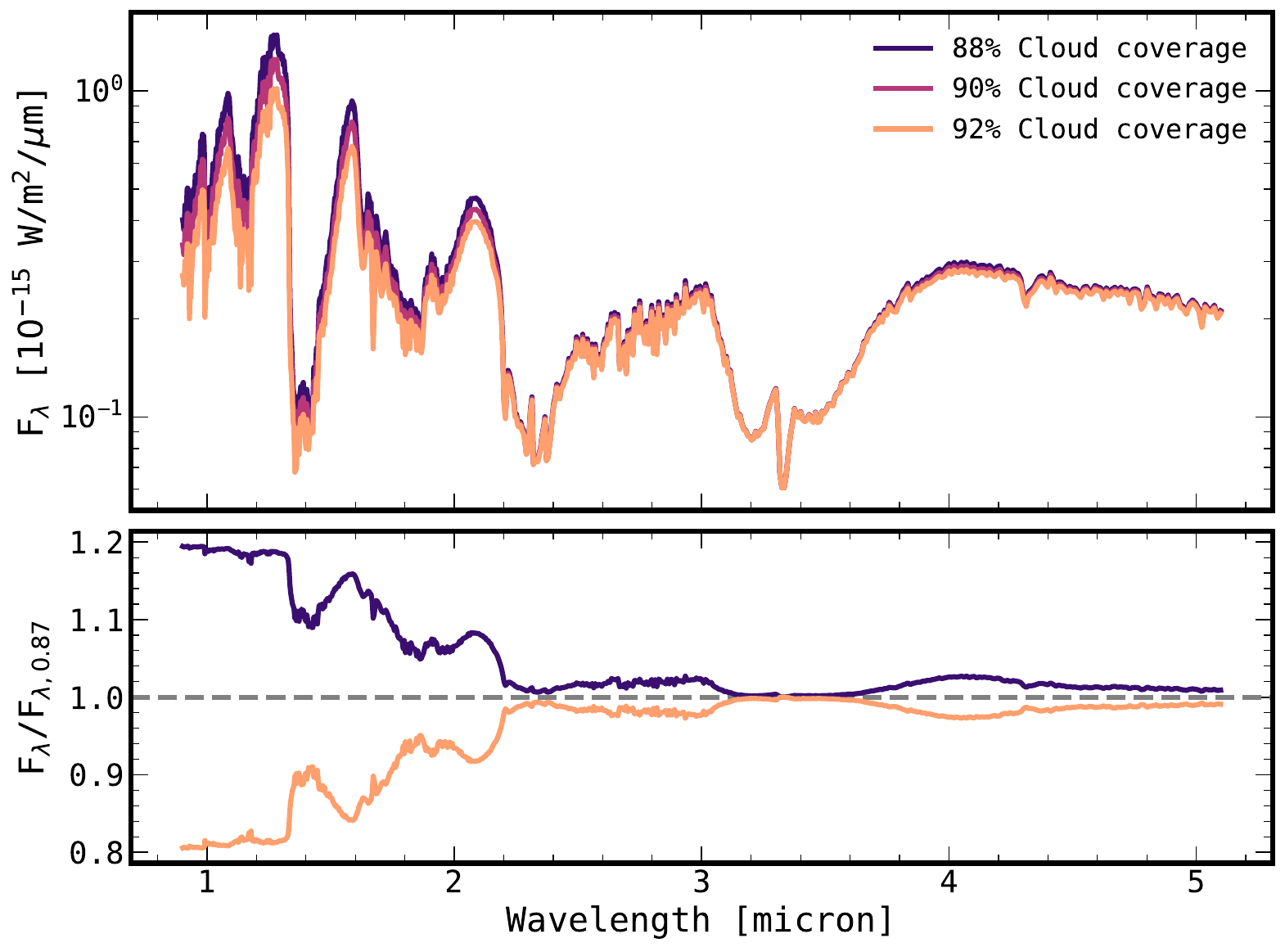}
    \caption{Cloud patchiness is thought be the primary driver of near IR variability. Here we show that changes in the silicate cloud patchiness of only a few percent have large impacts on the spectrum, with larger changes at short wavelengths. This is again clearly distinct from the spectral changes due to the thermal structure or composition.}
    \label{fig:patchy-cloud-model}
\end{figure}

Patchy cloud coverage is thought to be the primary driver of variability in L-T transition objects, and has been identified as a primary mechanism for the variability in SIMP 0136 \citep{mccarthy_multiple_2024}.
As the silicate clouds condense deeper in the atmosphere, the cloud deck occurs near the photosphere, leading to relatively substantial changes in the cloud coverage fraction over the surface.
We use the same equilibrium model as with the baseline for the temperature perturbation.
Test retrievals indicated that SIMP-0136 is around 90\% covered in silicate clouds, somewhat more than the 70\% found in \cite{vos_patchy_2023}.
We therefore varied the cloud patchiness by 2\% about this nominal value, as shown in Fig. \ref{fig:patchy-cloud-model}.
This change in the cloud coverage induces a 20\% change in the NIR flux, where the continuum absorption of the silicate clouds increases.
This degree of the variation demonstrates that we are sensitive to small changes in the cloudiness of the object, of the order of a percent.
There are also smaller changes induced in the K and M bands, where \water and CO absorption deep in the atmosphere are impacted by the presence of the silicate cloud layer.
The overall shape of this variation reproduces the patchy models of \cite{morley_spectral_2014}, although we find a greater impact on the spectrum for a smaller change in the cloud coverage, at a similar effective temperature.
While this patchiness is clearly idealised and does not account for degeneracies within the cloud model which may compensate for decreased cloud coverage with, for example, an increased mass fraction of the cloud, it highlights that the silicate clouds impart a unique spectral variation which cannot be replicated by varying the temperature structure or chemistry.

\newpage
\onecolumn
\section{Additional retrieval results}
\label{app:retrievals}
\begin{figure}[h!]
    \centering
    \includegraphics[width=0.9\linewidth]{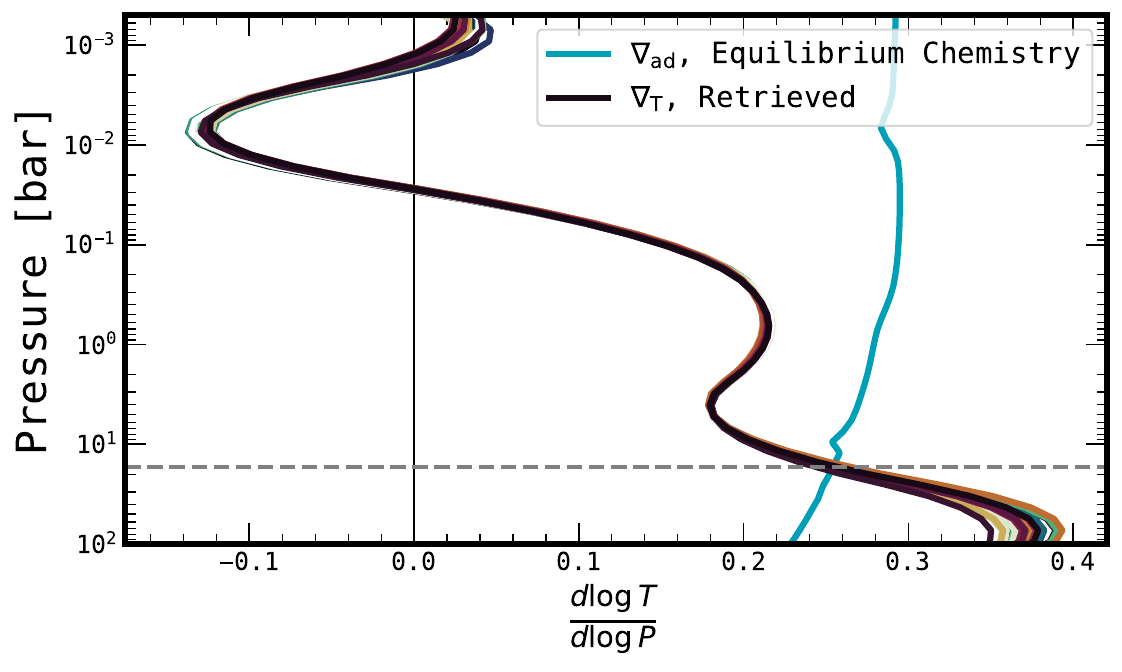}
    \caption{Temperature gradient as a function of pressure. At pressures higher than 17 bar, the retrieved temperature gradient is steeper than an adiabatic profile, and therefore convectively unstable, while at pressures lower than 17 bar it is shallower than an adiabatic profile and is convectively stable.}
    \label{fig:temp-gradient}
\end{figure}

\clearpage
\onecolumn
\begin{figure}
    \centering
    \includegraphics[width=\linewidth]{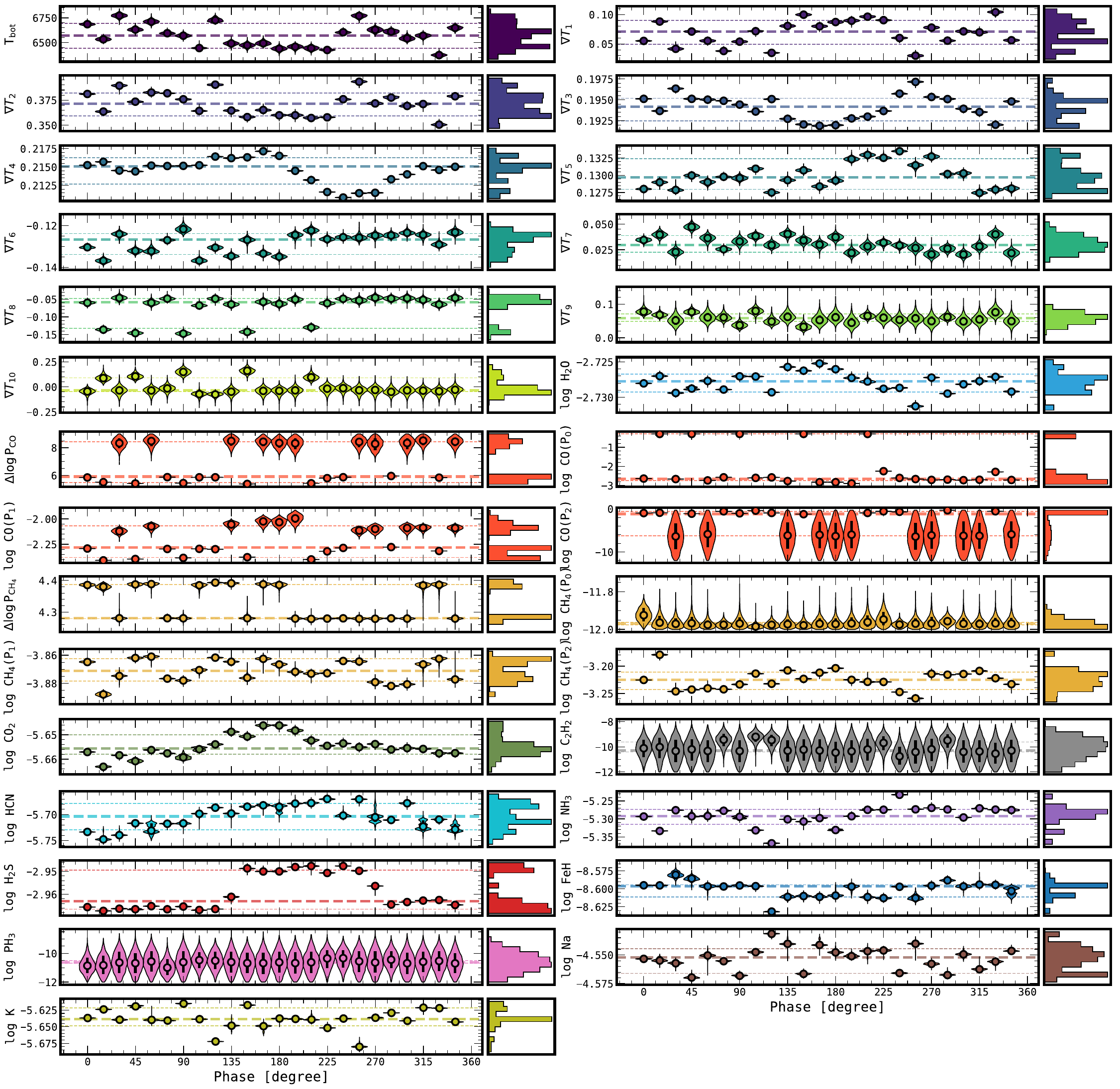}
    \caption{Measurements of all atmospheric parameters as a function of phase for the fixed cloud retrieval, which best reproduced the observed variability. Also indicated are the median and $\pm1\sigma$ confidence intervals for each set of measurements. }
    \label{fig:param_variation}
\end{figure}

\begin{figure*}
    \centering
    \includegraphics[width=0.88\linewidth]{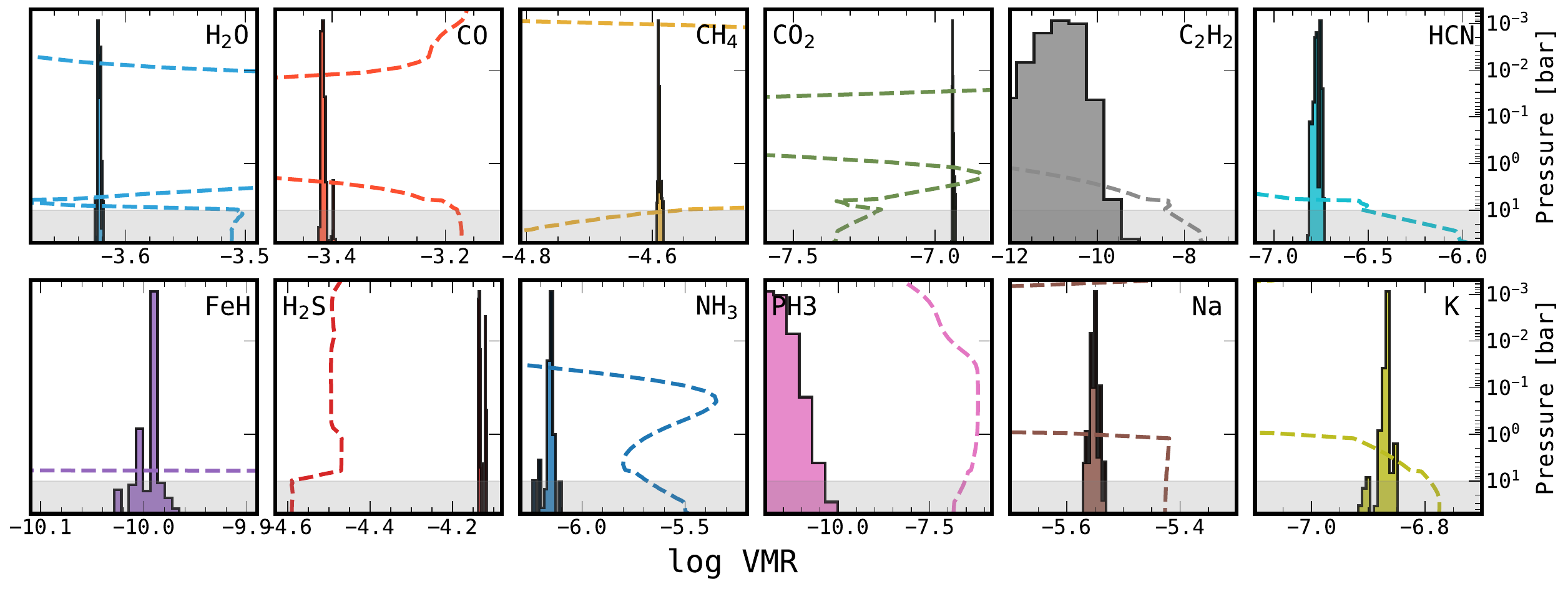}
    \caption{Combined volume mixing ratios of gas-phase species from all phases for the fixed-cloud retrieval setup. Shown in dashed lines are the abundance profiles in equilibrium for [M/H] = 0.18 and C/O=0.55. The grey shading indicates pressures at which the retrievals are insensitive to the chemical abundances. }
    \label{fig:vmr_hists}
\end{figure*}

\begin{figure*}
\centering
    \includegraphics[width=0.87\linewidth]{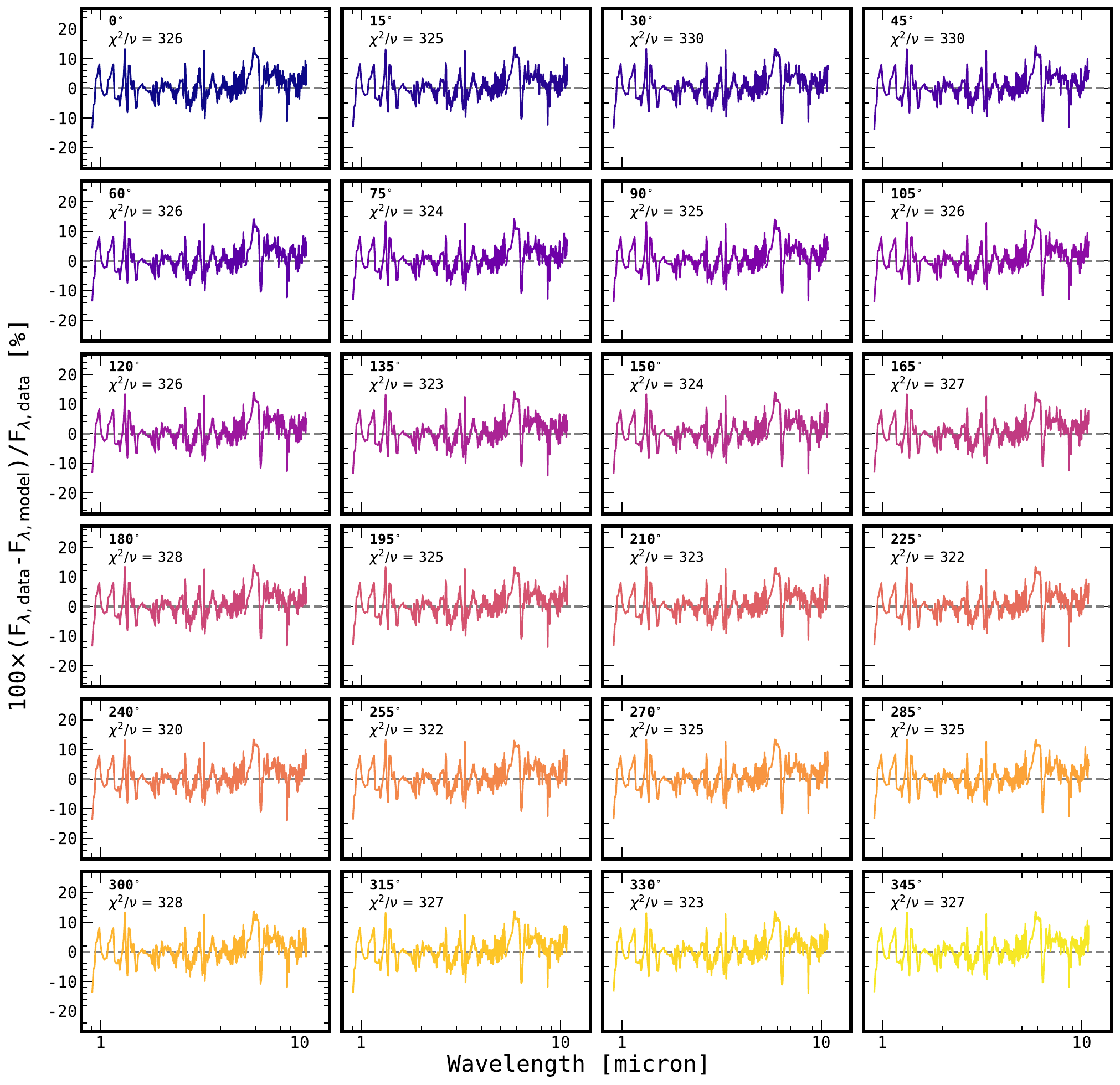}
    \caption{Residuals at all phases between the JWST data and the best-fit models for the fixed-cloud retrieval setup. Uncertainties on data too small to see at this scale, typically about 0.1\%. The structure present across phase indicates that the primary source of the residual discrepancy is systematic in nature, due to either incomplete modelling or artefacts from the data processing.}
    \label{fig:residuals_all}
\end{figure*}

\end{appendix}
\end{document}